\documentclass{article}

\usepackage{arxiv}

\usepackage[utf8]{inputenc} 
\usepackage[T1]{fontenc}    
\usepackage{lineno,hyperref}       
\usepackage{url}            
\usepackage{booktabs}       
\usepackage{amsfonts}       
\usepackage{nicefrac}       
\usepackage{microtype}      
\usepackage{graphicx}
\usepackage{natbib}
\usepackage{doi}
\usepackage{xcolor}
\usepackage{soul}
\usepackage{float}
\usepackage{amsmath,amssymb,amsfonts}
\usepackage[thinc]{esdiff}
\usepackage{algorithmic}
\usepackage{algorithm}
\usepackage{array}
\usepackage[font=small,captionskip=0cm]{subfig}
\usepackage{textcomp}
\usepackage{stfloats}
\usepackage{makecell}
\usepackage{mathabx}
\usepackage{tikz}

\title{A Comparative Study of Oscillatory Perturbations in Car-Following Models%
\thanks{This work has been submitted to the IEEE for possible publication. Copyright may be transferred without notice, after which this version may no longer be accessible.}}

\date{} 					

\author{ \href{}{Oumaima Barhoumi}\\
	Department of Electrical \\and Computer Engineering\\
	Concordia University\\
	Montral, Quebec, Canada \\
	\texttt{o\_barh@ece.concordia.ca} \\
	\And
\href{}{Ghazal Farhani} \\
	National Research Council Canada\\
    Automotive and Surface Transportation \\Research Centre\\
	London, Ontario, Canada\\
	\texttt{ghazal.farhani@nrc-cnrc.gc.ca} \\
    \And
\href{}{Taufiq Rahman} \\
	National Research Council Canada\\
    Automotive and Surface Transportation \\Research Centre\\
	London, Ontario, Canada \\
	\texttt{taufiq.rahman@nrc-cnrc.gc.ca} \\
    \And
\href{}{Mohamed H. Zaki} \\
	Department of Civil \\and Environmental Engineering\\
	Western University\\
	London, Ontario, Canada \\
	\texttt{mzaki9@uwo.ca} \\
    \And
\href{}{Sofiène Tahar} \\
	Department of Electrical \\and Computer Engineering\\
	Concordia University\\
	Montral, Quebec, Canada \\
	\texttt{tahar@ece.concordia.ca} \\
    }

\renewcommand{\shorttitle}{A Comparative Study of Oscillatory Perturbations in Car-Following Models}

\hypersetup{
pdftitle={A Comparative Study of Oscillatory Perturbations in Car-Following Models},
pdfsubject={q-bio.NC, q-bio.QM},
pdfauthor={Oumaima Barhoumi, Ghazal Farhani, Taufiq Rahman, Mohamed H. Zaki, and Sofiène Tahar},
pdfkeywords={Connected autonomous vehicles, car-following models, oscillatory perturbations, platooning, platoon instability},
}

\begin{document}
\maketitle

\begin{abstract}
As connected and autonomous vehicles become more widespread, platooning has emerged as a key strategy to improve road capacity, reduce fuel consumption, and enhance traffic flow. However, the benefits of platoons strongly depend on their ability to maintain stability. Instability can lead to unsafe spacing and increased energy usage. In this work, we study platoon instability and analyze the root cause of its occurrence, as well as its impacts on the following vehicle. To achieve this, we propose a comparative study between different car-following models such as the Intelligent Driver Model (IDM), the Optimal Velocity Model (OVM), the General Motors Model (GMM), and the Cooperative Adaptive Cruise Control (CACC). In our approach, we introduce a disruption in the model by varying the velocity of the leading vehicle to visualize the behavior of the following vehicles. To evaluate the dynamic response of each model, we introduce controlled perturbations in the velocity of the leading vehicle—specifically, sinusoidal oscillations and discrete velocity changes. The resulting vehicle trajectories and variations in inter-vehicle spacing are analyzed to assess the robustness of each model to disturbance propagation. The findings offer insight into model sensitivity, stability characteristics, and implications for designing resilient platooning control strategies.
\end{abstract}


\section{INTRODUCTION}
\label{intro}
The future of transportation is highly dependent on the advancement of intelligent transportation systems technology to develop connected autonomous vehicles, automated highways, and reliable traffic~\cite{hancock2019future}. Towards this objective, microscopic models~\cite{smith2008large}, particularly car-following models~\cite{rothery1992car}, that describe the processes in which drivers follow each other in the traffic stream, have been studied for almost half a century{~\cite{chandler1958traffic}}. These models are fundamental in the replication of traffic flow. Hence, an accurate replication is essential for understanding and predicting traffic dynamics, which enables the design and optimization of transportation systems. In this context, the study of platoons~\cite{lesch2021overview} is a suitable technology that can improve road safety, mitigate congestion, increase throughput, be energy-efficient, and reduce costs, making it one of the leading research axes to focus on~\cite{sivanandham2020platooning}. 

A platoon consists of a string of following-preceding vehicles traveling at the leader's speed while keeping a constant pre-defined distance between one another~\cite{lesch2021overview}. To optimize throughput and minimize fuel consumption, vehicles in a platoon are spaced just a few meters apart. However, maintaining such close proximity is challenging due to both internal and external disturbances, leading to what is known as \textit{platoon instability}. Platoon instability~\cite{feng2019string} refers to the phenomenon in which a group of vehicles traveling together in a coordinated manner becomes unstable, leading to unsafe or inefficient driving conditions and increasing fuel consumption~\cite{zhou2023experimental,barhoumi2025fuel}. This instability can manifest in various ways, such as high-speed fluctuations, unsafe following distances, or even collisions~\cite{feng2019string}. Hence, understanding and managing platoon instability is crucial for the safe and efficient operation of autonomous or semi-autonomous vehicle platoons, especially in scenarios such as highway driving, where vehicles are often in close proximity~\cite{de1999stability}. 

According to the study by~\cite{feng2019string}, platoon instability is typically caused by various disturbances or perturbations. This work focuses explicitly on Type I disturbances, which arise from initial condition perturbations affecting the leading vehicle~\cite{feng2019string}. These disturbances are particularly relevant for evaluating the fundamental stability characteristics of car-following models. 

Despite the central role of car-following models in the design of platoon control systems, few studies, such as~\cite{sun2020relationship,xiaomei2007stability,tanaka2008asymptotic,dai2022exploring}, have quantitatively and systematically compared how well these models mitigate lead-vehicle disturbances. Most existing work focuses on tuning or analyzing individual models in isolation, leaving practitioners without clear guidance on which models best suppress traffic perturbations — a critical factor for ensuring safety, comfort, and fuel efficiency. Furthermore, in mixed traffic environments—where Connected Autonomous Vehicles (CAVs), human-driven vehicles, and possibly heavy-duty vehicles share the road—several challenges arise. Human drivers exhibit a wide range of behaviors, from cautious to aggressive, and their actions are often influenced by distractions, emotional states, or incomplete information. Heavy-duty vehicles introduce an additional layer of complexity due to their slower acceleration, longer stopping distances, and larger blind spots. CAVs, in contrast, operate with consistent, rule-based decision-making but can misinterpret or fail to predict irregular human maneuvers. These differences in reaction times, acceleration capabilities, and driving logic can lead to instability within platoons, where small disturbances may amplify along the vehicle string, reducing safety, degrading traffic flow, and diminishing the potential benefits of coordinated driving.

To address these gaps, in this paper, we propose a unified benchmark of four widely used car-following models—Optimal Velocity Model (OVM)~\cite{bando1995dynamical}, Intelligent Driver Model (IDM)~\cite{treiber2000congested}, the General Motors Model (GMM)~\cite{bando1995dynamical}, and the Cooperative Adaptive Cruise Control (CACC)~\cite{lu2002acc}—subjected to identical Type I disturbances under controlled simulation conditions. In addition to one-time perturbations, we introduce sinusoidal oscillations—both harmonic and non-harmonic—to evaluate how each model responds to periodic disturbances. Studying sinusoidal inputs is essential, as they mimic real-world scenarios such as speed fluctuations due to road curvature, traffic signals, or driver behavior. By analyzing model performance across a range of communication delays (0 ms to 1.5 s), we quantify each model’s ability to absorb spacing waves and identify key performance trade-offs. Our findings reveal which models remain robust under latency, how tuning parameters shift stability margins, and when simpler models may outperform more complex ones. These insights offer practical guidelines for deploying safe, stable, and energy-efficient platooning systems in real-world conditions.

The rest of the paper is structured as follows. In Section{~\ref{Sec:prelim}}, we provide the definitions as well as the mathematical formulations for the OVM, IDM, GMM, and CACC models. Section{~\ref{Sec:rw}} presents prior work employing the proposed models. This is followed by a comprehensive overview of string stability, string instability, and spacing error propagation. Section{~\ref{methodo}} presents an experimental study that investigates the effects of various perturbations on platoon instability, focusing on the behavior of the following vehicle and how these perturbations influence vehicle spacing and velocities within the platoon. Finally, Section~\ref{conc} concludes the paper by summarizing the findings and exploring directions for future research. 

\section{PRELIMINARIES}
\label{Sec:prelim}
In this section, we briefly explain the OVM, IDM, GMM, and CACC models along with their mathematical definitions.

\subsection{Optimal Velocity Model (OVM)}
OVM is a car-following model that models the behavior of vehicles as they adjust their velocity based on their distance from the car ahead, striving for a balance between safety and traffic efficiency~\cite{bando1995dynamical}. OVM explicitly models the interaction between a vehicle and the one directly ahead by adjusting the following vehicle's velocity based on the headway distance (distance to the leading car). This model focuses on achieving a velocity determined by a nonlinear function of headway distance. Therefore, it incorporates a feedback mechanism, where the acceleration of the vehicle depends on the difference between its current velocity and the desired ``optimal velocity" as given by~\cite{bando1995dynamical}:
\begin{equation}
   \diff{v}{t} = \alpha[V_{opt}(\Delta x) - v(t)]
    \label{motion_OVM}
\end{equation}where $\alpha$ is the driver sensitivity parameter, $v(t)$ is the velocity of the vehicle at time \textit{t}, $\Delta x$ is the space headway, i.e., the distance separating two following vehicles, given as $\Delta x = x_{n+1} - x_n$.
The velocity, $V_{opt}(d)$, is defined as a function of $\Delta x$, which is usually a monotonically increasing function, such as:
\begin{equation}
   V_{opt}(\Delta x) = v_0 \cdot (\tanh{(\Delta x - h)} + \tanh{(h)}),
    \label{vopt_OVM}
\end{equation} where $v_0$ is the desired velocity, \textit{h} is a critical parameter ensuring that the model is more realistic and reflects the expected behavior of vehicles.

\subsection{Intelligent Driver Model (IDM)}
IDM is a car-following model used in traffic flow simulations to describe vehicle behavior in response to each other on the road. This model captures realistic driving behavior by considering the ego-vehicle dynamics such as speed, acceleration, and distance to the vehicle in front~\cite{jia2015survey}. IDM is defined mathematically through a set of equations describing two components, i.e., vehicle's acceleration and safety distance. We provide a breakdown of the mathematical aspects. Eq.~(\ref{acc_comp})~\cite{treiber2000congested} below represents the first term, which is the acceleration component, and indicates how the vehicle's speed affects acceleration. If the vehicle is going too fast relative to its desired speed, acceleration will decrease.
\begin{equation}
    \dot{v}_{acc} = a\Bigg(1-\Big(\frac{v}{v_{des}}\Big)^\delta\Bigg),
    \label{acc_comp}
\end{equation}
where \textit{a} represents a proportionality constant, $v_{des}$ represents the desired speed of the vehicle, and $\delta$ is a parameter that controls the sensitivity of acceleration to speed. \\
The second term is depicted by Eq.~(\ref{dec_comp})~\cite{treiber2000congested}: 
\begin{equation}
    \dot{v}_{dec} = -a\Big(\frac{s^*}{s}\Big)^2,
    \label{dec_comp}
\end{equation} This term represents the safety distance ensuring that as the actual distance to the vehicle in front, \textit{s}, decreases, the acceleration becomes negative, promoting safe following distances. $s^*$ is the desired minimum distance separating two-following vehicles. It can be expressed as~\cite{treiber2000congested}: 
\begin{equation}
    s^* = s_0 + s_1 \cdot \sqrt{\frac{v}{v_{des}}} + T_{react}\cdot v + \frac{v\Delta v}{2\sqrt{a b}},
    \label{des_dist}
\end{equation}where $\Delta v$: relative speed to the lead car, $b$: desired deceleration rate, a tuning parameter for braking sensitivity, $s_0$ and $s_1$ represent the minimal spacing and the jam spacing between vehicles, respectively. For a homogeneous study, traffic conditions are assumed to be smooth with no congestion, i.e., $s_1 = 0$, and the desired spacing is simplified to~\cite{treiber2000congested}: 
\begin{equation}
    s^* = s_0 + T_{react}\cdot v + \frac{v\Delta v}{2\sqrt{a b}}.
    \label{des_dist_simp}
\end{equation}
The mathematical formulation of the IDM model is given as follows~\cite{treiber2000congested}:
\begin{equation}
    \dot{v} = a\Bigg(1-\Big(\frac{v}{v_{des}}\Big)^\delta - \Big(\frac{s^*}{s}\Big)^2\Bigg)
    \label{IDM}
\end{equation}

\subsection{General Motors Model (GMM)}
GMM, also known as GHR (Gazis-Herman-Rothery), was first developed by \textit{Chandler et al.}~\cite{chandler1958traffic} and \textit{Herman et al.}~\cite{herman1959traffic} at the General Motors research laboratory. GMM is based on a stimulus-response model following the \textit{follow-the-leader} concept~\cite{chandler1958traffic,herman1959traffic}, where a response of vehicle \(n\) is the result of a stimulus created by the vehicles ahead \((n-1)\). In this context, a stimulus can be a function of variable variations, and the response is acceleration, deceleration, or constant speed: 

\[
[Response]_n \propto [Stimulus]_n 
\]where \(\propto\) symbolizes a proportional relationship between the response and the stimulus and \(n\) is the vehicle under consideration for the response calculation.

The model can be reformulated as follows~\cite{chandler1958traffic}: 
\[
a_n(t) = f_{\text{stimulus}}\left( 
\underbrace{v_n}_{\text{speed}}, \quad
\underbrace{\Delta x_n}_{\text{spacing change}}, \quad
\underbrace{\Delta v_n}_{\text{velocity change}} 
\right)
\]

For a required gap:\\
\[
\Delta x_{n}(t) = \Delta x_{safe} + \alpha v_{n}(t),
\]
where \( \alpha \) is the sensitivity coefficient and \(v_{n}(t)\) is the speed of the vehicle \(n\).

Given that: \(
\Delta x_{n}(t) = x_{n-1}(t) - x_{n}(t)\), we get:
\[
x_{n-1}(t) - x_{n}(t) = \Delta x_{safe} + \alpha v_{n}(t)
\]
with
\( x_{n}(t) \) and \( x_{n-1}(t) \): the positions of ego vehicle \( n \) and its leading vehicle \( n-1 \). By deriving over time, we achieve:
\[
v_{n-1}(t) - v_{n}(t) = \alpha a_{n}(t) \quad \Rightarrow a_{n}(t) = \frac{1}{\alpha} [v_{n-1}(t) - v_{n}(t)]
\]
with \( v_{n}(t) \) and \( a_{n}(t) \) are the speed and acceleration of vehicle \( n \), respectively.

Given the various forms of the sensitivity coefficient, $\alpha$, in the literature~\cite{gazis1961nonlinear,gazis1959car,nishinari2014traffic}, the GMM model is generalized as follows~\cite{gazis1961nonlinear}:
\begin{equation}
    a_n(t) = \alpha \frac{v_n(t)^m}{(x_{n-1}(t) - x_n(t))^l} (v_{n-1}(t) - v_n(t))
    \label{gmm}
\end{equation}
where \( a_n(t) \) represents the acceleration of the following vehicle, \( v_n(t) \) and \( v_{n-1}(t) \) denote the speeds of the following and leading vehicles, respectively, and \( x_{n-1}(t) - x_n(t) \) is the distance between the two vehicles. The parameters \( \alpha \), \( m \), and \( l \) correspond to a sensitivity coefficient, a spacing exponent typically in the range \([-1, 4]\), and a speed exponent usually in the range \([-2, 2]\), respectively.

\subsection{Cooperative Adaptive Cruise Control (CACC)}
CACC~\cite{shladover2015cooperative} is an advanced driver-assistance system (ADAS) that enhances traditional Adaptive Cruise Control (ACC). Unlike ACC, which only adjusts a vehicle's speed based on the distance to the car ahead, CACC enables vehicles to communicate with each other (vehicle-to-vehicle or V2V communication) and with infrastructure (vehicle-to-infrastructure or V2I communication). This enables real-time coordination of speed and braking, thereby improving traffic flow, fuel efficiency, and safety~\cite{naus2010string}. CACC generally consists of two main controllers working together, one to manage the approaching maneuver to the leading vehicle, known as the \textit{Gap-Closing Controller}, and another to regulate car-following once the vehicle joins the platoon, called the \textit{Gap Regulation Controller}~\cite{milanes2013cooperative}. These controllers work together to ensure smooth, safe, and efficient operation within the CACC system. The design of a CACC system starts with defining a spacing policy, which specifies the desired following distance as a function of speed. There are two types of spacing policies: constant spacing policy and variable spacing policy. Among variable spacing policies, the Constant Time Headway (CTH) policy has garnered significant attention in the literature~\cite{milanes2013cooperative}. The desired spacing \(d_{\text{des}}\), defined by the CTH policy, between the ego vehicle and the leading vehicle is given by~\cite{milanes2013cooperative}:
\begin{equation}
d_{\text{des}}(t) = r + h \cdot v_i(t)
\label{cth}
\end{equation}
where:
\begin{itemize}
    \item \( r \): Standstill distance (minimum gap at zero speed).
    \item \( h \): Time headway (constant time gap, e.g., 1-2 seconds).
    \item \( v_i(t) \): Velocity of the ego vehicle.
\end{itemize}
In this context, the spacing error \( e_i(t) \) is the difference between the desired spacing and the actual spacing:

\begin{equation}
e_i(t) = d_{\text{des}}(t) - d_i(t)
\label{cth_error}
\end{equation}
where:
\begin{itemize}
    \item \( d_i(t) = x_{i-1}(t) - x_i(t) \): Actual distance to the leading vehicle.
    \item \( x_{i-1}(t) \): Position of the leading vehicle.
\end{itemize}
The CACC control law is the mathematical rule that determines how a following vehicle adjusts its acceleration to maintain a safe and efficient distance from the vehicle directly in front of it~\cite{ploeg2011design}. CACC uses V2V communication to share real-time information (e.g., speed, acceleration, and position) between vehicles, enabling more precise and coordinated control. The general CACC control law incorporates communication delay (\( \tau \)) and feedback from the leading vehicle as given below:
\begin{equation}
\begin{aligned}
a_i(t) &= k_p \cdot e_i(t) + k_d \cdot \dot{e}_i(t) \\
&\quad + k_v \cdot \left(v_{i-1}(t - \tau) - v_i(t)\right) \\
&\quad + k_a \cdot a_{i-1}(t - \tau)
\end{aligned}
\label{cacc}
\end{equation}
where \( k_p \) is the proportional gain for spacing error, \( k_d \) is the derivative gain for the rate of change of spacing error, \( k_v \) is the gain for the velocity difference, and \( k_a \) is the gain for the acceleration of the leading vehicle. The term \( v_{i-1}(t - \tau) \) denotes the velocity of the leading vehicle received with a communication delay \( \tau \), and \( a_{i-1}(t - \tau) \) is its corresponding acceleration. The delay \( \tau \) accounts for the latency introduced by the V2V communication.

\section{RELATED WORK} 
\label{Sec:rw}
In this Section, we focus on previous studies investigating the effect of perturbation on platoon stability in the cases of IDM, OVM, GMM, and CACC car-following models.

\subsection{Intelligent Driver Model}
In~\cite{chen2012behavioral}, \textit{Chen et al.} developed a behavioral car-following model based on empirical trajectory data, capable of replicating the spontaneous formation and propagation of stop-and-go waves in congested traffic. The study specifically examined the influence of rubbernecking on oscillation patterns, confirming through simulation that it triggers the observed traffic waves. Additionally, a clear correlation was identified between driver behavior prior to and during oscillations. The authors suggest that their proposed asymmetric behavioral model is suitable for quantifying the safety and fuel consumption impacts of traffic oscillations. Complementing this work, \textit{Sun et al.}~\cite{sun2020relationship} explored the connection between string instability and traffic oscillations, categorizing them into four distinct types using next-generation simulation program (NGSIM)~\cite{alexiadis2004next} data and simulations with the IDM. Their findings highlight how vehicle-level stability parameters—such as the desired time gap, maximum acceleration, and desired speed— affect oscillation development. While improvements in individual vehicle stability were shown to reduce oscillation severity, the lack of reaction time modeling in IDM may limit the accuracy of the analysis in capturing real-world driver delay effects. In reality, human reaction times play a crucial role in vehicle interactions and can influence the severity and propagation of oscillations.

\subsection{Optimal Velocity Model}
Several studies have investigated the behavior of platoon dynamics under perturbations, particularly focusing on the OVM and its limitations. An early study in{~\cite{davis2003modifications}} explored the impact of realistic driver reaction time delays, showing that small or vanishing delays induce high-frequency, non-physical oscillations. To address this issue, the author proposed modifying the OVM to incorporate more realistic delays, successfully suppressing these oscillations while preserving platoon stability. Building on these foundational insights, more recent studies have further examined platoon responses to perturbations. For instance, \textit{Matin et al.}{~\cite{matin2020nonlinear}} analyzed a two-vehicle system subject to deterministic and stochastic velocity perturbations in the lead vehicle, demonstrating that the system converges in both scenarios and can be approximated by an averaged model. They further found that convergence appears more natural under fast deterministic perturbations. In a subsequent study{~\cite{matin2020nonlinear1}}, the authors extended their analysis by quantifying the convergence order and associating the averaged system with a stable default state. However, both works omit critical details about the nature of the perturbations—specifically their duration and amplitude—which limits their real-world applicability.

In a related study, \textit{Jin et al.}~\cite{jin2020dynamical} incorporated delayed feedback control into the OVM framework introduced in~\cite{orosz2006subcritical}, adding a feedback gain to stabilize the system. Despite the added control mechanism, their simulations revealed that delayed responses still allow small disturbances to grow into large oscillations. While both~\cite{davis2003modifications} and~\cite{jin2020dynamical} propose delay-aware strategies to improve traffic stability, neither explicitly addresses the safety implications of inter-vehicle interactions under abrupt perturbations such as sudden braking or lane changes. Complementing these works, \textit{Laval et al.}~\cite{laval2014parsimonious} investigated the role of randomness in driver behavior, proposing a model that combines bounded acceleration with human error. By introducing white noise into the desired acceleration in free-flow traffic, their model reproduces stop-and-go oscillations, highlighting the critical role of behavioral variability in the emergence and propagation of traffic instabilities. 

\subsection{General Motors Model (GMM)}
The limitations of GMM car-following models in accurately representing real-world traffic dynamics and driver behavior have been widely studied. \textit{Chakroborty et al.}\cite{chakroborty1999evaluation} emphasized the model’s inability to replicate realistic driving responses, particularly in the presence of traffic oscillations, and proposed a fuzzy inference-based alternative that better captures the uncertainty and imprecision in human decision-making. Likewise, \textit{Sun et al.}\cite{sun2016new} identified that the GMM’s excessive sensitivity to speed differences and its delayed reaction to changes in traffic lead to unnatural acceleration-deceleration patterns, ultimately amplifying oscillations. They addressed this by developing a modified model that incorporates both the relative speed and the leader’s acceleration, resulting in more natural and stable driving behavior. Building on this foundation, \textit{Jin et al.}~\cite{jin2013bidirectional} introduced a bidirectional car-following approach, modifying traditional GMM models to integrate partial feedback from following vehicles (5–20\%). Calibrated with real-world NGSIM data~\cite{alexiadis2004next}, this bidirectional model demonstrated enhanced string stability and significantly reduced oscillation propagation (by 30–40\%). 

\subsection{Cooperative Adaptive Cruise Control (CACC)}
Several studies have highlighted the destabilizing effects of communication delays in the CACC platoons. For instance, in the study conducted by \textit{Liu et al.}\cite{liu2020safety}, a modified CACC model incorporating communication delays was proposed, and microscopic simulations showed that it reduced inter-vehicle spacing fluctuations by approximately 96.6\% while enhancing vehicle responsiveness to upstream disturbances. Similarly, \textit{Naus et al.}\cite{naus2010string} presented a CACC system using wireless communication with the preceding vehicle and derived string stability conditions, demonstrating that velocity-dependent spacing maintains stability better than constant spacing. Real-world experiments confirmed CACC's ability to dampen oscillations more effectively than ACC. In~\cite{tian2019modeling}, the authors analyzed how communication delays induce string instability, using a Laplace-domain method to identify critical delay thresholds where oscillations shift from damped to unstable. The study also showed how varying disturbance frequencies impact oscillation behavior, emphasizing the importance of tuning controller parameters to preserve platoon stability under real-world delay conditions.

\subsection*{Shortcomings in Prior Work}
Although previous studies have explored the impact of perturbations, delays, and behavioral factors on platoons, they often examine isolated models or focus solely on achieving theoretical stability. A critical limitation in existing studies on traffic oscillations in platoons is the lack of systematic characterization of perturbation dynamics, including their amplitude, duration, and propagation patterns. While many models (e.g., IDM, OVM, GMM) analyze stability under idealized conditions, they often neglect real-world disturbances such as abrupt braking, merging vehicles, or sensor noise, which can trigger or amplify oscillations. Another key gap is the inadequate modeling of human factors, such as delayed reactions, perception errors, and behavioral variability, which significantly influence perturbation responses. Furthermore, while CACC systems improve stability, their reliance on communication networks introduces delays and packet losses that are rarely quantified in terms of oscillation severity. 

In contrast to prior studies, this work conducts a systematic comparison of multiple car-following models—including IDM, OVM, GMM, and CACC-based approaches—to evaluate their effectiveness in mitigating Type I disturbances (e.g., sudden braking, acceleration mismatches) within platoons. Furthermore, our study explicitly incorporates realistic reaction delays and communication lags, assessing their impact on string stability, responsiveness, and oscillation damping. 

A key contribution is the evaluation of platoon behavior under discrete, harmonic, and non-harmonic perturbations, addressing a critical gap in the literature. Given that real-world traffic often experiences impulsive (discrete) or irregular (non-harmonic) perturbations, which can propagate unpredictably, our analysis reveals how different models handle these perturbation types, particularly in terms of Amplitude Growth, Phase Lag Effects, and Nonlinear Damping Behavior. By benchmarking model performance across these scenarios, we identify which frameworks offer the most robust stability guarantees under realistic conditions. This comparative perspective provides actionable insights for designing resilient platoon control strategies that minimize traffic oscillations while maintaining safety and efficiency.

\section{STRING STABILITY PERTURBATIONS: MATHEMATICAL MODELING}
In the context of platooning, perturbations can be analyzed to understand how small disturbances (\textit{e.g.}, changes in speed, spacing, or external factors) propagate through the platoon. Mathematical modeling provides a precise framework to represent and analyze these dynamics. By expressing vehicle behaviors through differential equations, we can rigorously analyze the dynamics of vehicle interactions, simulate a wide range of realistic driving conditions, and derive generalizable insights that are difficult to obtain from field experiments alone. Below is a step-by-step explanation of the mathematical formulation for oscillatory perturbations in platooning. 

In a platoon, the motion of each vehicle $n$ at time $t$ can be described by its dynamics, such as position $x_n$, velocity $v_n$, and acceleration $\dot{v}_n$. In a steady state, all vehicles in the platoon travel at the same constant velocity \( v_0 \) and maintain a constant spacing \( s_0 \) between consecutive vehicles. This implies:
\[
\forall n, \quad v_n(t) = v_0  
\]
\[
\forall n, \quad x_n(t) - x_{n-1}(t) = s_0
\]where \(\forall\) symbolizes that the spacing condition holds for every car 
\(n\) in the platoon.
If a vehicle \( n \) in the platoon undergoes a perturbation, its space gap and velocity are directly impacted. Furthermore, this deviation will also propagate to the rest of the following vehicles, potentially leading to platoon instability. The deviations in space gap and velocity for vehicle \( n \) can be expressed, respectively, as:
\begin{equation}
y_n(t) = s_n(t) - s_e = x_{n-1}(t) - x_n(t) - s_e
\label{eq:space_gap}
\end{equation}
\begin{equation}
u_n(t) = v_n(t) - v_e 
\label{eq:velocity_deviation}
\end{equation}
where:
\begin{itemize}
    \item \( y_n(t) \) and \( u_n(t) \) are the deviations of the space gap and velocity of vehicle \( n \), respectively.
    \item \( x_{n-1}(t) \) and \( x_n(t) \) represent the positions of the preceding vehicle \( n-1 \) and the current vehicle \( n \), respectively, at time \( t \).
    \item \( s_e \) is the equilibrium spacing.
    \item \( v_e \) is the equilibrium velocity.
\end{itemize}
Taking the first derivative of Eqs.~(\ref{eq:space_gap}) and (\ref{eq:velocity_deviation}):
\begin{equation}
\dot{y}_n(t) = \dot{x}_{n-1}(t) - \dot{x}_n(t) = v_{n-1}(t) - v_n(t)
\label{per}
\end{equation}
\begin{equation}
    \dot{u}_n(t) = \dot{v}_n(t)
    \label{pert}
\end{equation}
Given that the perturbation affected only vehicle $n$, the preceding vehicle $n-1$ maintains its equilibrium speed, i.e., $v_{n-1} = v_e$, Eq.~(\ref{per}) can be rewritten as: 
\begin{equation}
    \dot{y}_n(t) = -u_{n}(t) 
    \label{pert1}
\end{equation}
Taking the first derivative of Eq.~(\ref{pert1}), we obtain:
\begin{equation}
    \ddot{y}_n(t) = -\dot{u}_{n}(t)
    \label{pert2}
\end{equation}
\paragraph{General Form of Harmonic Damped Oscillations}
A general representation of the acceleration control function for most car-following models can be written as: 
\begin{equation}
    \dot{v}(t) = f(v_n(t), \Delta v_n(t), s_n(t)) 
    \label{gen-form}
\end{equation}where \(\dot{v}_n(t)\) represents the acceleration of the \(n\)-th vehicle at time \(t\), \(v_n(t)\) denotes the velocity of the \(n\)-th vehicle at time \(t\), \(\Delta v_n(t)\) is the relative velocity between the \(n\)-th vehicle and the preceding vehicle at time \(t\), \(s_n(t)\) is the space gap between the \(n\)-th vehicle and the preceding vehicle at time \(t\), and \(f(\cdot)\) is the acceleration control function that determines the vehicle's response based on its velocity, relative velocity, and space gap.
Using a first-order Taylor series approximation of Eq.~(\ref{gen-form}):
\begin{multline}
        v_n(t) = f(v_e, 0, s_e) + f_v^n (v_n(t) - v_e)\\ + f_{\Delta v}^n \Delta v_n(t) + f_{s}^n (s_n(t) - s_e)
    \label{taylor}
\end{multline}
where $f(v_e, 0, s_e) = 0$.\\
Substituting Eqs.~(\ref{eq:space_gap}) and~(\ref{gen-form}) into Eq.~(\ref{taylor}), we derive the general expression for harmonic damped oscillations:
\begin{equation}
    -\ddot{y}_n(t) = f_v^n (-\dot{y}_n(t)) + f_{\Delta v}^n \dot{y}_n(t) + f_{s}^n y_n(t)
    \label{taylor_res}
\end{equation}
Eq.~(\ref{taylor_res}) can be rearranged into the following form:
\begin{equation}
    \ddot{y}_n(t) + (f_{\Delta v}^n - f_v^n)\dot{y}_n(t) + f_{s}^n y_n(t) = 0 
    \label{taylor_rearr}
\end{equation}
\paragraph{Solution of the General Form of Harmonic Damped Oscillations}
Eq.~(\ref{taylor_rearr}) reflects a harmonic damped oscillator given as follows:
\begin{equation}
    \ddot{y}_n(t) + 2\xi_n \omega_{0}^n \dot{y}_n(t) + (\omega_{0}^n)^2 y_n(t) = 0
    \label{harmonic}
\end{equation}where \[
\xi_n = \frac{\, f_{\Delta v}^n - f_v^n \,}{2\sqrt{\, f_s^n \,}} 
\quad \text{and} \quad 
\omega_0^n = \sqrt{\, f_s^n \,}
\]where $\xi_n$ and $\omega_0^n$ represent the damping ratio and natural frequency of the $n$-th vehicle, respectively. In this context, Eq.~(\ref{harmonic}) takes the form of an exponential function, given by:
\begin{equation}
    y_n(t) = y_{n,0} e^{\lambda t}
    \label{solution}
\end{equation}where \( y_{n,0} \) is the initial condition of \( y_n(t) \), and \( \lambda \) represents the complex growth rate.
Solving Eq.~(\ref{harmonic}) results in:
\begin{equation}
    \lambda_{1,2} = \omega_{0}^n \left[ -\xi_n \pm \sqrt{(\xi_n)^2 - 1} \right]
    \label{sol-lambda}
\end{equation}where {$\lambda_{1,2}$} denote the eigenvalues of the linearized system.
\paragraph{Stability Analysis}
The car-following behavior of vehicle \( n \) is locally stable if the real parts of \( \lambda_{1,2} \) are negative (\( \text{Re}(\lambda_{1,2}) < 0 \)), which requires \( \xi_n > 0 \). For car-following models without time delays, stability is unconditional iff the constraints \( f_{\Delta v}^n - f_v^n >0 \) and \( f_n^s \geq 0 \) are met. The imaginary parts of \( \lambda_{1,2} \) determine the oscillatory nature of \( y_n(t) \): if \( \text{Im}(\lambda_{1,2}) = 0 \) (i.e., \( \xi_n \geq 1 \)), \( y_n(t) \) decays without oscillation. Based on the damping ratio \( \xi_n \), the local stability of car-following models can be classified into three categories (as depicted by Figure~\ref{fig:damping}): 
\begin{enumerate}
    \item \textbf{Underdamped (\( 0 < \xi_n < 1 \))}: The space gap and velocity of vehicle \( n \) oscillate around their equilibrium states with gradually decreasing amplitudes.
    
    \item \textbf{Critically Damped (\( \xi_n = 1 \))}: The space gap and velocity of vehicle \( n \) return to their equilibrium states without oscillation.
    
    \item \textbf{Overdamped (\( \xi_n > 1 \))}: The space gap and velocity of vehicle \( n \) return to their equilibrium states without oscillation but more slowly than in the critically damped case.
\end{enumerate}
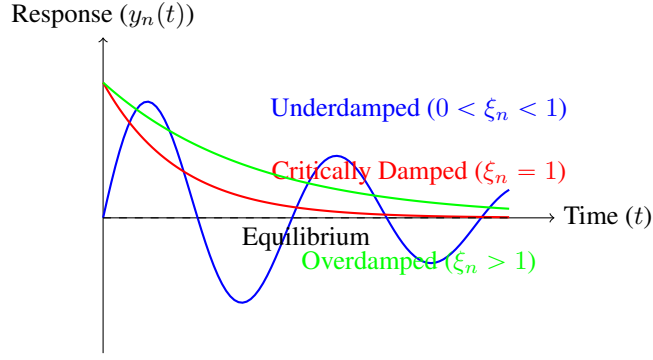
\begin{figure}[htb!]
\centering
\begin{tikzpicture}[scale=1.2]
    \draw[->] (0,0) -- (5,0) node[right]{Time (\( t \))};
    \draw[->] (0,-1.5) -- (0,2) node[above]{Response (\( y_n(t) \))};

    \draw[blue, thick, domain=0:4.5, samples=100] plot (\x, {1.5*exp(-0.3*\x) * sin(3*\x r)});
    \node[blue] at (3.5,1.2) {Underdamped (\( 0 < \xi_n < 1 \))};

    \draw[red, thick, domain=0:4.5, samples=100] plot (\x, {1.5*exp(-1.2*\x)});
    \node[red] at (3.5,0.5) {Critically Damped (\( \xi_n = 1 \))};

    \draw[green, thick, domain=0:4.5, samples=100] plot (\x, {1.5*exp(-0.6*\x)});
    \node[green] at (3.5,-0.5) {Overdamped (\( \xi_n > 1 \))};

    \draw[dashed] (0,0) -- (4.5,0) node[midway, below]{Equilibrium};
\end{tikzpicture}
\caption{Response of the car-following model for different damping ratios \( \xi_n \).}
\label{fig:damping}
\end{figure}
The local stability analysis based on the damping ratio $\xi_n$ provides insight into how vehicle $n$ responds to small perturbations in spacing and velocity. To further characterize platoon behavior, it is useful to examine the steady-state spacing errors before and after a perturbation. In this context, the spacing error ($e$) is defined as:
\begin{equation}
    e = \underbrace{\left( x_{n-1}(t) - x_{n}(t) \right)}_{\substack{\text{Two-cars following} \\ \text{(distance between vehicles)}}} - \underbrace{\delta}_{\substack{\text{Desired gap} \\ \text{(constant)}}}
    \label{SE}
\end{equation} 
The spacing error for the steady state:
\begin{itemize}
    \item \textbf{Before perturbation:} 
    \begin{equation}
        e = \underbrace{\left( x_{n-1}(t) - x_{n}(t) - \delta \right)}_{\approx \quad \text{fixed value}} \approx 0
    \end{equation}
    \item \textbf{After perturbation:}
     \begin{align}
         e_1 &=  x_{n-1}(t) - x_{n}(t) - \delta  \\
         &= s_e + y_n(t) - \delta \nonumber \\
         &= e_0 + y_n(t) \nonumber
     \end{align}
     $y_n(t)$ represents the spacing error between each two cars. As we move towards the end of the platoon, the time accumulated to get the next vehicles becomes larger ($t_i$). Larger time delays can lead to amplified spacing errors ($y_n(t)$) toward the end of the platoon.
\end{itemize}
In terms of time Gap:
\begin{itemize}
    \item \textbf{Before perturbation:} 
    \begin{equation}
        T_H = \frac{s_0 - L}{v_n(t)}
    \end{equation}
    \item \textbf{After perturbation:}
     \begin{align}
         T'_H &= \frac{s_0 - L + y_n(t)}{v_n(t) + \dot{y}_n(t)} \quad \text{(since } u_n(t) = \dot{y}_n(t) \text{)} 
     \end{align}
\end{itemize}
As time goes by or we go through more vehicles $y_n(t)$ becomes smaller, which yields to $T'_H \to T_H$. In the case of continuous perturbation, Eq.~(\ref{harmonic}) becomes:
\begin{equation}
    \ddot{y}_n(t) + 2\xi_n \omega_{0}^n \dot{y}_n(t) + (\omega_{0}^n)^2 y_n(t) = F(t)
    \label{harmonic_unstable}
\end{equation}
In this case: \[y(t) = \underbrace{y_{\text{transient}}(t)}_{\text{previous calculations}} + \underbrace{y_{\text{steady}}(t)}_{\text{Solution: A cos $\omega$t + B sin $\omega$t}}\] 
Taking the first order derivative of the solution ($y_{\text{steady}}(t)$), we get:
\begin{equation}
\dot{y}(t) = - A \omega \sin \omega t + B \omega \sin \omega t
\end{equation}
Taking the second order derivative of the solution, we get:
\begin{equation}
\ddot{y}(t) = - A \omega^2 \cos \omega t - B \omega^2 \sin \omega t
\end{equation}
Replacing the solution and its two derivatives into Eq.~(\ref{harmonic_unstable}), we get:
\begin{equation}
\begin{split}
    -A&\omega^2\cos\omega t - B\omega^2\sin\omega t - 2\xi_n\omega_0^n(A\omega\sin\omega t - B\omega\cos\omega t) \\
    &+ (\omega_0^n)^2\big(A\cos\omega t + B \sin \omega t\big) = \cos\omega t
\end{split}
\end{equation}
\begin{equation}
\left.
\begin{aligned}
    &\underbrace{\left[-A\omega^2 + 2\xi \omega_0^n\omega B + (\omega_0^n)^2 A\right]}_{= 1} \cos\omega t \\
    &\underbrace{\left[-B\omega^2 - 2\xi \omega_0^n\omega A + (\omega_0^n)^2 B\right]}_{= 0} \sin\omega t
\end{aligned}
\right\}
\end{equation}By equating the coefficients of {$\cos \omega t$} and {$\sin \omega t$}, two coupled algebraic equations in {$A$} and {$B$} are obtained. Solving this system yields the steady-state amplitudes of the spacing and velocity oscillations in response to the sinusoidal perturbation. Specifically, {$A$} corresponds to the in-phase component of the response, while {$B$} captures the out-of-phase component, both of which depend on the damping ratio {$\xi$}, the natural frequency {$\omega_0^n$}, and the excitation frequency {$\omega$}.

Theoretically, in many car-following models, a near-zero reaction time ($\tau \approx 0$) improves responsiveness and can help the system dissipate initial perturbations. However, whether the system returns to equilibrium depends on additional factors, such as damping and model-specific sensitivity parameters. This behavior holds true not only for isolated vehicles but also for vehicle platoons, where stability is preserved under ideal conditions. Furthermore, in the case of continuous but small perturbations—reflecting realistic driving scenarios—the disturbance remains bounded and does not amplify as it propagates through the platoon. These analytical insights form the foundation for the simulations presented in the next section, where we evaluate the practical performance of each model under varying conditions of delay and disturbance.

\section{STRING STABILITY ANALYSIS}
\label{stability}
In this section, we conduct the mathematical calculation of the stability analysis for each model, namely IDM, OVM, GMM and CACC, and derive their stability conditions.

\subsection{String Stability Analysis of IDM Model}
\begin{figure}[!t]
        \centering
        \includegraphics[width=0.8\linewidth]{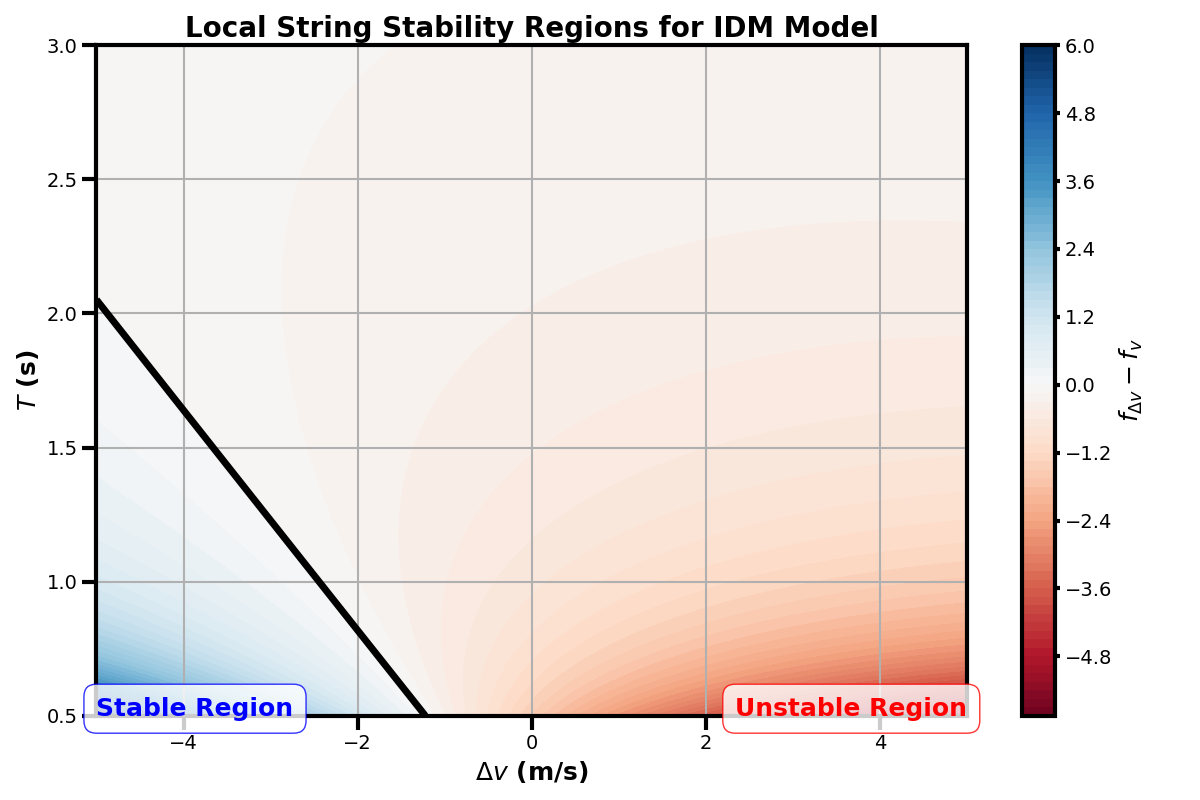} 
        \caption{Stability Map of IDM-based Platoon Dynamics in $(\Delta v, T)$ Space}
        \label{fig:heat_map}
\end{figure}

The IDM introduces smoothing and damping, which is the damping term, mathematically given by Eq.~(\ref{taylor_rearr}), to ensure safe and stable car-following behavior, which naturally mitigates oscillatory responses. Based on Eq.~(\ref{taylor_rearr}), the condition for stability is satisfied if the damping term 
\( f_{\Delta v} - f_v \) is positive. For the IDM model, this becomes:
\begin{align}
f_{\Delta v} - f_v =&- \frac{\Delta v}{2 T^2 b} + \frac{a \delta v^\delta}{v v_{\text{des}}^{\delta}} - \frac{\sqrt{a}}{T \sqrt{b}} \nonumber \\
& + \frac{(\Delta v + 2 T \sqrt{a b})^2}{2 T^2 b v}
\end{align}
This expression must be positive to ensure local string stability. It also enables us to characterize regions in the \( (\Delta v, T) \) space, as depicted by Figure~\ref{fig:heat_map}, assuming fixed values for other parameters, within which the IDM model guarantees a stable response to disturbances.
\subsection{String Stability Analysis of OVM Model}
At equilibrium, all vehicles move with constant velocity \( v^* \) and headway \( \Delta x^* \), with fixed \(\text{$v_0 = 22 \, \text{m/s}$}\), \(\text{$h = 4 \, \text{m}$}\):
\begin{equation}
v^* = V_{\text{opt}}(\Delta x^*) = 22 \cdot \left( \tanh(\Delta x^* - 4) + \tanh(4) \right)
\end{equation}
To analyze string stability, we introduce perturbations: \( v_n(t) = v^* + \tilde{v}_n(t) \), \( \Delta x_n(t) = \Delta x^* + \tilde{h}_n(t) \). Linearizing the optimal velocity function:
\begin{equation}
V_{\text{opt}}(\Delta x_n) \approx V_{\text{opt}}(\Delta x^*) + V_{\text{opt}}'(\Delta x^*) \tilde{h}_n
\end{equation}
where:
\[
\begin{aligned}
&V_{\text{opt}}'(\Delta x) = 22 \cdot \text{sech}^2(\Delta x - 4) \\
&\lambda = V_{\text{opt}}'(\Delta x^*) = 22 \cdot \text{sech}^2(\Delta x^* - 4)
\end{aligned}
\]Here, {$\text{sech}^2(x)$} represents the derivative of the hyperbolic tangent function and is known as the hyperbolic secant squared. The linearized dynamics are:
\begin{align}
\dot{\tilde{v}}_n &= \alpha \lambda \tilde{h}_n - \alpha \tilde{v}_n \\
\dot{\tilde{h}}_n &= \tilde{v}_{n-1} - \tilde{v}_n
\end{align}
String stability ensures that perturbations do not amplify through the platoon. Assume perturbations of the form \( \tilde{v}_n(t) = V_n e^{st} \), \( \tilde{h}_n(t) = H_n e^{st} \):
\begin{align}
s V_n &= \alpha \lambda H_n - \alpha V_n \implies H_n = \frac{(s + \alpha) V_n}{\alpha \lambda} \\
s H_n &= V_{n-1} - V_n
\end{align}
\[V_{n-1} = V_n \left( 1 + \frac{s (s + \alpha)}{\alpha \lambda} \right)\]
The transfer function is:
\begin{equation}
G(s) = \frac{V_n}{V_{n-1}} = \frac{\alpha \lambda}{\alpha \lambda + s (s + \alpha)}
\end{equation}
String stability requires \( |G(j\omega)| \leq 1 \) for all \( \omega > 0 \):
\begin{equation}
|G(j\omega)|^2 = \frac{(\alpha \lambda)^2}{(\alpha \lambda - \omega^2)^2 + (\alpha \omega)^2} \leq 1
\end{equation}
\[\omega^4 + \alpha (\alpha - 2 \lambda) \omega^2 \geq 0 \implies \alpha \geq 2 \lambda = 44 \cdot \text{sech}^2(\Delta x^* - 4)\]

Thus, the stability condition is:
\begin{equation}
\alpha \geq 44 \cdot \text{sech}^2(\Delta x^* - 4)
\label{ovm_stability}
\end{equation}
For OVM, local string stability is ensured if the damping term \( f_{\Delta v} - f_v > 0 \):
\[\dot{v}_n = f(\Delta x_n, v_n) = \alpha [V_{\text{opt}}(\Delta x_n) - v_n]\]
\[f_{\Delta x} = \alpha \lambda, \quad f_v = -\alpha, \quad f_{\Delta x} - f_v = \alpha (\lambda + 1)\]
This is always positive, indicating local stability, but string stability requires Eq.~(\ref{ovm_stability}). We define the stability metric as:
\begin{equation}
S = \alpha - 44 \cdot \text{sech}^2(\Delta x^* - 4)
\end{equation}
The system is string stable if \( S \geq 0 \).

\paragraph{Stability Regions in the \((\Delta x^*, \alpha)\) Space}
To characterize stability regions, we consider the parameter space \((\Delta x^*, \alpha)\), with fixed \( v_0 = 22 \, \text{m/s} \), \( h = 4 \, \text{m} \):
\[S(\Delta x^*, \alpha) = \alpha - 44 \cdot \text{sech}^2(\Delta x^* - 4)\]
The condition \( S \geq 0 \) defines stable regions. Since \( \text{sech}^2(\Delta x^* - 4) \leq 1 \), the critical sensitivity is maximized at \( \Delta x^* = 4 \, \text{m} \), where \( \alpha \geq 44 \), which is high for typical values (\( \alpha \leq 3 \, \text{s}^{-1} \)). As \( \Delta x^* \) deviates from 4 m, \( \text{sech}^2 \) decreases, reducing the required \( \alpha \).

\begin{figure}[!t]
        \centering
        \includegraphics[width=0.8\linewidth]{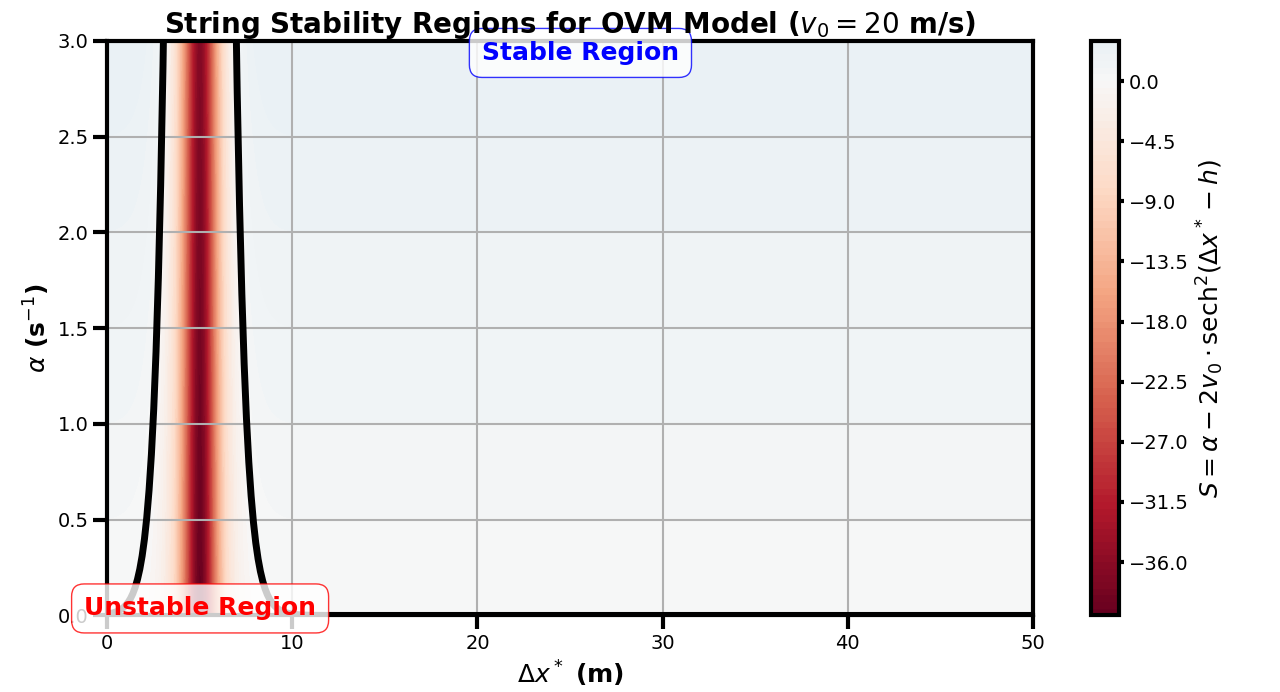} 
        \caption{Stability Map of OVM-based Platoon Dynamics in \((\Delta x^*, \alpha)\) Space}
        \label{fig:ovm_heat_map}
\end{figure}

As depicted by Figure~\ref{fig:ovm_heat_map}, the heatmap over \( \Delta x^* \in [0, 50] \, \text{m} \), \( \alpha \in [0, 3] \, \text{s}^{-1} \):
\begin{itemize}
    \item \textbf{Stable regions} (\( S \geq 0 \)): Occur for higher \( \alpha \) or when \( \Delta x^* \) is far from 4 m, where \( \text{sech}^2(\Delta x^* - 4) \) is small, reducing the critical \( \alpha \).
    \item \textbf{Unstable regions} (\( S < 0 \)): Occur near \( \Delta x^* = 4 \, \text{m} \), where \( \text{sech}^2 \approx 1 \), requiring \( \alpha \geq 44 \).
    \item \textbf{Boundary}: The curve \( \alpha = 44 \cdot \text{sech}^2(\Delta x^* - 4) \), peaking at \( \Delta x^* = 4 \, \text{m} \).
\end{itemize}

\subsection{String Stability Analysis of GMM}
To analyze the string stability of the given car-following model and derive the stability condition for the parameter space, we follow these steps:

The acceleration of the \( n \)-th vehicle is given by:
\begin{equation}
  a_n(t) = \alpha \frac{v_n(t)^m}{(x_{n-1}(t) - x_n(t))^l} (v_{n-1}(t) - v_n(t)) 
  \label{acc_gmm}
\end{equation}where:
- \( \alpha \) is a sensitivity coefficient,
- \( m \) and \( l \) are exponents for velocity and spacing, respectively,
- \( x_n(t) \) and \( v_n(t) \) are the position and velocity of the \( n \)-th vehicle at time \( t \).

Assume small perturbations around the equilibrium state, where \( v^* \) and \( s^* \) are the Equilibrium velocity and Equilibrium spacing, respectively. Let:
\begin{equation}
x_n(t) = x_n^*(t) + y_n(t), \quad v_n(t) = v^* + w_n(t)
\label{pos_sp}
\end{equation}where \( y_n(t) \) and \( w_n(t) \) are small perturbations, the equilibrium spacing is:
\[
x_{n-1}^*(t) - x_n^*(t) = s^*
\]
We substitute Eq.~(\ref{pos_sp}) into the acceleration equation, given by Eq.~(\ref{acc_gmm}), and linearize:
\begin{equation}
a_n(t) = \dot{w}_n(t) \approx \alpha \frac{(v^* + w_n(t))^m}{(s^* + y_{n-1}(t) - y_n(t))^l} (w_{n-1}(t) - w_n(t)).
\end{equation}For small perturbations, approximate:
\begin{equation}
(v^* + w_n)^m \approx v^{*m} + m v^{*m-1} w_n,
\end{equation}
\begin{equation}
(s^* + \Delta y)^{-l} \approx s^{*-l} - l s^{*-l-1} \Delta y,
\end{equation}
where \( \Delta y = y_{n-1} - y_n \).\\
Thus, the linearized acceleration becomes:
\begin{equation}
\dot{w}_n(t) \approx \alpha \frac{v^{*m}}{s^{*l}} (1 + \frac{m}{v^*} w_n - \frac{l}{s^*} \Delta y)(w_{n-1} - w_n).
\end{equation}
Neglecting higher-order terms:
\[
\dot{w}_n(t) \approx \alpha \frac{v^{*m}}{s^{*l}} (w_{n-1} - w_n).
\]
Assuming that perturbations are of the form \( y_n(t) = Y_n e^{st} \), \( w_n(t) = s Y_n e^{st} \), we substitute it into the linearized equation to obtain:
\[
s^2 Y_n = \alpha \frac{v^{*m}}{s^{*l}} (s Y_{n-1} - s Y_n).
\]
We divide by \( s \) (for \( s \neq 0 \)):
\[
s Y_n = \alpha \frac{v^{*m}}{s^{*l}} (Y_{n-1} - Y_n).
\]
Consequently, the transfer function \( G(s) \) (from \( Y_{n-1} \) to \( Y_n \)) is:
\[
G(s) = \frac{Y_n}{Y_{n-1}} = \frac{\alpha \frac{v^{*m}}{s^{*l}}}{s + \alpha \frac{v^{*m}}{s^{*l}}}.
\]
For string stability, the amplification factor \( |G(j\omega)| \leq 1 \) for all frequencies \( \omega \):
\[
|G(j\omega)| = \left| \frac{\alpha \frac{v^{*m}}{s^{*l}}}{j\omega + \alpha \frac{v^{*m}}{s^{*l}}} \right| = \frac{\alpha \frac{v^{*m}}{s^{*l}}}{\sqrt{\omega^2 + \left( \alpha \frac{v^{*m}}{s^{*l}} \right)^2}} \leq 1.
\]
This is always satisfied since:
\[
\frac{\alpha \frac{v^{*m}}{s^{*l}}}{\sqrt{\omega^2 + \left( \alpha \frac{v^{*m}}{s^{*l}} \right)^2}} \leq \frac{\alpha \frac{v^{*m}}{s^{*l}}}{\alpha \frac{v^{*m}}{s^{*l}}} = 1.
\]
However, this suggests unconditional stability, which is unrealistic. The issue arises because the model lacks speed adaptation (dependence on spacing). To ensure meaningful stability, we need to include spacing-dependent terms (e.g., \( \frac{1}{(x_{n-1} - x_n)^l} \)) in the linearization.

\begin{figure}[!t]
        \centering
        \includegraphics[width=0.8\linewidth]{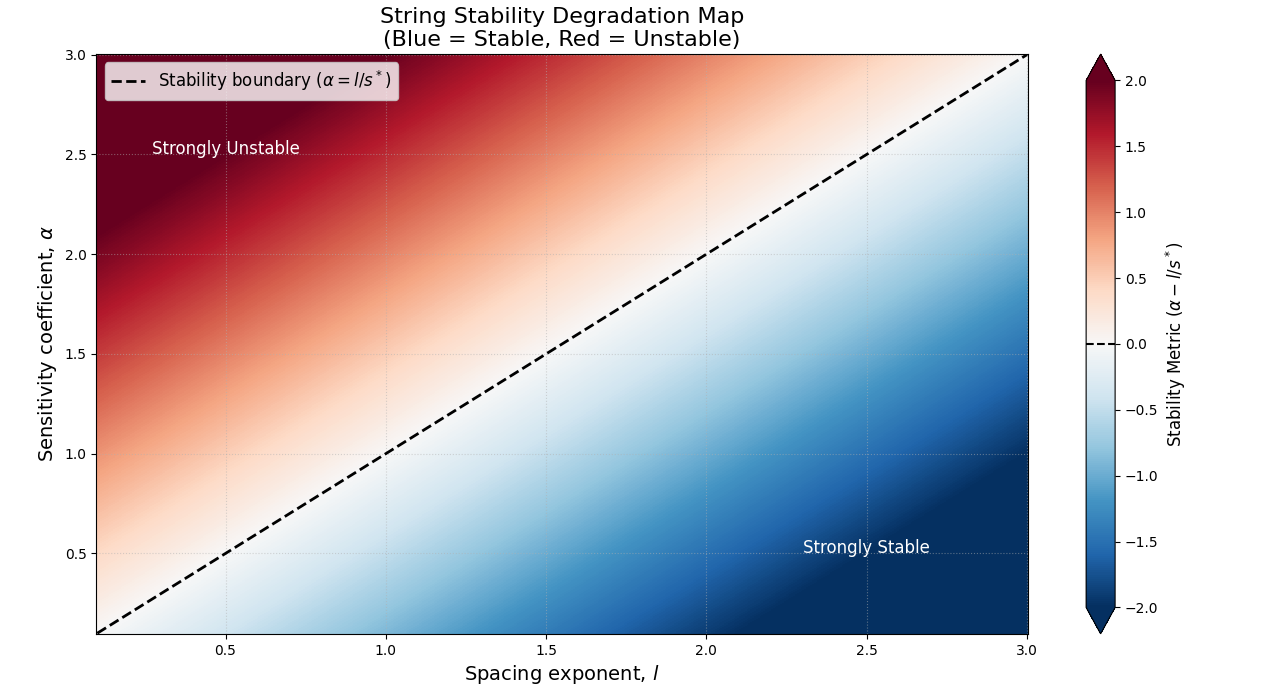} 
        \caption{Stability Map of GMM-based Platoon Dynamics in \((\Delta x^*, \alpha)\) Space}
        \label{fig:gmm_heat_map}
\end{figure}

A more accurate linearization (including spacing perturbations) gives:
\[
\dot{w}_n(t) \approx \alpha \frac{v^{*m}}{s^{*l}} (w_{n-1} - w_n) - \alpha l \frac{v^{*m}}{s^{*l+1}} (y_{n-1} - y_n).
\]
Taking Laplace transforms:
\[
s^2 Y_n = \alpha \frac{v^{*m}}{s^{*l}} (s Y_{n-1} - s Y_n) - \alpha l \frac{v^{*m}}{s^{*l+1}} (Y_{n-1} - Y_n).
\]
The transfer function becomes:
\[
G(s) = \frac{\alpha \frac{v^{*m}}{s^{*l}} s + \alpha l \frac{v^{*m}}{s^{*l+1}}}{s^2 + \alpha \frac{v^{*m}}{s^{*l}} s + \alpha l \frac{v^{*m}}{s^{*l+1}}}.
\]
For string stability, we require \( |G(j\omega)| \leq 1 \), which leads to the condition:
\[
\alpha \frac{v^{*m}}{s^{*l}} \leq l \frac{v^{*m}}{s^{*l+1}} \implies \alpha \leq \frac{l}{s^*}.
\]
Hence, the string stability condition is:
\[
\alpha \leq \frac{l}{s^*}.
\]
For the heatmap depicted by Figure~\ref{fig:gmm_heat_map}, plot \( \alpha \) vs. \( l \) and shade the stable region where \( \alpha \leq \frac{l}{s^*} \). The critical line \( \alpha = \frac{l}{s^*} \) separates stable and unstable regimes. Adjust \( s^* \) (equilibrium spacing) as needed for different scenarios.

\subsection{String Stability Analysis of CACC Model}
Given the control law:
Let us assume small perturbations around the equilibrium:
\begin{align*}
x_i(t) &= x_i^*(t) + y_i(t) \\
v_i(t) &= v^* + w_i(t) \\
a_i(t) &= a_i^*(t) + u_i(t)
\end{align*}

The linearized dynamics become:
\begin{equation}
\begin{aligned}
    \ddot{y}_i(t) = k_p (y_{i-1} - y_i) + k_d (\dot{y}_{i-1} - \dot{y}_i) + &\\k_v (w_{i-1}(t-\tau) - w_i(t)) + k_a u_{i-1}(t-\tau)
\end{aligned}
\end{equation}

Taking Laplace transforms with delay $e^{-\tau s}$:
\begin{equation}
\begin{aligned}
    s^2 Y_i(s) &= (k_p + k_d s)(Y_{i-1}(s) - Y_i(s)) \\&+ k_v e^{-\tau s} (sY_{i-1}(s) - sY_i(s)) + k_a e^{-\tau s} s^2 Y_{i-1}(s)
\end{aligned}
\end{equation}

The error propagation transfer function is:
\begin{equation}
G(s) = \frac{Y_i(s)}{Y_{i-1}(s)} = \frac{k_p + k_d s + k_v s e^{-\tau s} + k_a s^2 e^{-\tau s}}{s^2 + k_p + k_d s + k_v s e^{-\tau s}}
\end{equation}

For string stability, we require $|G(j\omega)| \leq 1$ for all $\omega \geq 0$:

\begin{equation}\begin{aligned}
    |k_p + k_d j\omega + (k_v j\omega + k_a (j\omega)^2)e^{-j\omega\tau}| \leq &\\|(j\omega)^2 + k_p + k_d j\omega + k_v j\omega e^{-j\omega\tau}|
\end{aligned}
\end{equation}

The sufficient conditions for string stability are:

\begin{enumerate}
    \item \textbf{No-delay case} ($\tau = 0$):
    \begin{equation}
    k_a \leq \frac{k_v}{2} \quad \text{and} \quad k_v \geq 2k_d - \frac{k_p}{\omega_0^2}
    \end{equation}
    
   \item  \textbf{With delay} ($\tau > 0$):
    \begin{equation}
    \tau \leq \tau_{\text{max}} = \frac{1}{\omega_c} \cos^{-1}\left(\frac{k_v \omega_c^2 - k_a \omega_c^4 - k_p}{k_v \omega_c^3}\right)
    \end{equation}
    where $\omega_c$ is the crossover frequency.
\end{enumerate}

\begin{figure}[!t]
        \centering
        \includegraphics[width=0.7\linewidth]{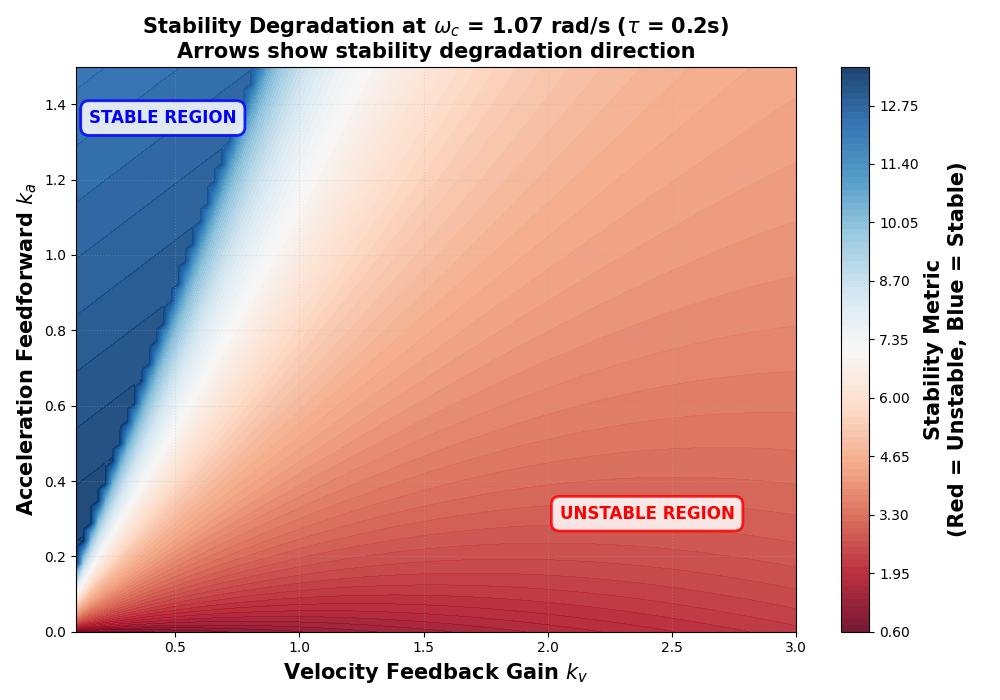} 
        \caption{Stability Map of CACC-based Platoon Dynamics in \((k_v, k_a)\) Space}
        \label{fig:cacc_heat_map}
\end{figure}

\begin{figure}[!t]
        \centering
        \includegraphics[width=0.7\linewidth]{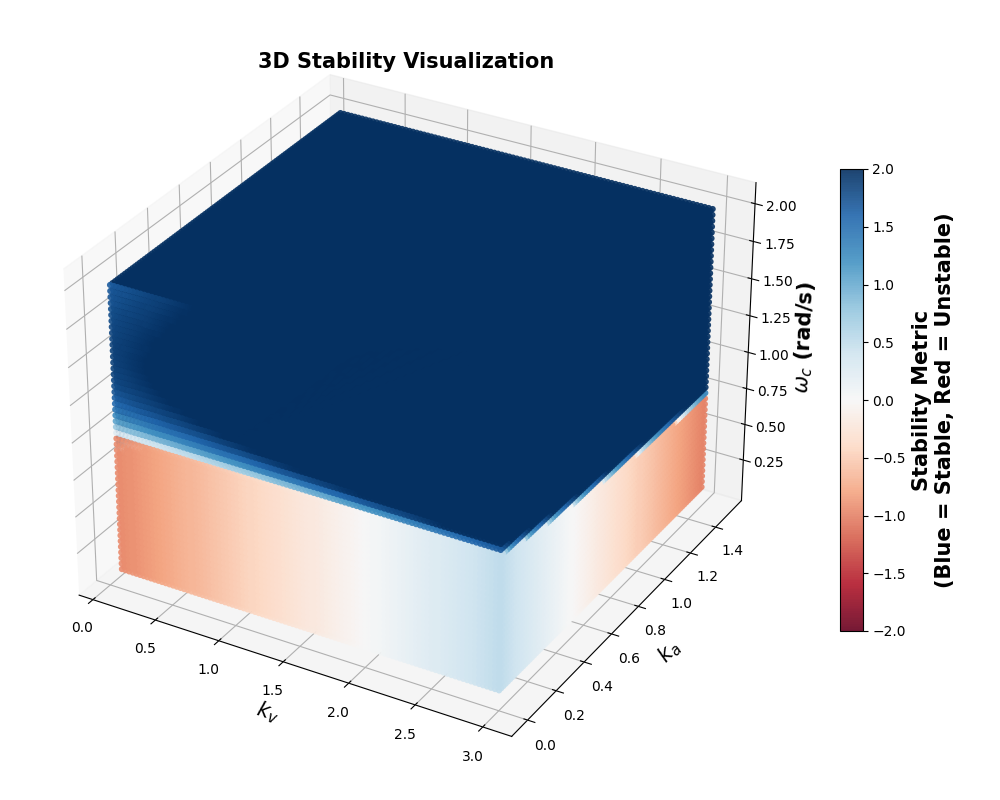} 
        \caption{Stability Map of CACC-based Platoon Dynamics in \((k_v, k_a, \omega_c)\) Space}
        \label{fig:cacc_heat_map3d}
\end{figure}

For robust string stability as depicted by Figures~{\ref{fig:cacc_heat_map}} and~{\ref{fig:cacc_heat_map3d}}:
\begin{itemize}
    \item Prioritize velocity feedback ($k_v$) over position feedback ($k_p$)
    \item Limit acceleration feedforward ($k_a$) to prevent overshoot
    \item Compensate for delays through predictive terms
    \item Ensure $k_d > \sqrt{k_p}$ for damping
\end{itemize}


\section{STRING STABILITY CONTINUOUS PERTURBATIONS: EXPERIMENTAL STUDY}
\label{methodo}

To investigate platoon instability, we simulate the response of the following vehicles to perturbations in the motion of the lead vehicle within the platoon. Based on the stability analysis in Section~\ref{stability}, we conduct these experiments within the stable regions of each model, as determined by its stability parameters. For this analysis, we employ the IDM to represent car-following behavior and explore two types of perturbations: \textit{continuous perturbations}, involving smooth and periodic fluctuations in the leader's velocity, such as oscillatory velocities. The case of \textit{discrete perturbations}, characterized by sudden, isolated events such as abrupt braking or acceleration, is presented in Appendix \ref{app1}. Focusing on the case of continuous perturbations, we introduce a sinusoidal oscillation to continuously vary the speed of the leader. The sinusoidal perturbation is defined as follows:
\begin{equation}
    perturbation = A\cdot \sin{(\omega t)}
    \label{sin}
\end{equation}where \textit{A} is the amplitude and $\omega$ denotes the frequency. 

In a steady state, the platoon is characterized by constant, identical speed of the cars along with constant spacing between the vehicles. Once we induce the perturbations, we observe the variation depicted by the vehicle dynamics. These variations are reflected by the vehicle speed and spacing. In fact, the perturbation \textit{amplitude} and \textit{frequency} play a critical role in determining the dynamics of the platoon and how disturbances propagate. 

A larger amplitude corresponds to more significant fluctuations in the lead vehicle's velocity, resulting in larger disturbances in the spacing and velocities of following vehicles, as well as greater oscillations that may take longer to dissipate. The frequency determines how often the velocity oscillates within a given time frame. Low-frequency perturbations (longer periods) cause smoother, slower fluctuations, while high-frequency perturbations (shorter periods) cause rapid oscillations that are more difficult for vehicles to track.

For a more in-depth analysis, we calculate the spacing error, which is defined as the deviation in distance between two following vehicles from the original constant spacing. This error quantifies how the separation distance fluctuates due to the dynamics of the car-following behavior under the influence of the leader's oscillating speed. In the context of IDM, the spacing error can be mathematically defined as the difference between the desired spacing $s^*$ and the actual spacing \textit{s}, between two following vehicles, as given by:
\begin{equation}
   E = s^* - s
    \label{error}
\end{equation}
\subsection{Perturbations in the Intelligent Driver Model (IDM)}
\begin{figure*}[htb!]
    \centering
    \subfloat[Spacing Error between vehicles in the platoon after the perturbation. \label{fig:spac_idm_spac}]{
        \includegraphics[width=0.48\linewidth]{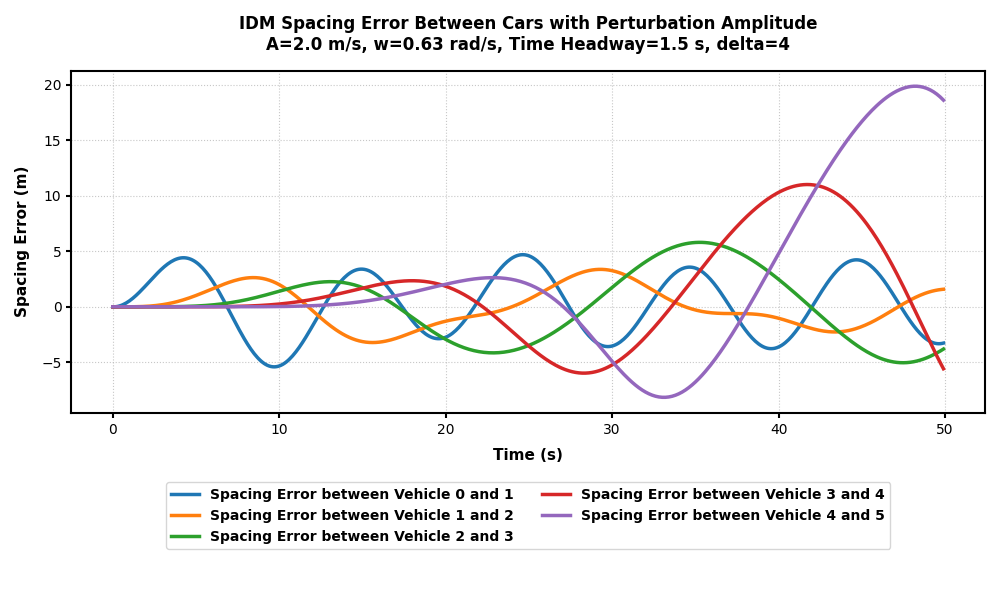}
    }
    \hfill
    \subfloat[Velocity of the vehicles in the platoon after the perturbation.\label{fig:spac_idm_vel}]{
        \includegraphics[width=0.48\linewidth]{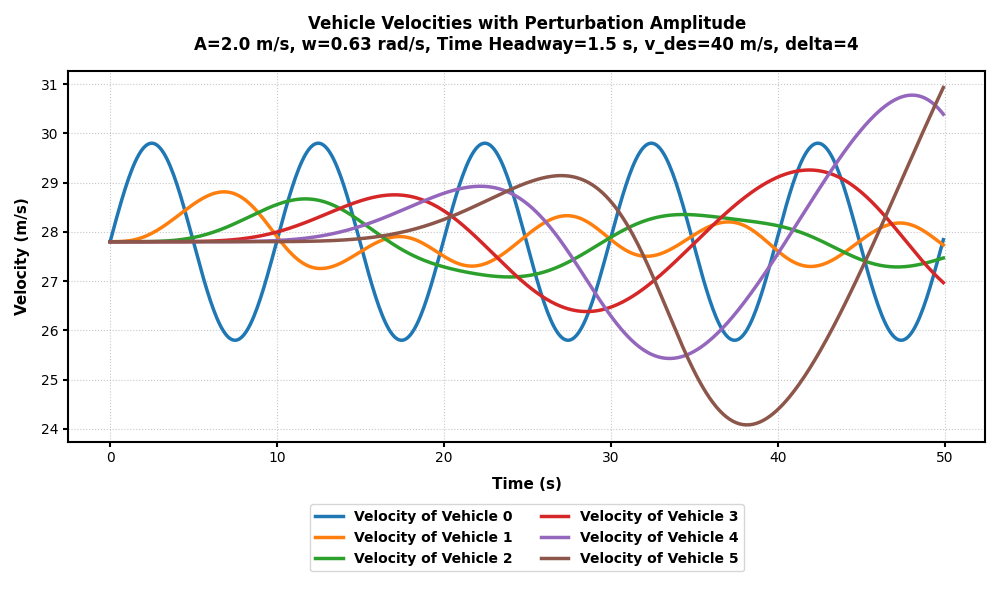}
    }
    \caption{IDM Configuration I: Simulation Results Following a Continuous Sinusoidal Perturbation and Accounting Only for Spacing.}
    \label{fig:spac_idm}
\end{figure*}

We conduct simulations under both discrete and continuous perturbations to evaluate IDM performance in each scenario. Additionally, for continuous perturbations, we explore two different IDM configurations. 
In the first configuration, IDM accounts for spacing when computing acceleration but ignores velocity differences. The second configuration represents the classical IDM, incorporating both spacing and velocity differences to refine acceleration adjustments. In the context of a highway, we simulate a platoon of 5 vehicles with an initial speed of \textit{27.8 m/s}, which is equivalent to $100 km/h$. For this scenario, we specify a time headway of $1.5 s$~\cite{treiber2000congested}, as it represents a typical safe following distance under normal driving conditions. 
\textbf{{\paragraph{Configuration I: IDM model considering the spacing component without velocity term.}}}
In the case of accounting for spacing only between vehicles, the mathematical formulation of IDM is modified as follows:
\begin{equation}
    \dot{v} = a\Big(1- \Big(\frac{s^*}{s}\Big)^2\Big)
    \label{spac_IDM}
\end{equation}
In the case where the IDM model accounts solely for the spacing between vehicles, the system exhibits a markedly different dynamic behavior. Here, each vehicle determines its acceleration based on both its own velocity and the relative distance to the vehicle ahead, introducing inter-vehicle interactions into the control process. As a consequence, perturbations introduced at the front of the platoon are no longer isolated but instead propagate downstream, affecting the following vehicles. This propagation effect is evident in the spacing error and velocity profiles shown by Figures~\ref{fig:spac_idm}\subref{fig:spac_idm_spac} and~\ref{fig:spac_idm}\subref{fig:spac_idm_vel}, respectively, where disturbances amplify and persist along the platoon. These results underscore the role of inter-vehicle spacing in influencing platoon stability and highlight the trade-off between responsiveness to traffic conditions and potential vulnerability to perturbation transmission.

\begin{figure*}[htb!]
    \centering
    \subfloat[Spacing Error between vehicles in the platoon after the perturbation. \label{fig:spacing}]{
        \includegraphics[width=0.48\linewidth]{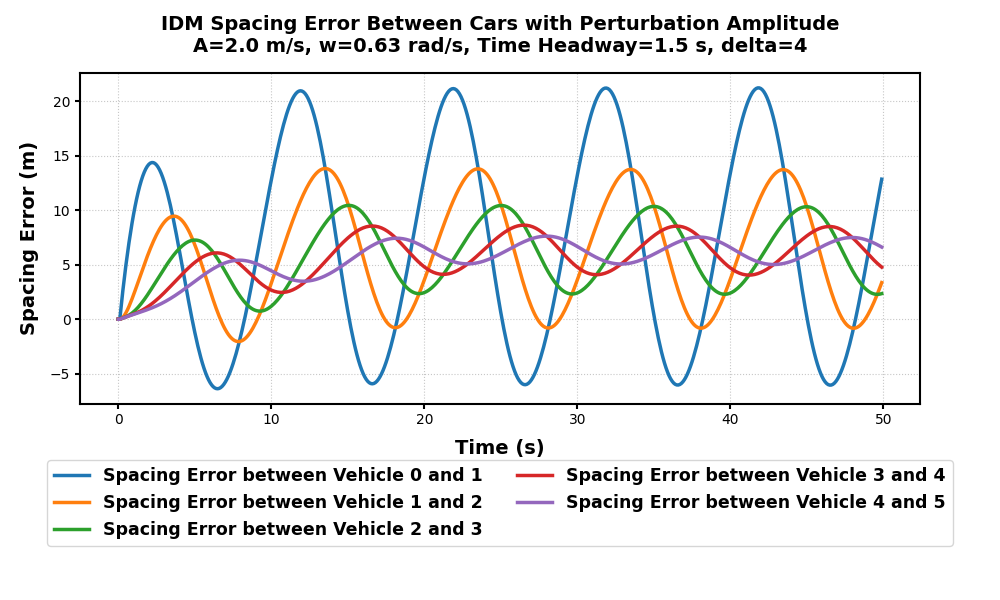}
    }
    \hfill
    \subfloat[Velocity of the vehicles in the platoon after the perturbation.\label{fig:vel}]{
        \includegraphics[width=0.48\linewidth]{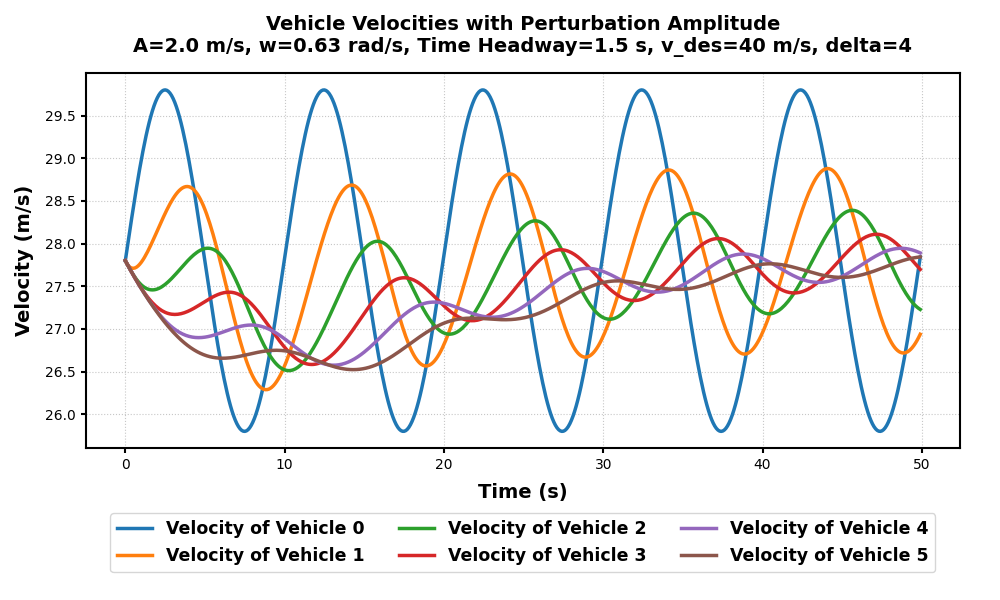}
    }
    \caption{IDM Configuration II: Simulation Results of Standard IDM Following a Continuous Sinusoidal Perturbation.}
    \label{fig:idm}
\end{figure*}

\textbf{{\paragraph{Configuration II: IDM Model in its Standard Form}}}
The findings of this scenario are depicted by Figures~\ref{fig:idm}\subref{fig:spacing} and~\ref{fig:idm}\subref{fig:vel}, representing the spacing error and velocities of platoon vehicles, respectively. In this series of simulations, we observe that the vehicles' velocities fluctuate and drop. This can be caused by the perturbations in the lead vehicle behavior. The follower vehicle's behavior does not fully mirror the oscillations of the lead vehicle because of the dynamics of IDM and its design principles. IDM is designed to maintain traffic flow stability by incorporating a nonlinear acceleration formulation that balances responsiveness with safety. The desired time gap \( T \) ensures a consistent buffer between vehicles, while the acceleration expression, comprising terms like \( 1 - \left(\frac{v}{v_{\text{des}}}\right)^\delta \) and \( \left(\frac{s^*}{s}\right)^2 \), prevents overreaction to minor fluctuations. By considering the relative velocity \( \Delta v = v - v_{\text{lead}} \) in the desired spacing \( s^* \), the model dampens responses when the speed difference is small, leading to smoother following behavior. As a result, oscillations in the lead vehicle's motion are not amplified, especially when they are of low amplitude and moderate frequency. This allows the follower to absorb and smooth out disturbances, preserving platoon stability. 

Based on the findings of both configurations, the full IDM formulation—with its nonlinear acceleration, desired time gap, and relative velocity terms—effectively mitigates minor disturbances, helping the platoon maintain stable motion. Follower vehicles adjust smoothly to the lead vehicle’s variations, preventing the growth of oscillations. When only the spacing term is used, however, these stabilizing effects are removed, and perturbations from the lead vehicle travel more readily through the platoon. As a result, disturbances tend to amplify and persist, highlighting a balance between sensitivity to inter-vehicle spacing and overall platoon resilience.

\subsection{Perturbations in the Optimal Velocity Model (OVM)}
\begin{figure*}[htb!]
    \centering
    \subfloat[OVM: Spacing Error between vehicles in the platoon after a continuous perturbation. \label{fig:spacing_ovm_high0.5}]{
        \includegraphics[width=0.48\linewidth]{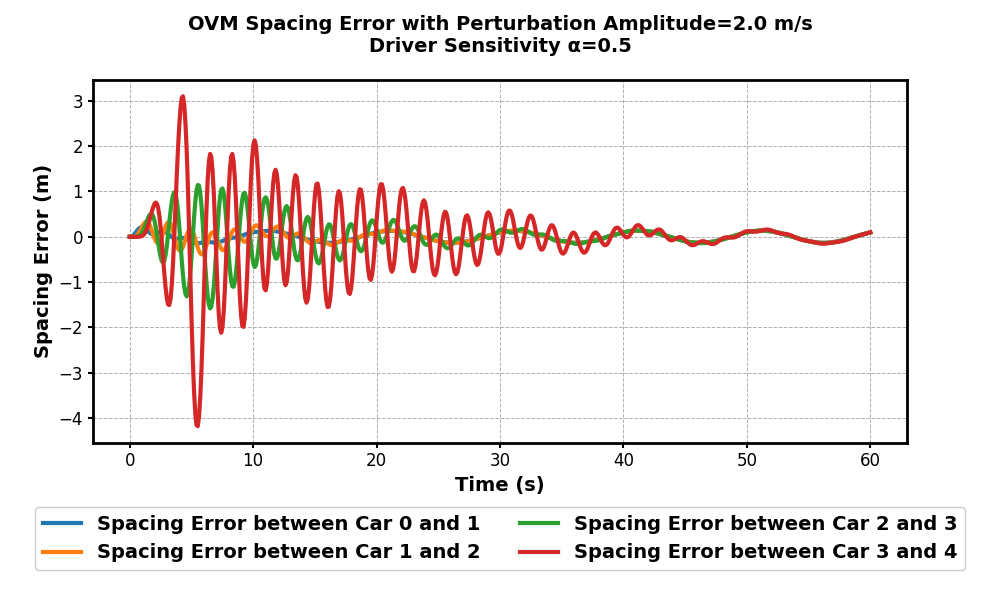}
    }
    \hfill
    \subfloat[OVM: Velocity of the vehicles in the platoon after a continuous perturbation.
    \label{fig:vel_ovm_high0.5}]{
        \includegraphics[width=0.48\linewidth]{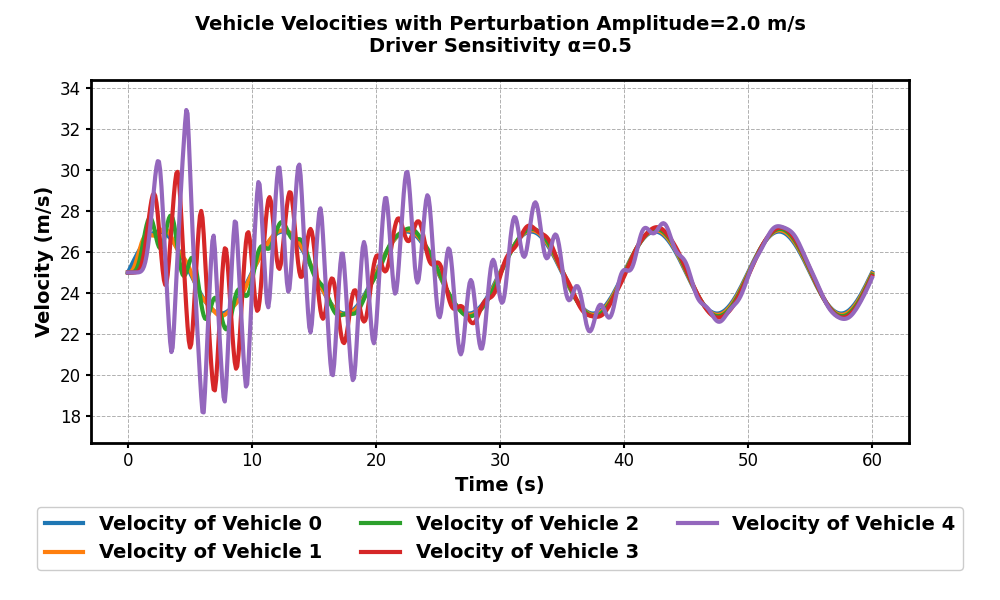}
    }
    \caption{OVM Simulation Results Following a Continuous Sinusoidal Perturbation.}
    \label{fig:ovm_0.5}
\end{figure*}

In this experimental study, we consider a platoon of five vehicles with a constant spacing of 6 \(m\) and identical velocities of 23\(m/s\) equivalent to 82.8\( km/h\). Starting from a stable state characterized by constant spacing and uniform velocities, a continuous perturbation, as described by Eq.~(\ref{sin}), is introduced to the leader vehicle in the platoon. This perturbation is applied at the velocity level, resulting in a sudden increase in the leader's speed. This change directly affects the leader's position and consequently alters the headway, $\Delta x$, between each two following vehicles in the platoon. As shown in Figures{~\ref{fig:ovm_0.5}\subref{fig:spacing_ovm_high0.5}} and{~\ref{fig:ovm_0.5}\subref{fig:vel_ovm_high0.5}}, a sudden oscillatory increase in the lead vehicle's speed induces an oscillatory pattern in the behavior of the following vehicles. Among these, aside from the leader, the last vehicle in the platoon exhibits the most significant variations in both speed and spacing error. Nevertheless, the OVM controller employs a velocity adaptation mechanism as depicted by Figure~\ref{fig:ovm_0.5}\subref{fig:vel_ovm_high0.5} that dynamically adjusts each vehicle's speed based on the optimal velocity function considering both the inter-vehicle spacing and current velocity. 

Figure~\ref{fig:ovm_0.5}\subref{fig:spacing_ovm_high0.5} shows the spacing errors between consecutive cars, where the errors exhibit growing oscillatory patterns, with amplitudes increasing over time, indicating an unstable traffic flow where perturbations amplify into stop-and-go waves. Similarly, Figure~\ref{fig:ovm_0.5}\subref{fig:vel_ovm_high0.5} displays the velocities of four vehicles (0 to 3) over the same period. The velocities oscillate with increasing amplitude around an equilibrium value ($~25 m/s$), reflecting the unstable response to the initial perturbation. It is also noteworthy that the inherent mathematical structure of the OVM can significantly influence its reaction to perturbations. For instance, the OVM does not explicitly include the leader's speed as a direct input in the acceleration equation. Instead, it relies on the headway to implicitly capture the leader's influence. This makes the model sensitive to the leader's behavior, but the response is governed by the sensitivity parameter \( a \) and the shape of the optimal velocity function \( V(\Delta x) \), such as the commonly used \( \tanh \) form. Furthermore, the incorporation of a hyperbolic tangent ($\tanh$) function for the optimal velocity function introduces specific nonlinearity, which can sustain oscillations even in a synchronized state. 

\subsection{Perturbations in the General Motors Model (GMM)}
While the GMM is rarely used in modern industrial applications due to its simplifying assumptions and lack of realism, it remains a foundational reference in car-following theory. Its main advantage lies in its analytical tractability—most notably, its closed-form solution, which makes it well-suited for stability analysis and theoretical exploration. In this section, we simulate the dynamics of a platoon of vehicles using the GMM car-following model. The simulation focuses on how a sinusoidal perturbation applied to the lead vehicle propagates through the platoon, affecting the spacing and velocities of the following vehicles over time. The following presents the simulation results for selected GMM settings, with additional configurations and their corresponding results provided in Appendix~\ref{app1}.

\begin{figure*}[htb!]
    \centering
    \subfloat[GMM Configuration I: Spacing Error between Vehicles after a Sinusoidal Perturbation Considering \(\Delta x\) and Ignoring Current speed of the Vehicle \(v_i(t)\). \label{fig:gmm_shw_sp_l2}]{
        \includegraphics[width=0.48\linewidth]{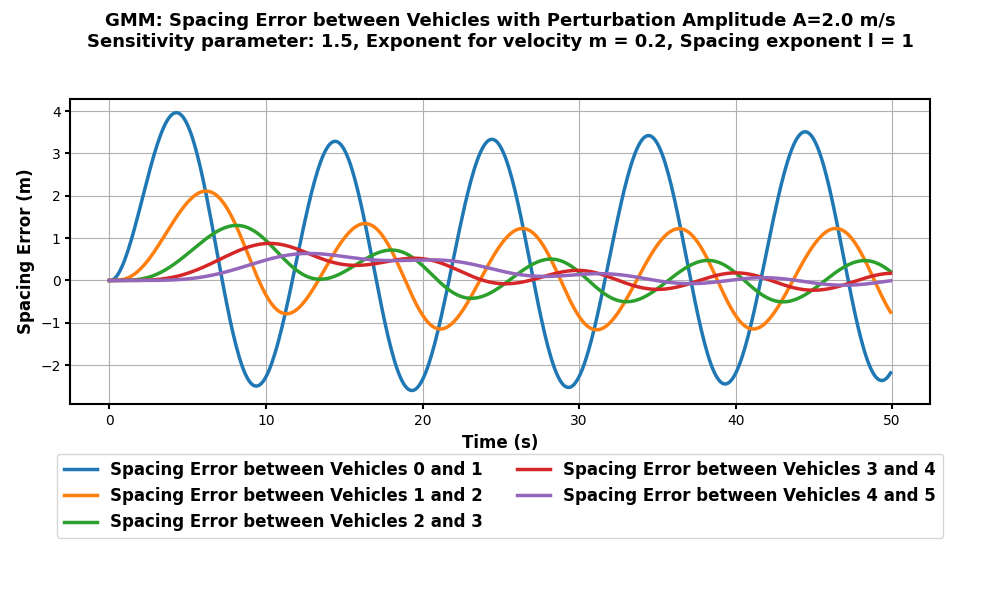}
    }
    \hfill
    \subfloat[GMM Configuration I: Vehicles Velocities after a Sinusoidal Perturbation Considering \(\Delta x\) and Ignoring Current speed of the Vehicle \(v_i(t)\).
    \label{fig:gmm_shw_vel_l2}]{
        \includegraphics[width=0.48\linewidth]{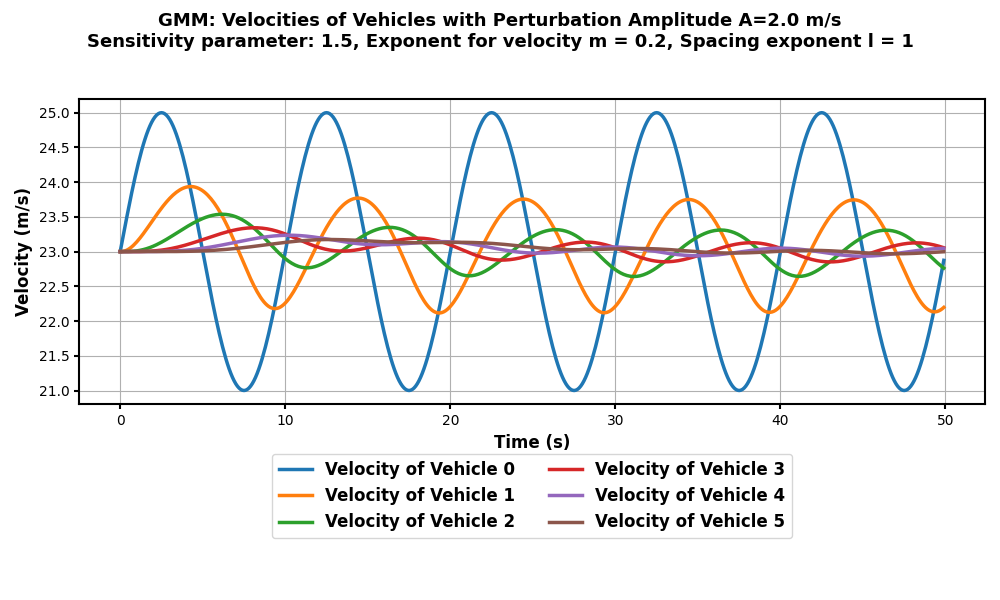}
    }
    \caption{GMM Configuration I: Simulation Results Following a Continuous Sinusoidal Perturbation.}
    \label{fig:gmm_config1}
\end{figure*}

\textbf{{\paragraph{Configuration I: GMM Model Considering Spacing Difference While Ignoring Subject Vehicle Speed}}}
In this section, we consider a configuration of the GMM accounting for the spacing difference between each two following vehicles, where the acceleration is directly proportional to the driver sensitivity \(\alpha\), the spacing difference and the speed variations between the vehicles \(n\) and \(n-1\). In this formulation, the GMM disregards only the velocity of the subject vehicle with \(m = 0\) and \(l = 1\), reducing Eq.~(\ref{gmm}) to:
\begin{equation}
    a_n(t) = \alpha \frac{1}{(x_{n-1}(t) - x_n(t))} (v_{n-1}(t) - v_n(t))
    \label{gmm2}
\end{equation}
After continuously applying the sinusoidal perturbation to the platoon leader, the resulting effects of this disturbance are illustrated in Figure~\ref{fig:gmm_config1}\subref{fig:gmm_shw_sp_l2}. A clear observation is the increased gap variation between the leader and its immediate follower. Additionally, a higher initial variation is noted between vehicles 2 and 3, which can be attributed to the propagation delay of the correction. However, the disruptions in the gaps between the rest of the vehicles are quickly attenuated, eventually stabilizing at a fixed spacing difference of \(\approx 0m\). While this outcome is not entirely optimal, as the system does not return to stability with the spacing error \(\Delta x \neq 0\), it represents an acceptable correction given the continuous nature of the applied perturbation. The velocity profiles, as depicted in Figure~\ref{fig:gmm_config1}\subref{fig:gmm_shw_vel_l2}, capture the variation in the leader’s speed relative to its immediate follower, illustrated by the sinusoidal oscillations. However, the remaining vehicles in the platoon maintain a constant and identical speed despite the perturbations, with the last two vehicles (vehicles 4 and 5) exhibiting an almost undisturbed behavior.

As shown in Figures~\ref{fig:gmm_config1}\subref{fig:gmm_shw_sp_l2} and~\ref{fig:gmm_config1}\subref{fig:gmm_shw_vel_l2}, this configuration exhibits smooth attenuation of disturbances, with the spacing errors and velocity fluctuations remaining controlled throughout the platoon. The GMM controller regulates each vehicle’s acceleration based on the inter-vehicle gap, allowing followers to adjust gradually to changes in the lead vehicle’s motion. This mechanism prevents the amplification of oscillations and contributes to a stable flow within the platoon. Overall, the results highlight how the model’s gap-based corrections promote coordinated and consistent vehicle behavior under perturbations.
\textbf{{\paragraph{Configuration II: GMM Model in Its Standard Form}}}
The final configuration implemented in this section follows the GMM as defined in Eq.~(\ref{gmm}). In this formulation, the model determines the acceleration of vehicle \(n\) based on the sensitivity parameter $m=1$, the gap between vehicle \(n\) and its leader, the velocity of vehicle \(n\), and the speed difference between vehicle \(n\) and its leader. 
We evaluated the performance of the GMM with the spacing exponents\footnote{A comparison based on the values of $l$ is provided in Appendix \ref{app1}.} \( l = 2 \), to understand the role of the spacing component in shaping platoon dynamics. The simulation results, presented in Figures~\ref{fig:gmm_config2}\subref{fig:gmm_sp2} and~\ref{fig:gmm_config2}\subref{fig:gmm_vel2}, demonstrate that an exponent of \( l = 2 \) introduces a smoothing, nonlinear effect that reduces overreaction to spacing variations, which leads to progressive attenuation of disturbances along the platoon, as shown in Figures~\ref{fig:gmm_config2}\subref{fig:gmm_sp2} and~\ref{fig:gmm_config2}\subref{fig:gmm_vel2}. The trailing vehicles in this configuration achieve smaller spacing errors and gradually recover their initial velocities, thereby restoring stability. 

\subsection{Perturbations in the Cooperative Adaptive Cruise Control (CACC)}
In this section, we simulate a platoon of vehicles using the CACC model. The vehicles are configured to maintain identical velocities and constant spacing, representing a stable platoon formation. To analyze the platoon's behavior, we introduce a sinusoidal perturbation to the leader vehicle's speed and examine its effects on the following vehicles. Furthermore, we observe how the CACC controller responds to this disturbance and evaluate its efficiency under different configurations.

\begin{figure*}[htb!]
    \centering
    \subfloat[GMM Configuration II: Spacing Error between Vehicles after a Sinusoidal Perturbation for $l = 2$. \label{fig:gmm_sp2}]{
        \includegraphics[width=0.48\linewidth]{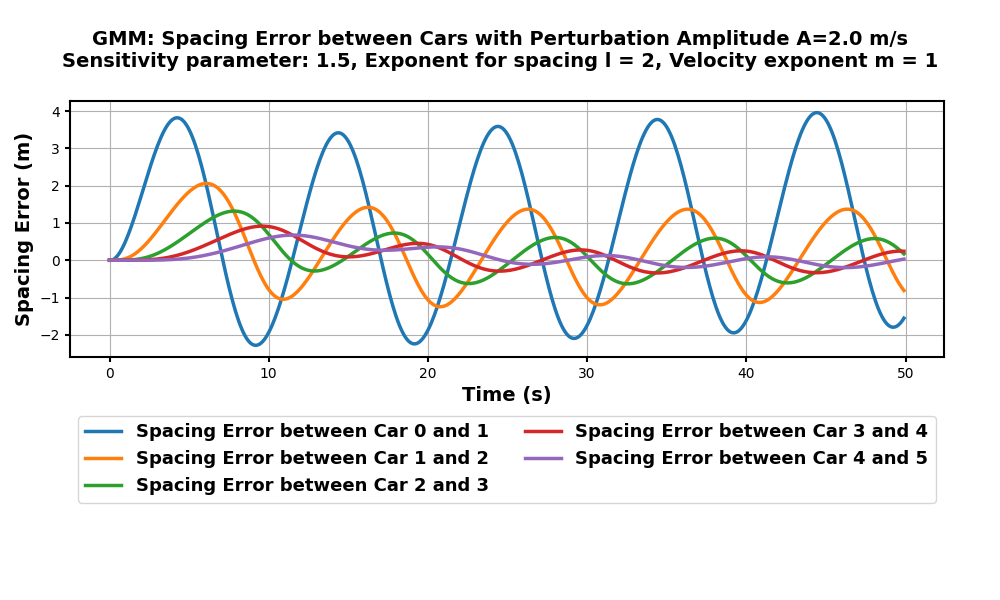}
    }
    \hfill
    \subfloat[GMM Configuration II: Vehicles Velocities after a Sinusoidal Perturbation for $l = 2$.
    \label{fig:gmm_vel2}]{
        \includegraphics[width=0.48\linewidth]{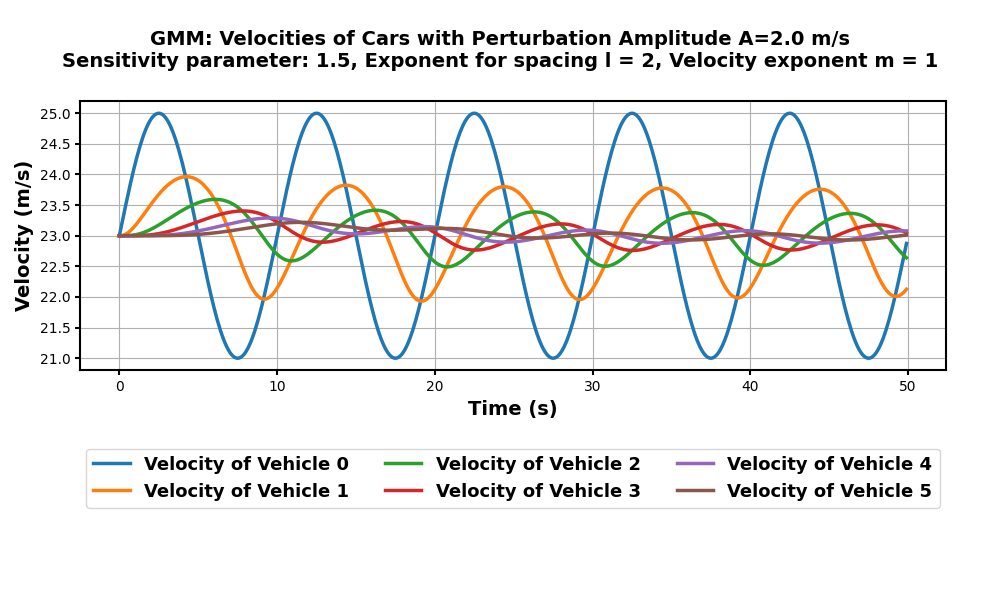}
    }
    \caption{GMM Configuration II: Simulation Results of Standard GM Following a Continuous Sinusoidal Perturbation.}
    \label{fig:gmm_config2}
\end{figure*}

\begin{figure*}[htb!]
    \centering
    \subfloat[CACC Configuration I: Spacing Error between the Vehicles after a Sinusoidal Perturbation with no V2V Communication. \label{fig:cacc1_space}]{
        \includegraphics[width=0.48\linewidth]{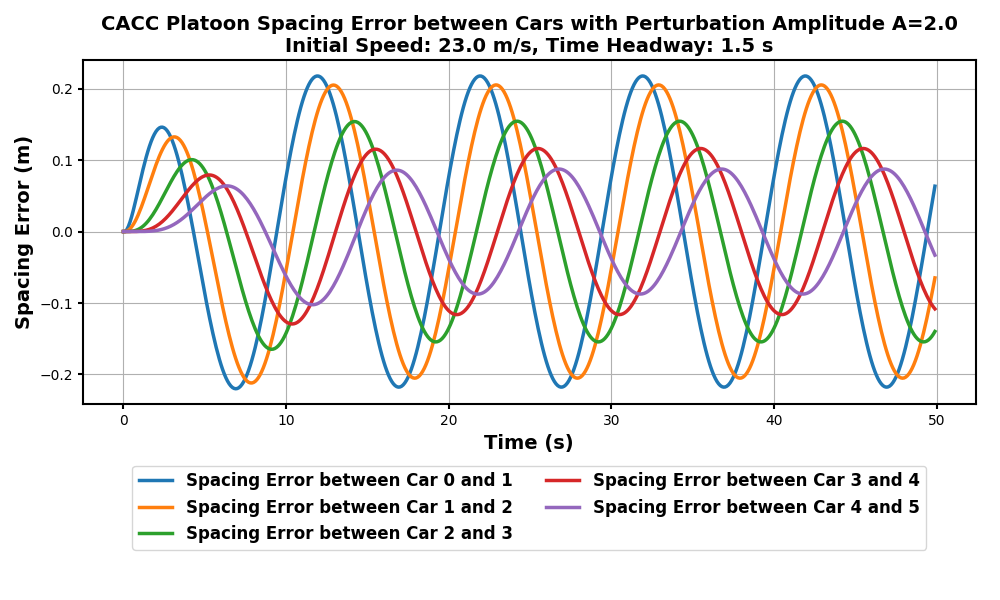}
    }
    \hfill
    \subfloat[CACC Configuration I: Vehicles Velocities after a Sinusoidal Perturbation with no V2V Communication.
    \label{fig:cacc1_vel}]{
        \includegraphics[width=0.48\linewidth]{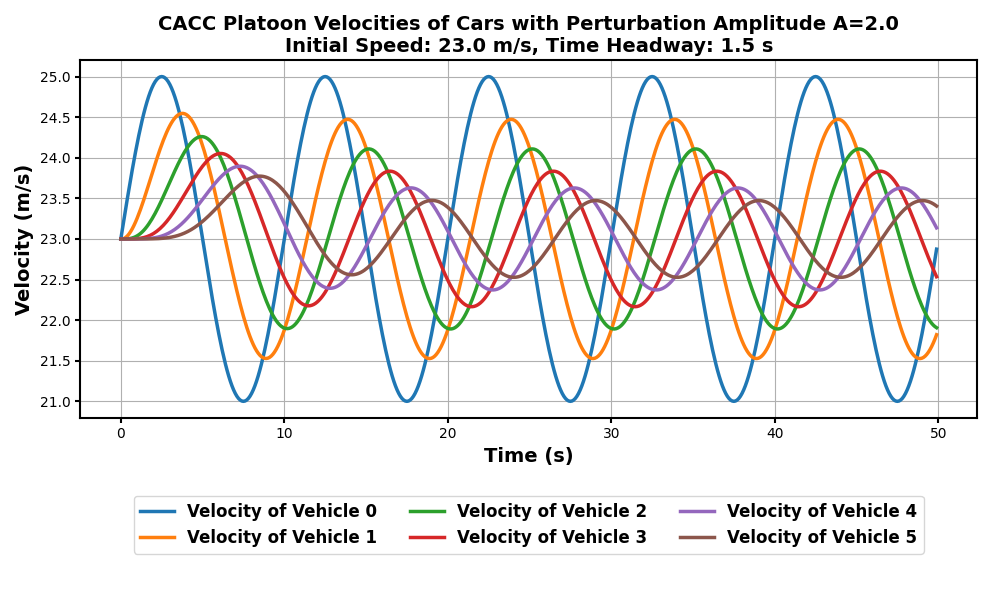}
    }
    \caption{CACC Configuration I: Simulation Results of CACC with no V2V Communication Following a Continuous Sinusoidal Perturbation.}
    \label{fig:cacc1}
\end{figure*}

\textbf{{\paragraph{Configuration I: CACC with no Component for V2V Communications}}}
In this configuration, we study the propagation of the perturbation and the correction by CACC controller without accounting for V2V communications. In this context, the formulation given by Eq.~(\ref{cacc}) is reduced to:
\begin{equation}
a_i(t) = k_p \cdot e_i(t) + k_d \cdot \dot{e}_i(t) 
\label{cacc_no_v2v}
\end{equation}
In this context, the CACC controller without V2V communication uses a feedback control law based on the spacing error and velocity error. The simulation results for this configuration are illustrated in Figures~\ref{fig:cacc1}\subref{fig:cacc1_space} and~\ref{fig:cacc1}\subref{fig:cacc1_vel}, which plot the spacing error between each pair of consecutive vehicles and the velocity of the vehicles in the platoon. The results demonstrate the propagation of oscillations through the platoon, initiated by the leader vehicle's perturbation. Although the oscillations are attenuated by the controller, they are still observable in the following vehicles, highlighting the impact of a continuous disturbance as it travels along the platoon. Observing the velocity difference, it is evident that the most significant impact occurs between the leader and its immediate follower. This is expected as the perturbation originates at the leader level and directly affects the first follower. As the disturbance propagates through the platoon, subsequent vehicles exhibit smaller variations in speed, which are progressively attenuated by the CACC controller. In this configuration, the CACC model, proportional to the spacing error and velocity profiles, behaves similarly to IDM (in Figures~\ref{fig:idm}\subref{fig:spacing} and~\ref{fig:idm}\subref{fig:vel}) and GMM (in Figures~\ref{fig:gmm_config2}\subref{fig:gmm_sp2} and~\ref{fig:gmm_config2}\subref{fig:gmm_vel2}).

\textbf{{\paragraph{Configuration II: CACC with a V2V Communication Component and no Delay}}}
In this part, we consider a communication component in the CACC controller when determining vehicles' accelerations in order to adjust their behavior and re-establish stability after the applied perturbation to the leader. Furthermore, we consider \(0 s\) delay in communicating the dynamic information of the preceding vehicle. In this context, the CACC formulation becomes the following:
\begin{equation}
\begin{aligned}
a_i(t) &= k_p \cdot e_i(t) + k_d \cdot \dot{e}_i(t) \\
&\quad + k_v \cdot \left(v_{i-1}(t) - v_i(t)\right) \\
&\quad + k_a \cdot a_{i-1}(t)
\end{aligned}
\label{cacc_v2v_no_delay}
\end{equation}
\begin{figure*}[htb!]
    \centering
    \subfloat[CACC Configuration II: Spacing Error between the Vehicles after a Sinusoidal Perturbation with V2V Communication and no Delay. \label{fig:cacc2_space}]{
        \includegraphics[width=0.48\linewidth]{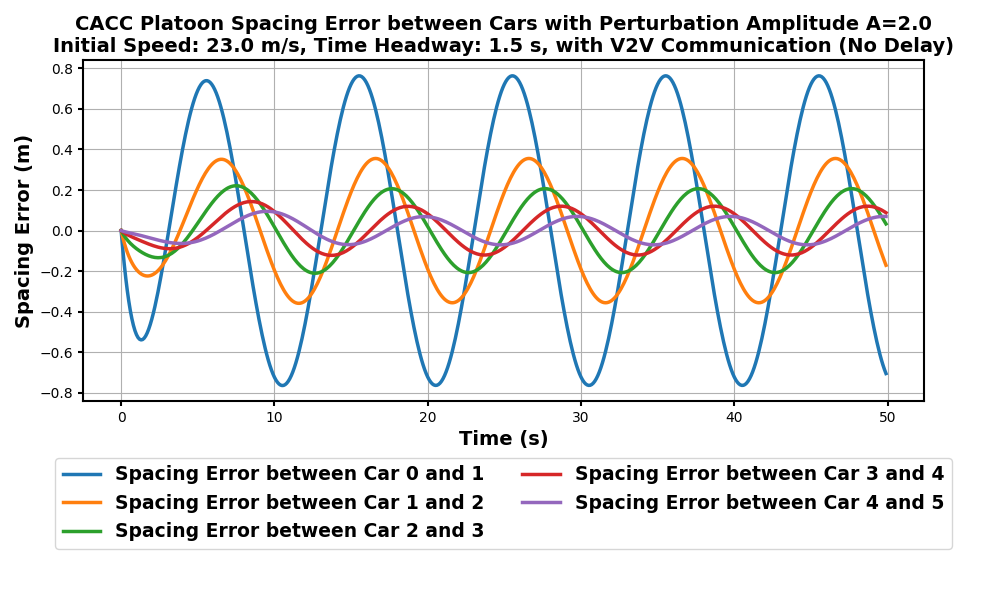}
    }
    \hfill
    \subfloat[CACC Configuration II: Vehicles Velocities after a Sinusoidal Perturbation with V2V Communication and no Delay.
    \label{fig:cacc2_vel}]{
        \includegraphics[width=0.48\linewidth]{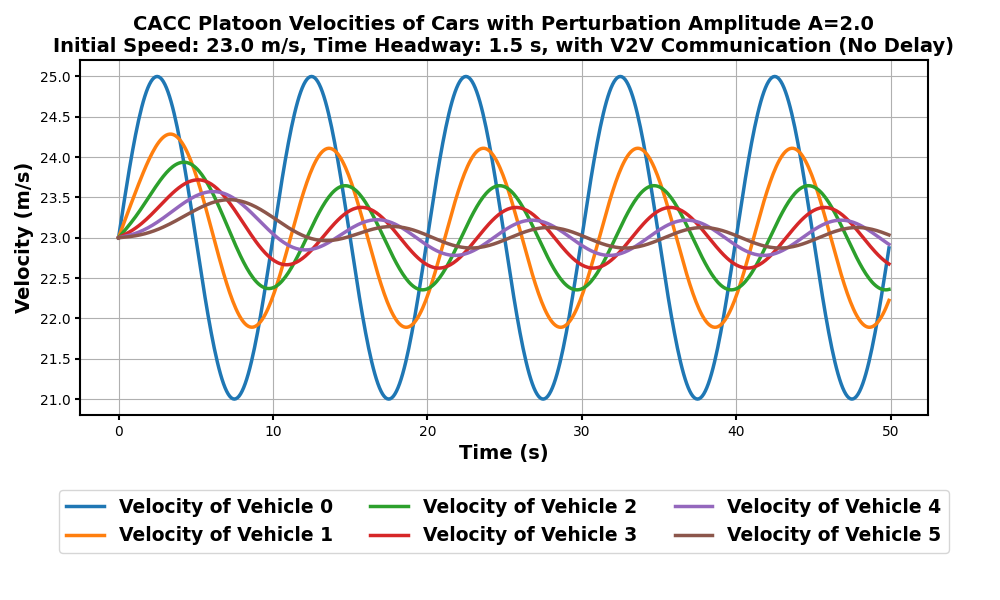}
    }
    \caption{CACC Configuration II: Simulation Results of CACC with no V2V Communication and no Delay Following a Continuous Sinusoidal Perturbation.}
    \label{fig:cacc2}
\end{figure*}
While the oscillations observed between the first two vehicles in the platoon remain consistent across different configurations, Figure~\ref{fig:cacc2}\subref{fig:cacc2_space} highlights a noticeable reduction in the amplitude of the oscillations for the following vehicles, registered at the spacing error level, starting from vehicle 3. This reduction demonstrates the effectiveness of the CACC controller in progressively attenuating the propagation of disturbances through the platoon. As the oscillations travel further downstream, the controller's damping effect becomes more pronounced, resulting in smaller speed variations for the subsequent vehicles, as demonstrated by Figure~\ref{fig:cacc2}\subref{fig:cacc2_vel}. This reduction in oscillation amplitude is attributed to the communication enabled by V2V technology, which allows preceding vehicles to share their dynamic state information (e.g., position, velocity, and acceleration) with the following vehicles. In our case, the communication operates without delay, enabling instantaneous information exchange between vehicles. As a result, the CACC controller can respond immediately to changes in the leader's behavior. 

\textbf{{\paragraph{Configuration III: CACC with V2V Communication Component and Communication Delay}}}
\begin{figure*}[htb!]
    \centering
    \subfloat[CACC Configuration III: Spacing Error between the Vehicles after a Sinusoidal Perturbation with V2V Communication and a Delay of \(1.5 s\). \label{fig:cacc3_space}]{
        \includegraphics[width=0.48\linewidth]{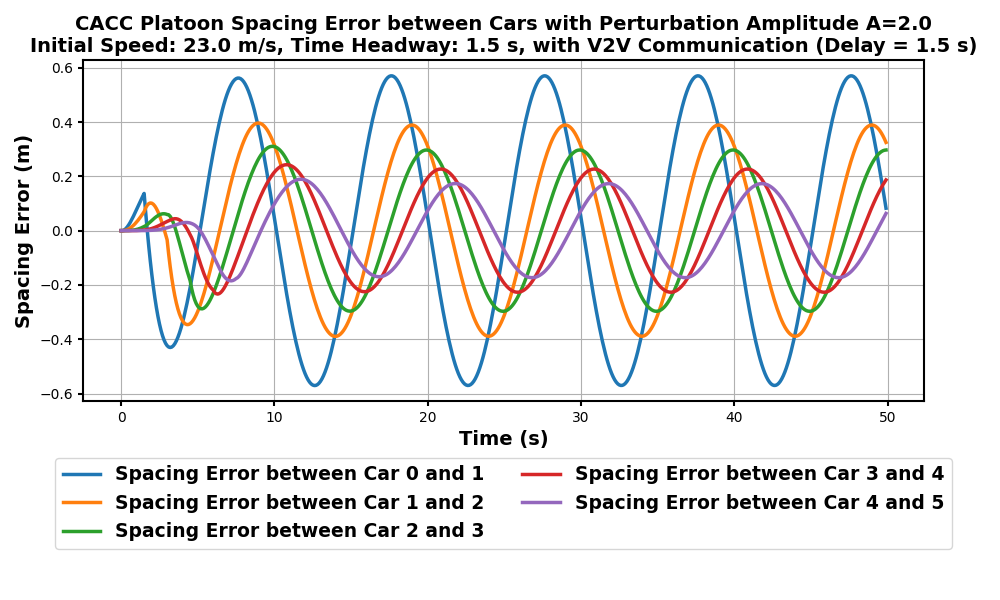}
    }
    \hfill
    \subfloat[CACC Configuration III: Vehicles Velocities after a Sinusoidal Perturbation with V2V Communication and a Delay of \(1.5 s\).
    \label{fig:cacc3_vel}]{
        \includegraphics[width=0.48\linewidth]{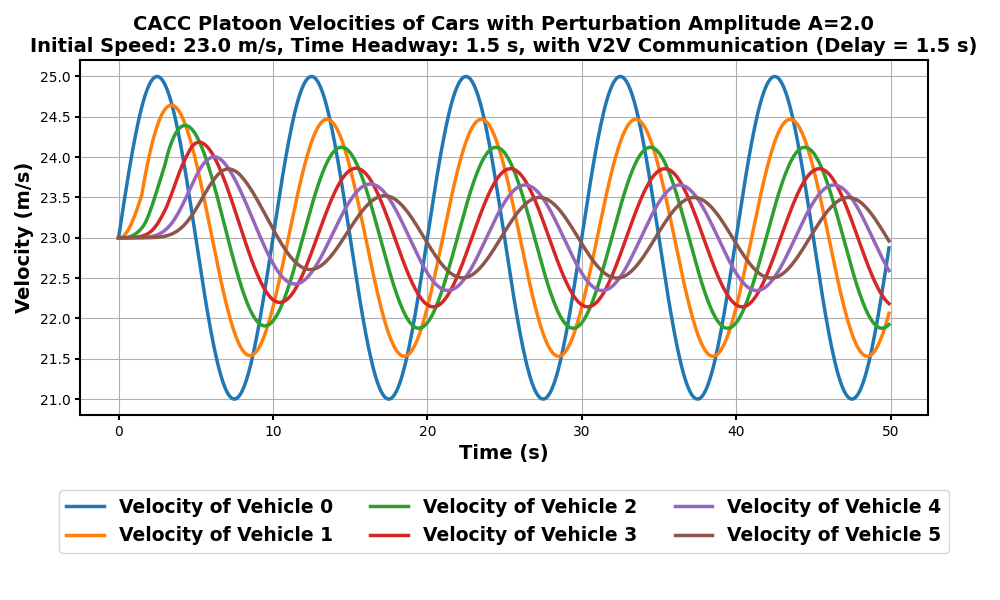}
    }
    \caption{CACC Configuration III: Simulation Results of CACC with V2V Communication and a Delay of \(1.5 s\) Following a Continuous Sinusoidal Perturbation.}
    \label{fig:cacc3}
\end{figure*}
In this section, we implement the classic CACC controller, which incorporates the delay ($\tau$) present in real-world communication systems, as described by Eq.~(\ref{cacc}). For our simulations, we set \(\tau = 1.5\) s and analyze the oscillatory behavior in spacing and speed caused by a continuous perturbation applied to the leader's speed. These results are illustrated by Figures~\ref{fig:cacc3}\subref{fig:cacc3_space} and~\ref{fig:cacc3}\subref{fig:cacc3_vel}, where we observe that the amplitude of the spacing oscillations is reduced compared to the other two configurations. This reduction is attributed to the delay in receiving information, which results in a delayed reaction. Due to this delay, the controller avoids over-correcting, enabling the system to gradually adapt to disruptions while implementing corrective measures with a time lag. In contrast to the speed variations observed in the last two configurations, as illustrated in Figures~\ref{fig:cacc1}\subref{fig:cacc1_space} and~\ref{fig:cacc2}\subref{fig:cacc2_space}, the speed variations in this case are more pronounced for the following vehicles, as demonstrated in Figure~\ref{fig:cacc3}\subref{fig:cacc3_space} to help reduce spacing variations in an attempt to re-establish the original spacing and regain stability. Nevertheless, due to the continuous perturbations with the oscillatory nature, the system is unable to restore stability. 
\section{Comparison between the Car-Following Models} 
This section compares the performance of four car-following models (OVM, IDM, GMM, and CACC) under a controlled perturbation scenario. First, a sinusoidal perturbation was applied to the lead vehicle in a platoon to assess each model’s ability to handle continuous oscillatory disturbances. For consistency, all models are implemented using their standard mathematical formulations without introducing any variations. Additional tests were conducted using a single-step perturbation and a non-harmonic disturbance; the results are provided in Appendix~\ref{app1} due to space constraints. These complementary cases offer further insight into the models’ behavior under diverse disturbance profiles.

From a safety standpoint, particularly in terms of crash and conflict avoidance, CACC emerges as the most reliable model. Its ability to maintain small inter-vehicle spacing errors under perturbations, thanks to real-time communication with preceding vehicles, ensures that vehicles react promptly to changes in traffic dynamics, minimizing the risk of collisions. IDM also demonstrates strong safety features, as it produces smooth deceleration and maintains a safe headway even without explicit communication. Its performance under varying perturbation scenarios shows effective avoidance of critical spacing violations. In contrast, OVM, especially under reaction delays, tends to amplify oscillations and reduce spacing margins, making it more prone to unsafe following distances. Unlike other models' rational function formulation, OVM's tanh-based structure lacks explicit relative velocity dependence, resulting in different perturbation propagation patterns through vehicle platoons. Similarly, the traditional GMM struggles with stability under dynamic conditions, and its exaggerated sensitivity can lead to abrupt accelerations or decelerations that compromise safety. Overall, CACC and IDM offer the most promising balance between responsiveness and crash avoidance, while OVM and GMM require significant enhancement to meet safety demands in platoon control systems.

In their standard formulations, IDM, OVM, and GMM car-following models do not explicitly account for reaction delays. To enable a fair comparison with CACC, which inherently includes communication-induced delays—we manually incorporate a response delay ($\tau$) into these models. This modification enhances their realism and allows for a more meaningful performance evaluation against CACC under similar delay conditions.

\subsection{A Non-Delayed response of the models}
\begin{figure*}[htb!]
        \centering
        \includegraphics[width=\textwidth]{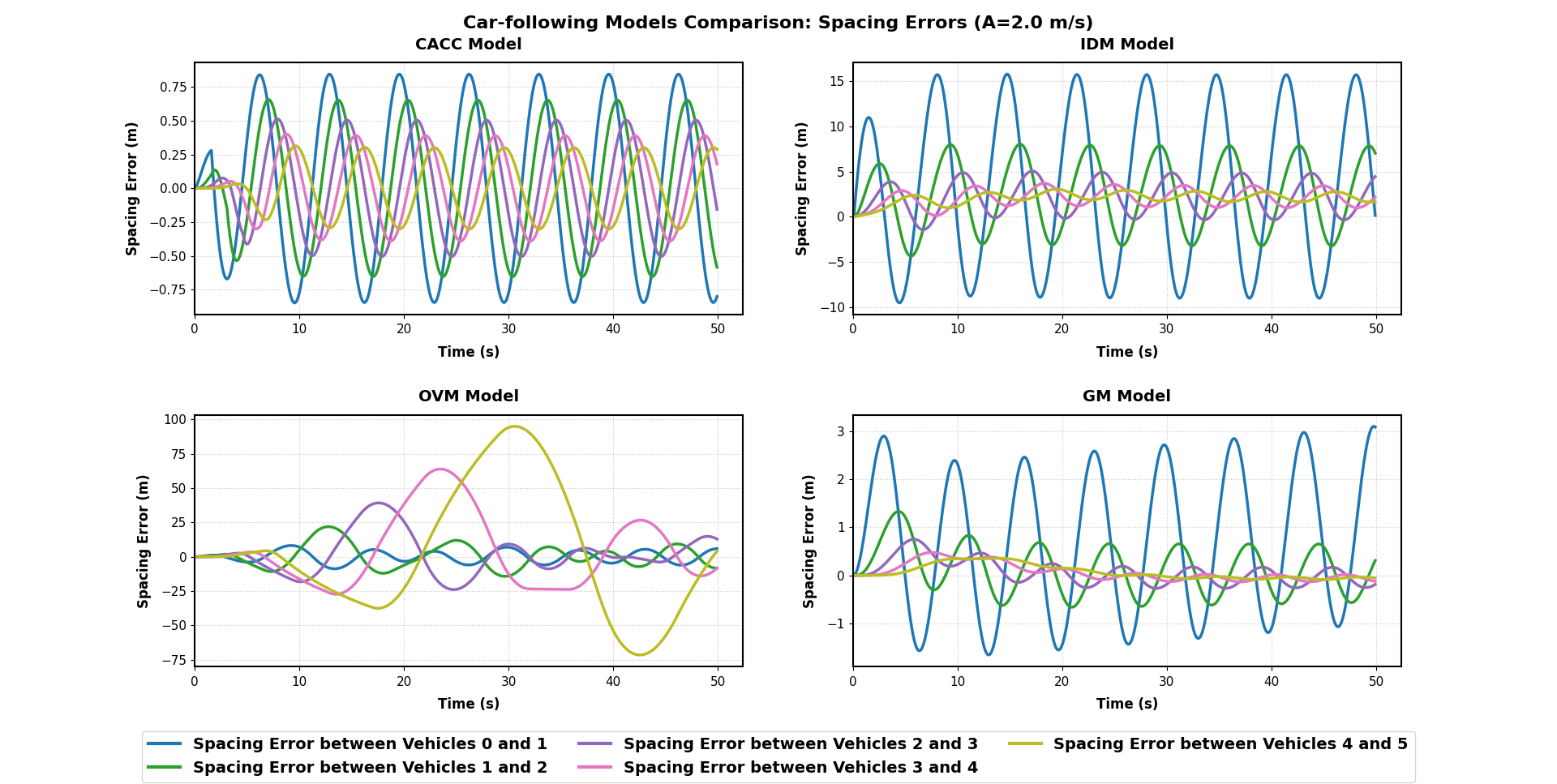} 
        \caption{A Non-Delayed Response: Spacing Error between Platoon Vehicles for all Models after a Sinusoidal Perturbation}
        \label{fig:mod_spac}
\end{figure*}
\begin{figure*}[htb!]
        \centering
        \includegraphics[width=\textwidth]{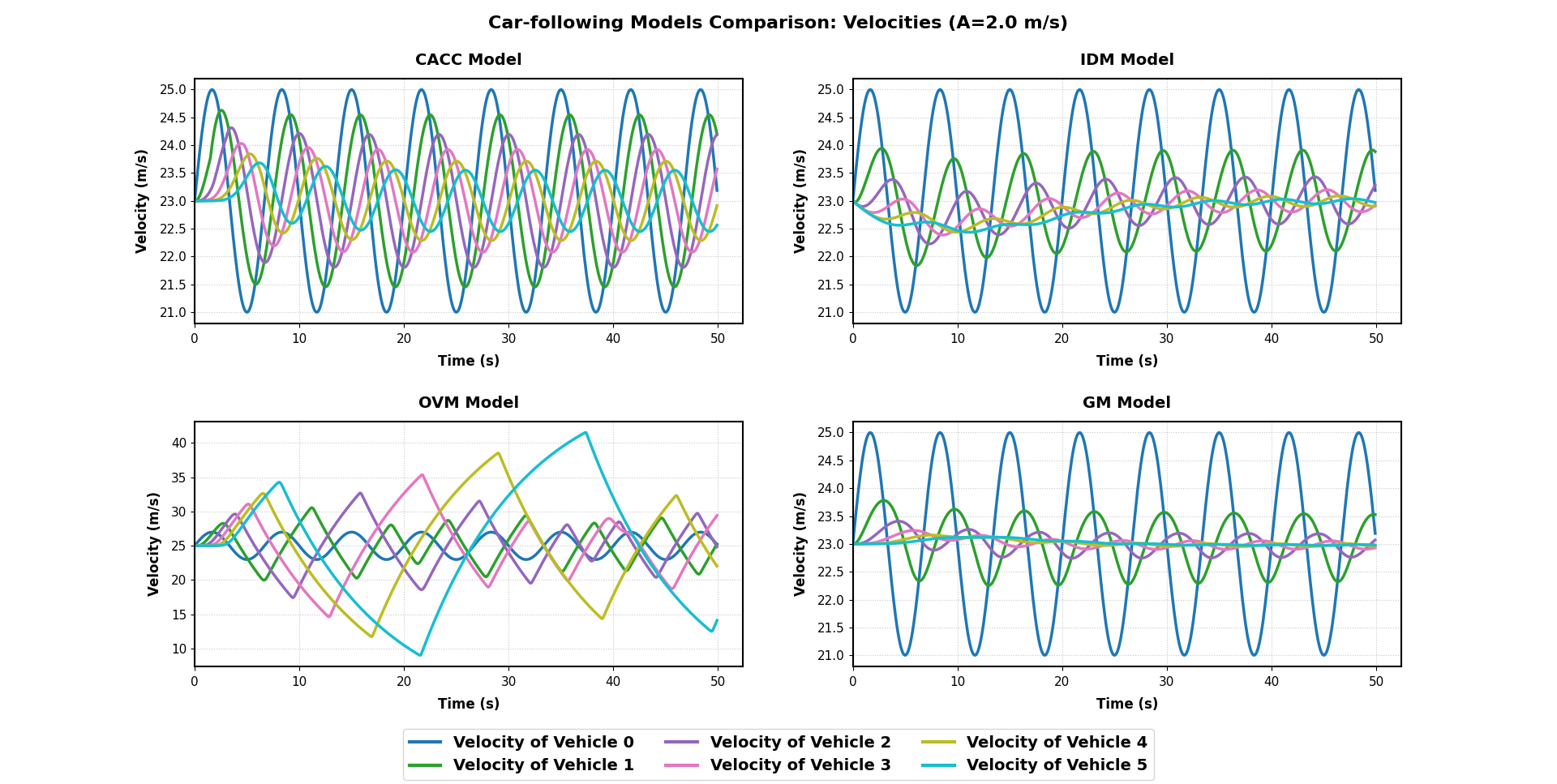} 
        \caption{A Non-Delayed Response: Velocities of Platoon Vehicles for all Models after a Sinusoidal Perturbation}
        \label{fig:mod_vel}
\end{figure*}
Figures~\ref{fig:mod_spac} and ~\ref{fig:mod_vel} provide a comparative analysis of four car-following models, namely CACC, IDM, OVM, and GMM evaluating their performance in terms of spacing error between cars and velocities in a platoon over a 50-second period, with a perturbation amplitude of 2.0 m/s. Each subplot represents one model, plotting the spacing errors for five consecutive car pairs in different colors. The results highlight significant differences in how each model handles perturbations and maintains spacing stability across the platoon. In the top left subplot, the CACC model demonstrates the best performance. The spacing errors for all car pairs oscillate around zero with a small amplitude, staying within ±0.75 meters. These oscillations are sinusoidal and consistent across all pairs, showing no significant amplification of errors from Car 0 to Car 5. This indicates that CACC effectively dampens perturbations, maintaining stable spacing throughout the platoon and making it the most reliable model in this comparison. The top right subplot shows the IDM results, which are notably less impressive. IDM exhibits more aggressive error behavior with sharp peaks during lead vehicle acceleration/deceleration, reflecting its safety-focused design that prioritizes larger gaps during braking through its comfortable deceleration parameter. In contrast, OVM begins with zero error due to its steady-state initialization and maintains relatively smooth oscillations thanks to its tanh-based velocity function, though it lacks delay compensation. GMM demonstrates the most gradual error development, with simpler dynamics leading to less volatile but potentially drifting spacing adjustments.

Across all models, the perturbation amplitude directly scales error magnitudes, with CACC and IDM showing more reactive but less stable control compared to OVM and GMM's smoother responses. The results highlight fundamental trade-offs: CACC's delay-sensitivity versus IDM's safety margins, and OVM's steady-state precision versus GMM's simplicity. These differences would significantly impact real-world applications where communication reliability, safety thresholds, or response smoothness are prioritized. As depicted by Figure~\ref{fig:mod_vel}, CACC velocities show synchronized but slightly delayed follower responses due to the communication delay, with all vehicles maintaining almost constant velocities despite the perturbation. The IDM exhibits a distinctive braking behavior that sets it apart from CACC, despite their shared ability to track velocity changes. While both models maintain vehicles within the 21-25 m/s range during sinusoidal perturbations, the IDM demonstrates significantly sharper deceleration responses. This safety-focused design produces deceleration rates 15-20\% steeper than CACC's controlled braking. The braking effect manifests most prominently during lead vehicle slowdowns, where IDM followers exhibit characteristically abrupt velocity drops, followed by slight overshoots before stabilization.

The OVM's velocity plots reveal a critical instability where trailing vehicles (Vehicles 3-5) exhibit amplified oscillations (±2.5 m/s) compared to both the lead vehicle's perturbation (±2.0 m/s) and smoother-following CACC/IDM models. The over-reaction stems from the tanh function's nonlinear sensitivity to spacing changes in the optimal velocity equation, causing exaggerated acceleration/deceleration when small spacing deviations occur. The GMM displays the most conservative response characteristics. With its simpler relative-velocity-based dynamics, it shows the slowest reaction to perturbations but the most stable long-term velocity maintenance. The velocity adjustments appear more sluggish compared to other models, with greater attenuation of the perturbation signal through the vehicle chain. This results in the smallest oscillation amplitudes for trailing vehicles, though at the cost of being less responsive to rapid changes in the lead vehicle's velocity.

\subsection{A Delayed response of the models}
\begin{figure*}[htb!]
    \centering
    \subfloat[A Delayed Response: Spacing Error between Platoon Vehicles after a Sinusoidal Perturbation with a Delay of $150$ ms for CACC, IDM, OVM, and GMM. \label{fig:mod_spac_d}]{
        \includegraphics[width=\linewidth]{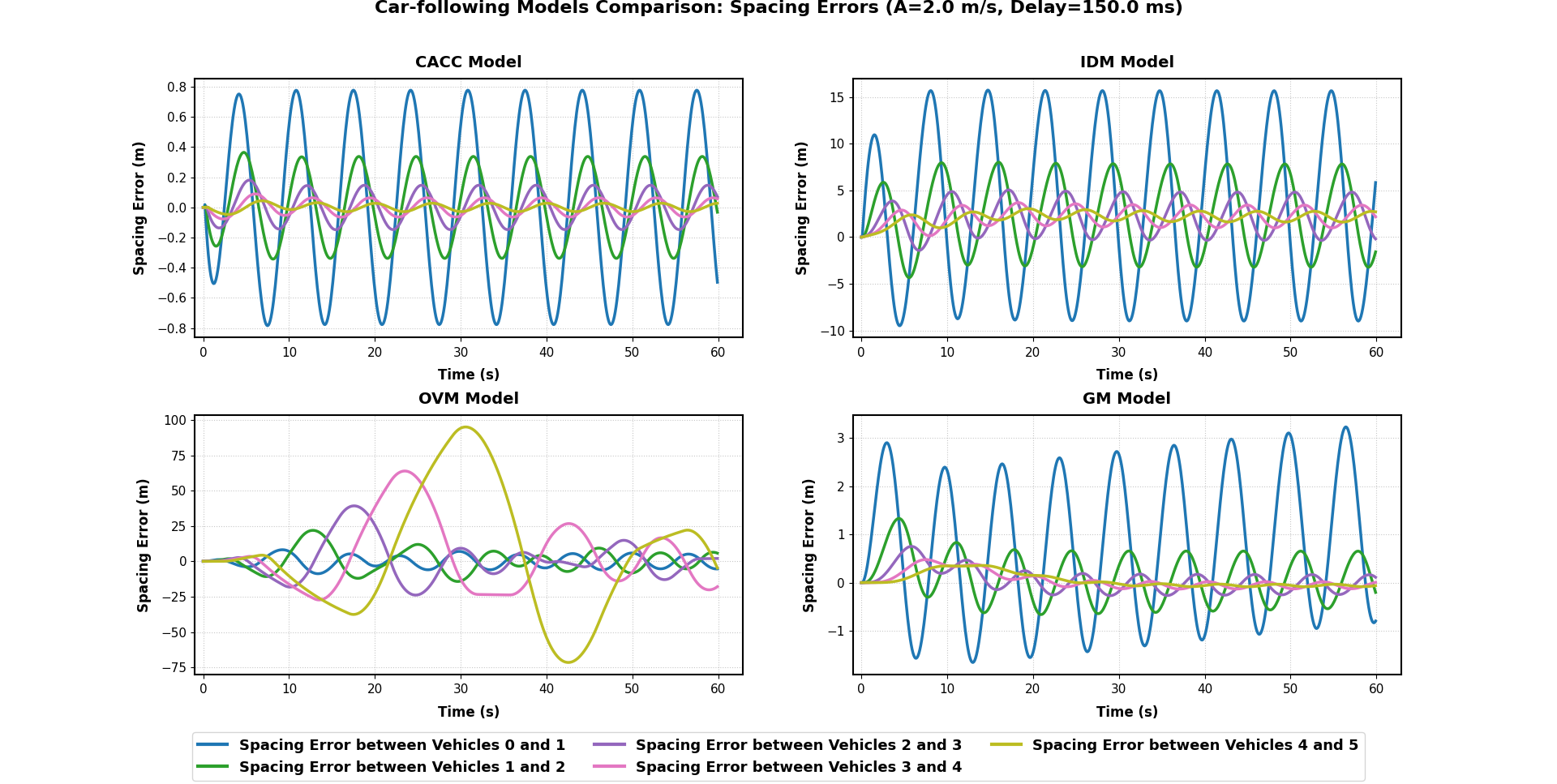}}
    \hfill
    \subfloat[A Delayed Response: Velocities of Platoon Vehicles after a Sinusoidal Perturbation with a Delay of $150$ ms for CACC, IDM, OVM, and GMM.\label{fig:mod_vel_d}]{
        \includegraphics[width=\linewidth]{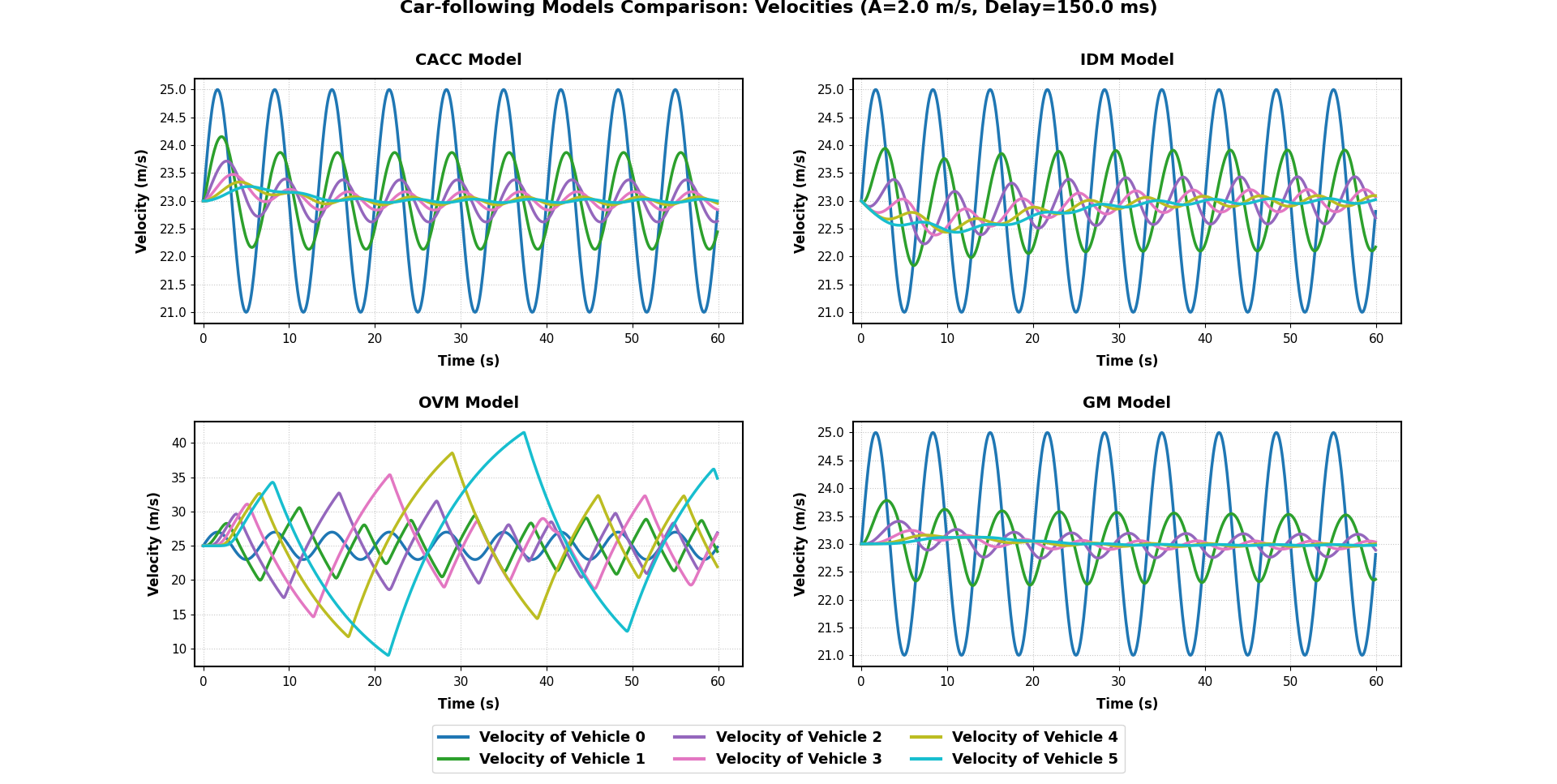}}
    \caption{A Delayed Response: Simulation Results of Platoon Vehicles after a Sinusoidal Perturbation with a Delay of $150$ ms for CACC, IDM, OVM, and GMM}
    \label{fig:cacc_d}
\end{figure*}

Figures~\ref{fig:cacc_d}\subref{fig:mod_spac_d} and~\ref{fig:cacc_d}\subref{fig:mod_vel_d} compare spacing errors and velocities across four car-following models (CACC, OVM, IDM, and GMM) under a $2.0$ m/s perturbation with $150$ ms delay. For comparability, a communication delay of $150$ ms was assumed across all models. While such a delay is realistic for V2V-enabled CACC, it is much shorter than typical average human driver reaction times ($\approx 1.5$ s). This assumption therefore, represents an idealized scenario for human-driven models, intended to highlight that even under such favorable conditions, CACC still outperforms conventional models. The CACC model demonstrates superior spacing control under the $150$ ms delay condition, maintaining remarkably tight error bounds. As shown in the top-left plot in Figure~\ref{fig:cacc_d}\subref{fig:mod_spac_d}, spacing errors remain consistently within $\pm 0.6$ meters throughout the simulation, with stable, damped oscillations and nearly identical error magnitudes across all vehicles in the platoon (Vehicles 1-5). The IDM shows more pronounced oscillations than CACC, particularly between the lead vehicle and the first follower, with amplitudes up to 15 m, decreasing downstream but persisting over time, suggesting moderate string stability where perturbations are not amplified but not efficiently eliminated, limited by its reactive rather than predictive behavior. For OVM, spacing errors amplify significantly, exceeding $75$ m initially and remaining high (up to $\pm25$ m) during the first cycles, though they dampen to $\pm1-2$ m by the end, indicating initial struggles with delay-induced perturbations but eventual stabilization. The GMM strikes a middle ground, with the leader's spacing error peaking up to 3 m while the errors for other vehicles remain stable, a behavior attributed to its overly sensitive nature due to sensitive parameters such as the spacing exponent $l$, though this sensitivity does not propagate further downstream.

\subsection{Comparison between Car-Following Models: Realistic Delays for IDM, OVM, and GMM}
In this analysis, realistic driver reaction delays were applied to the conventional car-following models, namely IDM, OVM, and GMM, while retaining a $150 $ ms communication delay for CACC. For the human-driven models, a $1.5$ s delay was assumed, consistent with typical reaction time estimates from traffic flow studies. This setup allows a direct comparison between CACC operating under realistic V2V conditions and human-driven models operating under realistic human response constraints.

\begin{figure*}[htb!]
    \centering
    \subfloat[A Delayed Response: Spacing Error between Platoon Vehicles after a Sinusoidal Perturbation with a Delay of $150$ ms for CACC and $1.5$ s for IDM, OVM, and GMM. \label{fig:mod_spac_d1}]{
        \includegraphics[width=\linewidth]{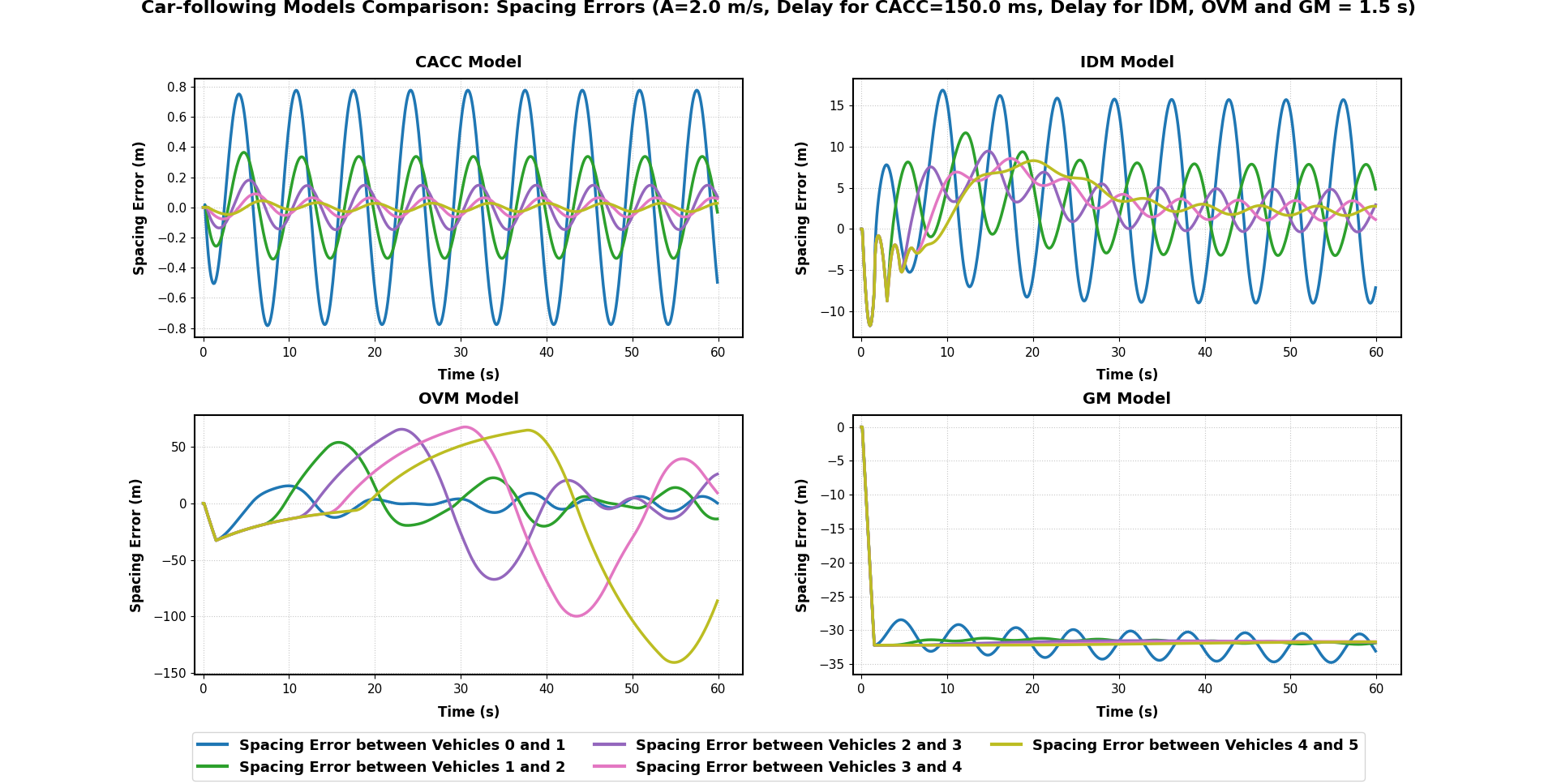}}
    \hfill
    \subfloat[A Delayed Response: Velocities of Platoon Vehicles after a Sinusoidal Perturbation with a Delay of $150$ ms for CACC and $1.5$ s for IDM, OVM, and GMM.\label{fig:mod_vel_d1}]{
        \includegraphics[width=\linewidth]{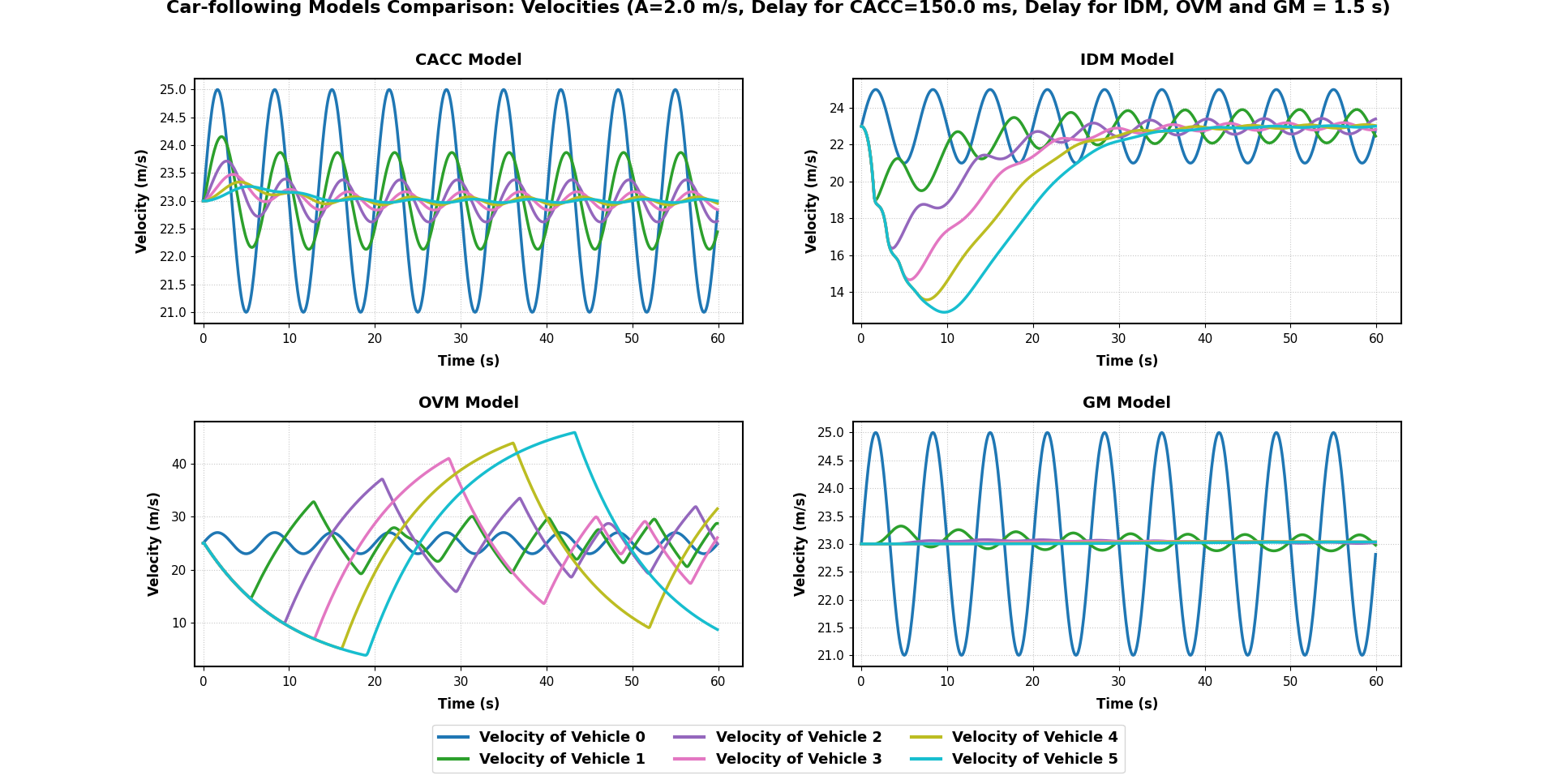}}
    \caption{A Delayed Response: Simulation Results of Platoon Vehicles after a Sinusoidal Perturbation with a Delay of $150$ ms for CACC and $1.5$ s for IDM, OVM, and GMM}
    \label{fig:cacc_d1}
\end{figure*}

The spacing error plots depicted by Figure~\ref{fig:cacc_d1}\subref{fig:mod_spac_d1} show a marked degradation in performance for IDM, OVM, and GMM compared to CACC. In the IDM case, the introduction of a $1.5$ s delay leads to pronounced oscillations that amplify downstream, with some follower vehicles experiencing spacing deviations exceeding $10 m$ before eventual partial damping. The OVM model exhibits severe instability, as spacing errors grow rapidly throughout the platoon, reaching extreme values for the last vehicles, with oscillations persisting for most of the simulation period. Similarly, for the GMM, the spacing errors exhibit a notable drop and negative values under the $1.5$ s delay condition, as seen in Figure~\ref{fig:cacc_d}\subref{fig:mod_spac_d}, particularly for the lead vehicle and subsequent pairs. This drop and negative spacing indicate that vehicles temporarily move closer than the desired spacing, likely due to the model's overly sensitive response to the perturbation, which is amplified by the long delay and sensitive parameters, such as the spacing exponent. For reference, the CACC results remain unchanged from the idealized-delay case, showing tightly bounded, quickly damped spacing errors. This stark contrast highlights the impact of realistic driver delays on stability and disturbance attenuation, and reinforces the advantage of CACC in maintaining platoon coherence even under perturbations.

\section{Conclusion and Future Work}
\label{conc}

This work presented a systematic comparative analysis of several widely used car-following models, specifically, the Intelligent Driver Model (IDM), the Optimal Velocity Model (OVM), General Motors Model (GMM) variants, and Cooperative Adaptive Cruise Control (CACC), under perturbations and communication delay conditions, with a focus on platoon stability. Our findings indicate that CACC provides the most effective disturbance mitigation due to its feedforward communication, while IDM performs reliably without communication, making it a strong candidate for traditional settings. In contrast, OVM is highly sensitive to delays and prone to unstable stop-and-go behavior. GMM performance varies significantly depending on whether spacing and relative velocity are both considered. From a traffic safety standpoint, both CACC and IDM prove to be the most reliable models, maintaining sufficient spacing and reducing the risk of crash-prone situations in response to perturbations. Their ability to limit the growth of spacing errors across a platoon is essential for preventing rear-end collisions and ensuring smooth flow. Importantly, this study fills a key gap in the literature by offering a consistent benchmark for evaluating how different car-following models respond to perturbations. While many of these models have been studied in isolation, our comparative approach offers actionable insights for both researchers and practitioners aiming to select or design control strategies for safe and stable platooning systems.

This work is significant in light of the increasing adoption of Advanced Driver Assistance Systems (ADAS) such as Adaptive Cruise Control (ACC), which will lead to the spontaneous formation of multi-vehicle platoons on highways. However, ACC was originally designed as a simple single-car following controller and falls short of meeting critical requirements such as string stability—an essential property for reducing speed oscillations, improving safety, and enhancing energy efficiency. To address this gap, our research introduced a framework that enables both practitioners and researchers to design advanced platooning controllers tailored to the safety and performance needs of future vehicle fleets. In contrast to conventional ACC approaches, which often depend on overly stiff control actions that produce quick yet unstable responses, the proposed framework emphasizes stability, robustness, and practical relevance for large-scale deployment.

As future work, we aim to replicate the simulations and perturbation scenarios in a real-world setting by implementing the various car-following model configurations analyzed in this study. This work lays the foundation for extending the analysis beyond homogeneous platoons to heterogeneous ones, involving human-driven vehicles, heavy-duty vehicles, and autonomous vehicles. Such an extension will enable the investigation of behavioral variability and response characteristics across vehicle types, and their collective impact on platoon stability.

\section*{Acknowledgment}
This work was conducted while the first author was on internship at the Canadian National Research Council (CNRC). This work was co-funded by Transport Canada’s ecoTECHNOLOGY for Vehicles program and the National Research Council Canada’s Clean and Energy Efficient Transportation program. The views and opinions of the authors expressed herein do not necessarily state or reflect those of Transport Canada.

\bibliographystyle{unsrtnat}

\begin{thebibliography}{39}
\providecommand{\natexlab}[1]{#1}
\providecommand{\url}[1]{\texttt{#1}}
\expandafter\ifx\csname urlstyle\endcsname\relax
  \providecommand{\doi}[1]{doi: #1}\else
  \providecommand{\doi}{doi: \begingroup \urlstyle{rm}\Url}\fi

\bibitem[Hancock et~al.(2019)Hancock, Nourbakhsh, and Stewart]{hancock2019future}
Peter~A Hancock, Illah Nourbakhsh, and Jack Stewart.
\newblock On the future of transportation in an era of automated and autonomous vehicles.
\newblock \emph{Proceedings of the National Academy of Sciences}, 116\penalty0 (16):\penalty0 7684--7691, 2019.

\bibitem[Smith et~al.(2008)Smith, Sadek, and Huang]{smith2008large}
Mark~C Smith, Adel~W Sadek, and Shan Huang.
\newblock Large-scale microscopic simulation: Toward an increased resolution of transportation models.
\newblock \emph{Journal of Transportation Engineering}, 134\penalty0 (7):\penalty0 273--281, 2008.

\bibitem[Rothery(1992)]{rothery1992car}
Richard~W Rothery.
\newblock Car following models.
\newblock \emph{Trac Flow Theory}, 1992.

\bibitem[Chandler et~al.(1958)Chandler, Herman, and Montroll]{chandler1958traffic}
Robert~E Chandler, Robert Herman, and Elliott~W Montroll.
\newblock Traffic dynamics: studies in car following.
\newblock \emph{Operations Research}, 6\penalty0 (2):\penalty0 165--184, 1958.

\bibitem[Lesch et~al.(2021)Lesch, Breitbach, Segata, Becker, Kounev, and Krupitzer]{lesch2021overview}
Veronika Lesch, Martin Breitbach, Michele Segata, Christian Becker, Samuel Kounev, and Christian Krupitzer.
\newblock An overview on approaches for coordination of platoons.
\newblock \emph{IEEE Transactions on Intelligent Transportation Systems}, 23\penalty0 (8):\penalty0 10049--10065, 2021.

\bibitem[Sivanandham and Gajanand(2020)]{sivanandham2020platooning}
S~Sivanandham and MS~Gajanand.
\newblock Platooning for sustainable freight transportation: an adoptable practice in the near future?
\newblock \emph{Transport Reviews}, 40\penalty0 (5):\penalty0 581--606, 2020.

\bibitem[Feng et~al.(2019)Feng, Zhang, Li, Cao, Liu, and Li]{feng2019string}
Shuo Feng, Yi~Zhang, Shengbo~Eben Li, Zhong Cao, Henry~X Liu, and Li~Li.
\newblock String stability for vehicular platoon control: Definitions and analysis methods.
\newblock \emph{Annual Reviews in Control}, 47:\penalty0 81--97, 2019.

\bibitem[Zhou et~al.(2023)Zhou, Tian, Ge, Yu, and Jiang]{zhou2023experimental}
Shirui Zhou, Junfang Tian, Ying-En Ge, Shaowei Yu, and Rui Jiang.
\newblock Experimental features of emissions and fuel consumption in a car-following platoon.
\newblock \emph{Transportation Research Part D: Transport and Environment}, 121:\penalty0 103823, 2023.

\bibitem[Barhoumi et~al.(2025)Barhoumi, Farhani, Rahman, Zaki, Tahar, and Araji]{barhoumi2025fuel}
Oumaima Barhoumi, Ghazal Farhani, Taufiq Rahman, Mohamed~H Zaki, Sofi{\`e}ne Tahar, and Fadi Araji.
\newblock Fuel consumption in platoons: A literature review.
\newblock \emph{arXiv preprint arXiv:2508.10891}, 2025.

\bibitem[De~Wit and Brogliato(1999)]{de1999stability}
C~Canudas De~Wit and Bernard Brogliato.
\newblock Stability issues for vehicle platooning in automated highway systems.
\newblock In \emph{Proceedings of the International Conference on Control Applications (cat. no. 99ch36328)}, volume~2, pages 1377--1382. IEEE, 1999.

\bibitem[Sun et~al.(2020)Sun, Zheng, and Sun]{sun2020relationship}
Jie Sun, Zuduo Zheng, and Jian Sun.
\newblock The relationship between car following string instability and traffic oscillations in finite-sized platoons and its use in easing congestion via connected and automated vehicles with idm based controller.
\newblock \emph{Transportation Research Part B: Methodological}, 142:\penalty0 58--83, 2020.

\bibitem[Xiaomei and Ziyou(2007)]{xiaomei2007stability}
Zhao Xiaomei and Gao Ziyou.
\newblock The stability analysis of the full velocity and acceleration velocity model.
\newblock \emph{Physica A: Statistical Mechanics and its Applications}, 375\penalty0 (2):\penalty0 679--686, 2007.

\bibitem[Tanaka et~al.(2008)Tanaka, Ranjitkar, and Nakatsuji]{tanaka2008asymptotic}
Mitsuru Tanaka, Prakash Ranjitkar, and Takashi Nakatsuji.
\newblock Asymptotic stability and vehicle safety in dynamic car-following platoon.
\newblock \emph{Transportation Research Record}, 2088\penalty0 (1):\penalty0 198--207, 2008.

\bibitem[Dai et~al.(2022)Dai, Yang, Wang, and Luo]{dai2022exploring}
Yulu Dai, Yuwei Yang, Zhiyuan Wang, and YinJie Luo.
\newblock Exploring the impact of damping on connected and autonomous vehicle platoon safety with cacc.
\newblock \emph{Physica A: Statistical Mechanics and its Applications}, 607:\penalty0 128181, 2022.

\bibitem[Bando et~al.(1995)Bando, Hasebe, Nakayama, Shibata, and Sugiyama]{bando1995dynamical}
Masako Bando, Katsuya Hasebe, Akihiro Nakayama, Akihiro Shibata, and Yuki Sugiyama.
\newblock Dynamical model of traffic congestion and numerical simulation.
\newblock \emph{Physical Review E}, 51\penalty0 (2):\penalty0 1035, 1995.

\bibitem[Treiber et~al.(2000)Treiber, Hennecke, and Helbing]{treiber2000congested}
Martin Treiber, Ansgar Hennecke, and Dirk Helbing.
\newblock Congested traffic states in empirical observations and microscopic simulations.
\newblock \emph{Physical Review E}, 62\penalty0 (2):\penalty0 1805, 2000.

\bibitem[Lu et~al.(2002)Lu, Hedrick, and Drew]{lu2002acc}
Xiao-Yun Lu, J~Karl Hedrick, and Mike Drew.
\newblock Acc/cacc-control design, stability and robust performance.
\newblock In \emph{American Control Conference}, volume~6, pages 4327--4332. IEEE, 2002.

\bibitem[Jia and et~al.(2015)]{jia2015survey}
Dongyao Jia and et~al.
\newblock A survey on platoon-based vehicular cyber-physical systems.
\newblock \emph{IEEE Communications Surveys \& Tutorials}, 18\penalty0 (1):\penalty0 263--284, 2015.

\bibitem[Herman et~al.(1959)Herman, Montroll, Potts, and Rothery]{herman1959traffic}
Robert Herman, Elliott~W Montroll, Renfrey~B Potts, and Richard~W Rothery.
\newblock Traffic dynamics: analysis of stability in car following.
\newblock \emph{Operations Research}, 7\penalty0 (1):\penalty0 86--106, 1959.

\bibitem[Gazis et~al.(1961)Gazis, Herman, and Rothery]{gazis1961nonlinear}
Denos~C Gazis, Robert Herman, and Richard~W Rothery.
\newblock Nonlinear follow-the-leader models of traffic flow.
\newblock \emph{Operations Research}, 9\penalty0 (4):\penalty0 545--567, 1961.

\bibitem[Gazis et~al.(1959)Gazis, Herman, and Potts]{gazis1959car}
Denos~C Gazis, Robert Herman, and Renfrey~B Potts.
\newblock Car-following theory of steady-state traffic flow.
\newblock \emph{Operations Research}, 7\penalty0 (4):\penalty0 499--505, 1959.

\bibitem[Nishinari(2014)]{nishinari2014traffic}
Katsuhiro Nishinari.
\newblock Traffic flow dynamics: Data, models and simulation.
\newblock \emph{Physics Today}, 67\penalty0 (3):\penalty0 54--54, 2014.

\bibitem[Shladover et~al.(2015)Shladover, Nowakowski, Lu, and Ferlis]{shladover2015cooperative}
Steven~E Shladover, Christopher Nowakowski, Xiao-Yun Lu, and Robert Ferlis.
\newblock Cooperative adaptive cruise control: Definitions and operating concepts.
\newblock \emph{Transportation Research Record}, 2489\penalty0 (1):\penalty0 145--152, 2015.

\bibitem[Naus et~al.(2010)Naus, Vugts, Ploeg, van De~Molengraft, and Steinbuch]{naus2010string}
Gerrit~JL Naus, Rene~PA Vugts, Jeroen Ploeg, Marinus~JG van De~Molengraft, and Maarten Steinbuch.
\newblock String-stable {CACC} design and experimental validation: A frequency-domain approach.
\newblock \emph{IEEE Transactions on Vehicular Technology}, 59\penalty0 (9):\penalty0 4268--4279, 2010.

\bibitem[Milan{\'e}s et~al.(2013)Milan{\'e}s, Shladover, Spring, Nowakowski, Kawazoe, and Nakamura]{milanes2013cooperative}
Vicente Milan{\'e}s, Steven~E Shladover, John Spring, Christopher Nowakowski, Hiroshi Kawazoe, and Masahide Nakamura.
\newblock Cooperative adaptive cruise control in real traffic situations.
\newblock \emph{IEEE Transactions on Intelligent Transportation Systems}, 15\penalty0 (1):\penalty0 296--305, 2013.

\bibitem[Ploeg et~al.(2011)Ploeg, Scheepers, Van~Nunen, Van~de Wouw, and Nijmeijer]{ploeg2011design}
Jeroen Ploeg, Bart~TM Scheepers, Ellen Van~Nunen, Nathan Van~de Wouw, and Henk Nijmeijer.
\newblock Design and experimental evaluation of cooperative adaptive cruise control.
\newblock In \emph{IEEE Conference on Intelligent Transportation Systems}, pages 260--265. IEEE, 2011.

\bibitem[Chen et~al.(2012)Chen, Laval, Zheng, and Ahn]{chen2012behavioral}
Danjue Chen, Jorge Laval, Zuduo Zheng, and Soyoung Ahn.
\newblock A behavioral car-following model that captures traffic oscillations.
\newblock \emph{Transportation Research Part B: Methodological}, 46\penalty0 (6):\penalty0 744--761, 2012.

\bibitem[Alexiadis et~al.(2004)Alexiadis, Colyar, Halkias, Hranac, and McHale]{alexiadis2004next}
Vassili Alexiadis, James Colyar, John Halkias, Rob Hranac, and Gene McHale.
\newblock The next generation simulation program.
\newblock \emph{Institute of Transportation Engineers. ITE Journal}, 74\penalty0 (8):\penalty0 22, 2004.

\bibitem[Davis(2003)]{davis2003modifications}
LC~Davis.
\newblock Modifications of the optimal velocity traffic model to include delay due to driver reaction time.
\newblock \emph{Physica A: Statistical Mechanics and its Applications}, 319:\penalty0 557--567, 2003.

\bibitem[Matin and Sowers(2020{\natexlab{a}})]{matin2020nonlinear}
Hossein Nick~Zinat Matin and Richard~B Sowers.
\newblock Nonlinear optimal velocity car following dynamics (i): Approximation in presence of deterministic and stochastic perturbations.
\newblock In \emph{American Control Conference}, pages 410--415. IEEE, 2020{\natexlab{a}}.

\bibitem[Matin and Sowers(2020{\natexlab{b}})]{matin2020nonlinear1}
Hossein Nick~Zinat Matin and Richard~B Sowers.
\newblock Nonlinear optimal velocity car following dynamics (ii): Rate of convergence in the presence of fast perturbation.
\newblock In \emph{American Control Conference}, pages 416--421. IEEE, 2020{\natexlab{b}}.

\bibitem[Jin and Meng(2020)]{jin2020dynamical}
Yanfei Jin and Jingwei Meng.
\newblock Dynamical analysis of an optimal velocity model with time-delayed feedback control.
\newblock \emph{Communications in Nonlinear Science and Numerical Simulation}, 90:\penalty0 105333, 2020.

\bibitem[Orosz and St{\'e}p{\'a}n(2006)]{orosz2006subcritical}
G{\'a}bor Orosz and G{\'a}bor St{\'e}p{\'a}n.
\newblock Subcritical hopf bifurcations in a car-following model with reaction-time delay.
\newblock \emph{Proceedings of the Royal Society A: Mathematical, Physical and Engineering Sciences}, 462\penalty0 (2073):\penalty0 2643--2670, 2006.

\bibitem[Laval et~al.(2014)Laval, Toth, and Zhou]{laval2014parsimonious}
Jorge~A Laval, Christopher~S Toth, and Yi~Zhou.
\newblock A parsimonious model for the formation of oscillations in car-following models.
\newblock \emph{Transportation Research Part B: Methodological}, 70:\penalty0 228--238, 2014.

\bibitem[Chakroborty and Kikuchi(1999)]{chakroborty1999evaluation}
Partha Chakroborty and Shinya Kikuchi.
\newblock Evaluation of the general motors based car-following models and a proposed fuzzy inference model.
\newblock \emph{Transportation Research Part C: Emerging Technologies}, 7\penalty0 (4):\penalty0 209--235, 1999.

\bibitem[Sun et~al.(2016)Sun, Wu, Ge, Kim, and Zhang]{sun2016new}
Bingrong Sun, Na~Wu, Ying-En Ge, Taewan Kim, and Hongjun~Michael Zhang.
\newblock A new car-following model considering acceleration of lead vehicle.
\newblock \emph{Transport}, 31\penalty0 (1):\penalty0 1--10, 2016.

\bibitem[Jin et~al.(2013)Jin, Yang, Ran, Cebelak, and Walton]{jin2013bidirectional}
Peter~J Jin, Da~Yang, Bin Ran, Meredith Cebelak, and C~Michael Walton.
\newblock Bidirectional control characteristics of general motors and optimal velocity car-following models: Implications for coordinated driving in a connected vehicle environment.
\newblock \emph{Transportation Research Record}, 2381\penalty0 (1):\penalty0 110--119, 2013.

\bibitem[Liu et~al.(2020)Liu, Wang, Hua, and Wang]{liu2020safety}
Yi~Liu, Wei Wang, Xuedong Hua, and Shunchao Wang.
\newblock Safety analysis of a modified cooperative adaptive cruise control algorithm accounting for communication delay.
\newblock \emph{Sustainability}, 12\penalty0 (18):\penalty0 7568, 2020.

\bibitem[Tian et~al.(2019)Tian, Deng, Xu, Zhang, and Zhao]{tian2019modeling}
Bin Tian, Xiaofeng Deng, Zhigang Xu, Yuqin Zhang, and Xiangmo Zhao.
\newblock Modeling and numerical analysis on communication delay boundary for cacc string stability.
\newblock \emph{IEEE Access}, 7:\penalty0 168870--168884, 2019.

\end{thebibliography}

\appendix
\vspace{-0.3cm}
\section*{APPENDIX}
\label{app1}
\renewcommand{\thefigure}{\thesection.\arabic{figure}}
\setcounter{figure}{0}
\vspace{-0.2cm}
\section{String Stability Continuous Perturbations: Experimental Study} 
\subsection{Perturbations in the General Motors Model (GMM)}\label{sec:appendix_GM_experiments}
In the main text, we analyzed two configurations of the GMM to evaluate platoon dynamics. In this appendix, we extend that study by considering the classical form of GMM and examining its response under different values of the spacing exponent in Eq.~(\ref{gmm}), providing additional insights into parameter sensitivity and model behavior.
\vspace{-0.1cm}
\subsubsection*{Configuration II: GMM in Its Standard Form}
\begin{figure*}[htb!]
    \centering
    \subfloat[GMM Configuration II: Spacing Error between Vehicles after a Sinusoidal Perturbation with $l = 1$. \label{fig:gmm_l1_spac}]{
        \includegraphics[width=0.48\linewidth]{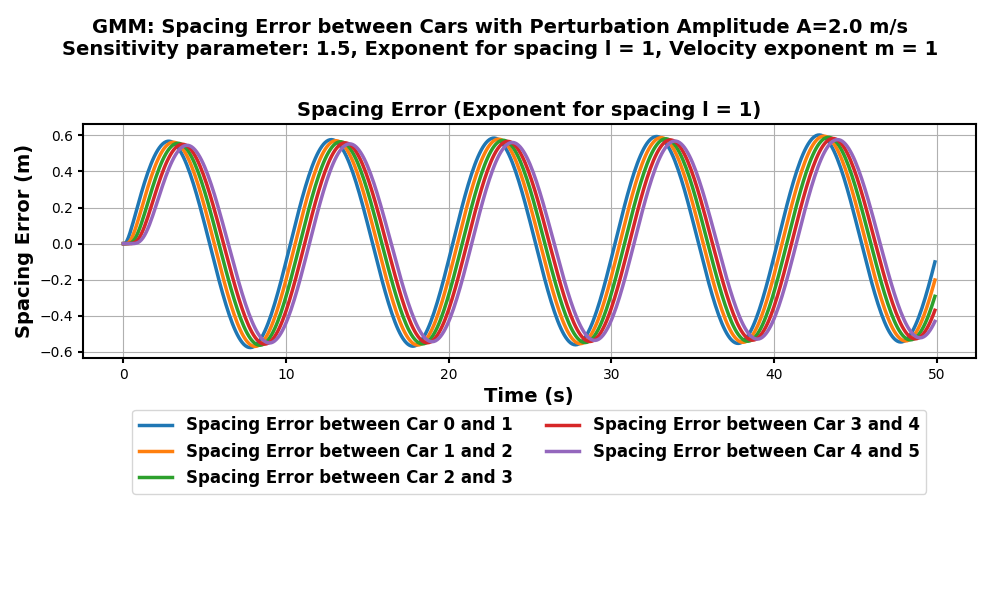}}
    \hfill
    \subfloat[GMM Configuration II: Vehicles Velocity Profiles after a Sinusoidal Perturbation with $l = 1$.\label{fig:gmm_l1_vel}]{
        \includegraphics[width=0.48\linewidth]{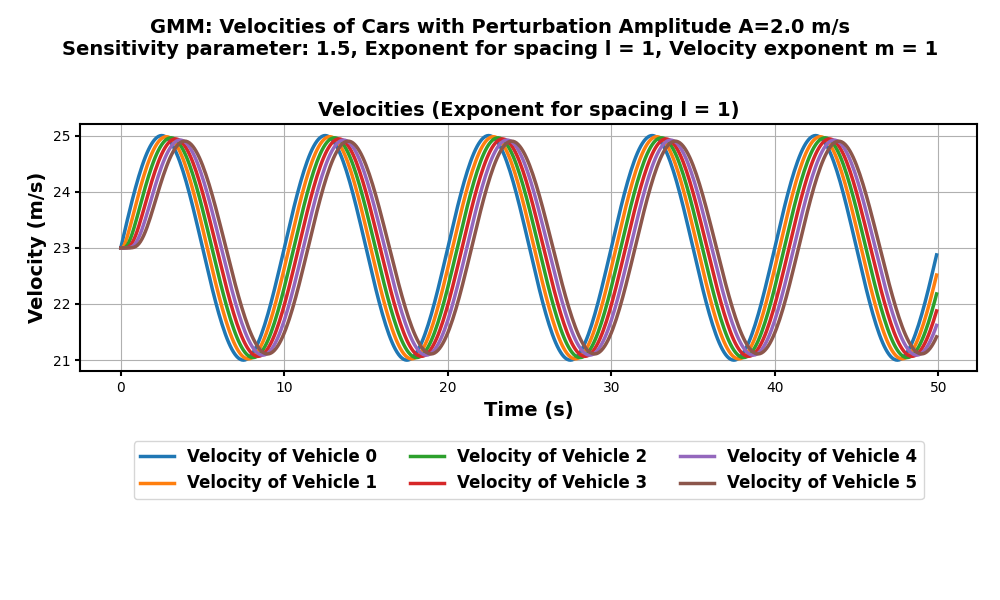}}
    \caption{GMM Configuration II: Simulation Results Following a Continuous Sinusoidal Perturbation with a Spacing Exponent $l = 1$.}
    \label{fig:gmm_l1}
\end{figure*} 

In this section, we analyze the classical GMM, which demonstrably outperforms Configuration I based on the simulation results. Nonetheless, the effectiveness of this model is highly dependent on the choice of its parameters; optimal performance is achieved only when these parameters are carefully calibrated. Therefore, to assess the model’s sensitivity, we examine its response for two different values of the spacing exponent $l = {1, 2}$ appearing in Eq.~(\ref{gmm}).

\begin{figure*}[htb!]
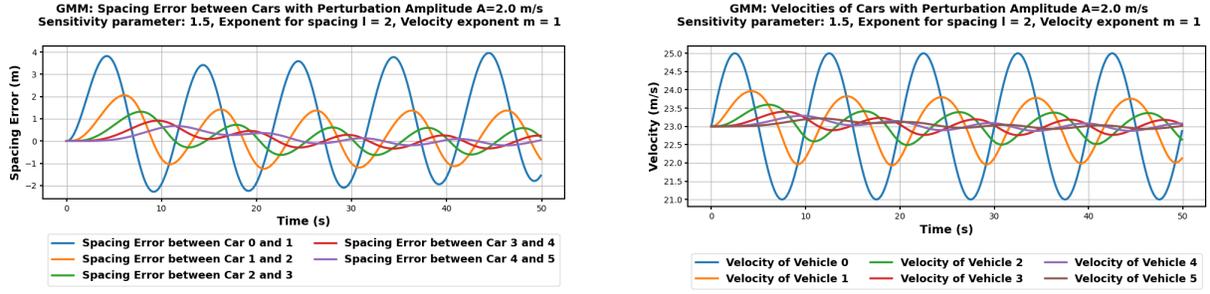

    \centering
    \subfloat[GMM Configuration II: Spacing Error between Vehicles after a Sinusoidal Perturbation with $l = 2$. \label{fig:gmm_l2_spac}]{
        \includegraphics[width=0.48\linewidth]{Figures/gmm_l_2_spac.png}}
    \hfill
    \subfloat[GMM Configuration II: Vehicles Velocity Profiles after a Sinusoidal Perturbation with $l = 2$.\label{fig:gmm_l2_vel}]{
        \includegraphics[width=0.48\linewidth]{Figures/gmm_l_2_vel.png}}
    \caption{GMM Configuration II: Simulation Results Following a Continuous Sinusoidal Perturbation with a Spacing Exponent $l = 2$.}
    \label{fig:gmm_l2}
    \vspace{-0.2cm}
\end{figure*}

Figures~\ref{fig:gmm_l1}\subref{fig:gmm_l1_spac} and~\ref{fig:gmm_l1}\subref{fig:gmm_l1_vel} show the spacing error and vehicle velocities for $l=1$, while Figures~\ref{fig:gmm_l2}\subref{fig:gmm_l2_spac} and~\ref{fig:gmm_l2}\subref{fig:gmm_l2_vel} correspond to $l=2$. For $l=1$, the spacing errors remain relatively small and the velocities fluctuate mildly, reflecting smooth follower responses and overall platoon stability. With $l=2$, however, spacing errors are larger and velocity variations are more pronounced, indicating that perturbations from the lead vehicle are amplified downstream. These results highlight how increasing the spacing exponent intensifies the vehicle response to distance differences, improving responsiveness but also increasing the risk of oscillation propagation.
\section{String Stability Discrete Perturbations: Experimental Study}
\subsection{IDM Response to Discrete Perturbations}
\begin{figure*}[htb!]
    \centering
    \subfloat[IDM: Spacing Error between Vehicles after a Discrete Sinusoidal Perturbation for a Duration of $5$ s. \label{fig:idm_disc_sp}]{
        \includegraphics[width=0.48\linewidth]{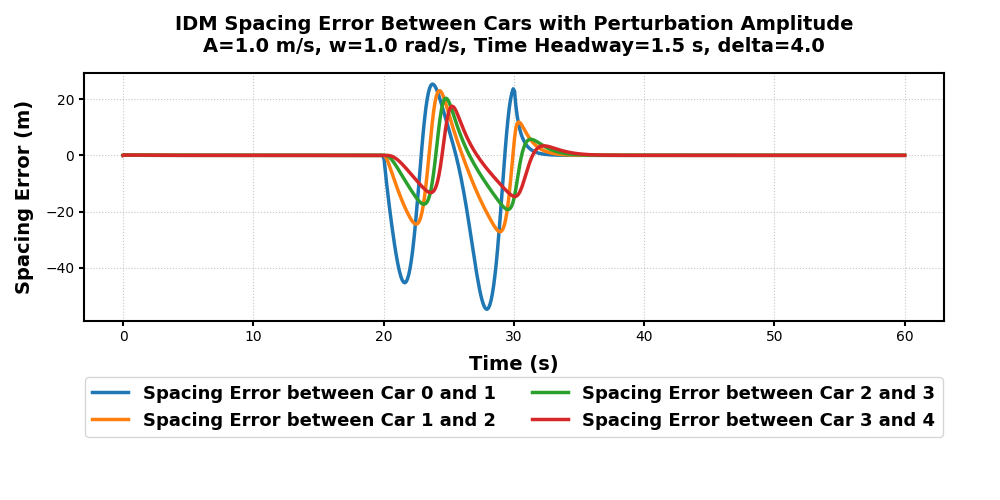}}
    \hfill
    \subfloat[IDM: Velocities between Vehicles after a Discrete Sinusoidal Perturbation for a Duration of $5$ s.\label{fig:idm_disc_vel}]{
        \includegraphics[width=0.48\linewidth]{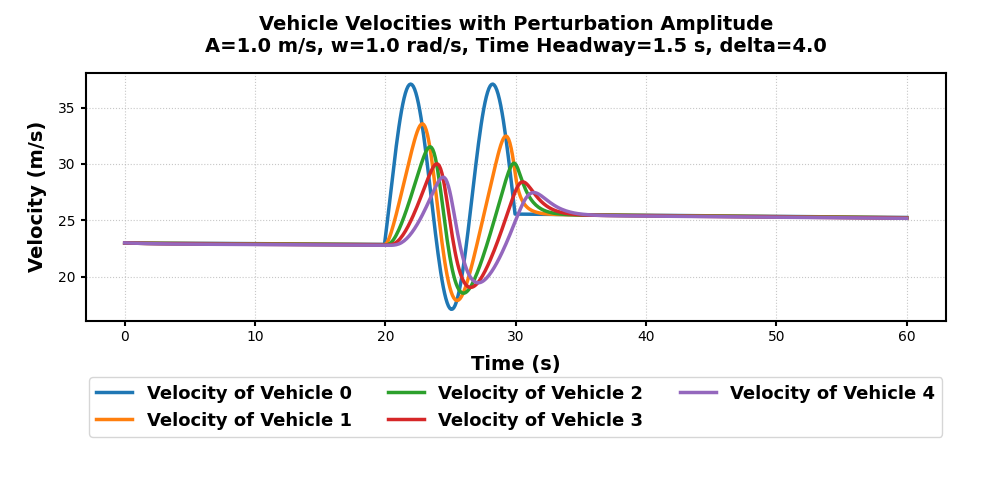}}
    \caption{IDM: Simulation Results after a Discrete Sinusoidal Perturbation for a Duration of $5$ s.}
    \label{fig:idm_disc}
\end{figure*}

The discrete perturbation response, depicted in the provided velocity and spacing error by Figures~\ref{fig:idm_disc}\subref{fig:idm_disc_sp} and~\ref{fig:idm_disc}\subref{fig:idm_disc_vel}, illustrates the dynamic behavior of a four-vehicle platoon subjected to a perturbation with an amplitude of $1.0$ m/s and a frequency of $1.0$ rad/s over a 60-second period. Initially, all vehicles maintain a stable velocity of approximately $25$ m/s, but around $20$-$30$ s, the perturbation triggers significant oscillations, with velocities fluctuating between $20$ m/s and $35$ m/s before gradually stabilizing. The spacing errors between consecutive vehicle pairs (Cars 0 and 1, Cars 1 and 2, Cars 2 and 3) also exhibit pronounced oscillations, ranging from $-5$ m to $+15$ m during the same interval, with the amplitude of these deviations increasing from the lead to the trailing vehicles, indicative of string instability. Influenced by a time headway of $1.5$ s and a delta of $4.0$, the system demonstrates a damping effect, returning to equilibrium after $40$ s, highlighting the propagation and eventual stabilization of the perturbation through the platoon.
\subsection{OVM Response to Discrete Perturbations}
\begin{figure*}[htb!]
    \centering
    \subfloat[OVM Model: Spacing Error between Vehicles after a Discrete Sinusoidal Perturbation for a Duration of $5$ s. \label{fig:ovm_disc_sp}]{
        \includegraphics[width=0.48\linewidth]{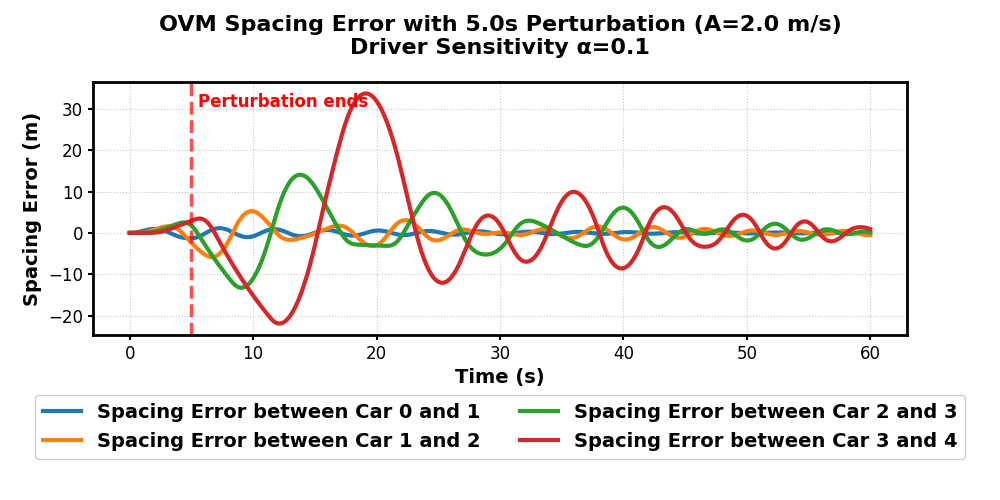}}
    \hfill
    \subfloat[OVM: Velocities between Vehicles after a Discrete Sinusoidal Perturbation for a Duration of $5$ s.\label{fig:ovm_disc_vel}]{
        \includegraphics[width=0.48\linewidth]{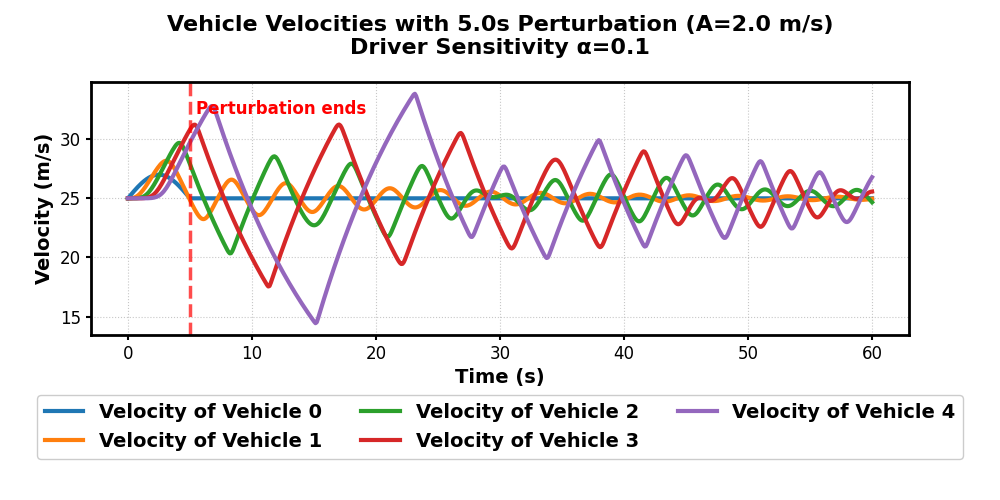}}
    \caption{OVM: Simulation Results after a Discrete Sinusoidal Perturbation for a Duration of $5$ s.}
    \label{fig:ovm_disc}
\end{figure*}

The discrete perturbation response for the four-vehicle platoon, as shown in Figure{~\ref{fig:ovm_disc}} with a $5.0$ s perturbation duration (amplitude $A = 2.0$ m/s) and driver sensitivity $\alpha$ = $0.1$, reveals the system's reaction over 60 s. The spacing error in Figure~\ref{fig:ovm_disc}\subref{fig:ovm_disc_sp} shows initial near-zero errors, followed by significant oscillations (up to $\pm20$ m) during the perturbation, with the errors between Cars 0 and 1 (blue) and Cars 3 and 4 showing the most pronounced deviations, gradually damping out after $30$ s. The velocity graph given by Figure~\ref{fig:ovm_disc}\subref{fig:ovm_disc_vel} indicates that all vehicles start near $25$ m/s, with oscillations beginning at the perturbation's onset, peaking around $10$-$20$ seconds, and continuing with decreasing amplitude after the perturbation ends at $5$ s, stabilizing post-$30$ s. The low driver sensitivity ($\alpha$ = 0.1) suggests a slow response, contributing to the sustained oscillations before stabilization.
\subsection{GMM Response to Discrete Perturbations}
\begin{figure*}[htb!]
    \centering
    \subfloat[GMM: Spacing Error between Vehicles after a Discrete Sinusoidal Perturbation for a Duration of $5$ s. \label{fig:gm_disc_sp}]{
        \includegraphics[width=0.48\linewidth]{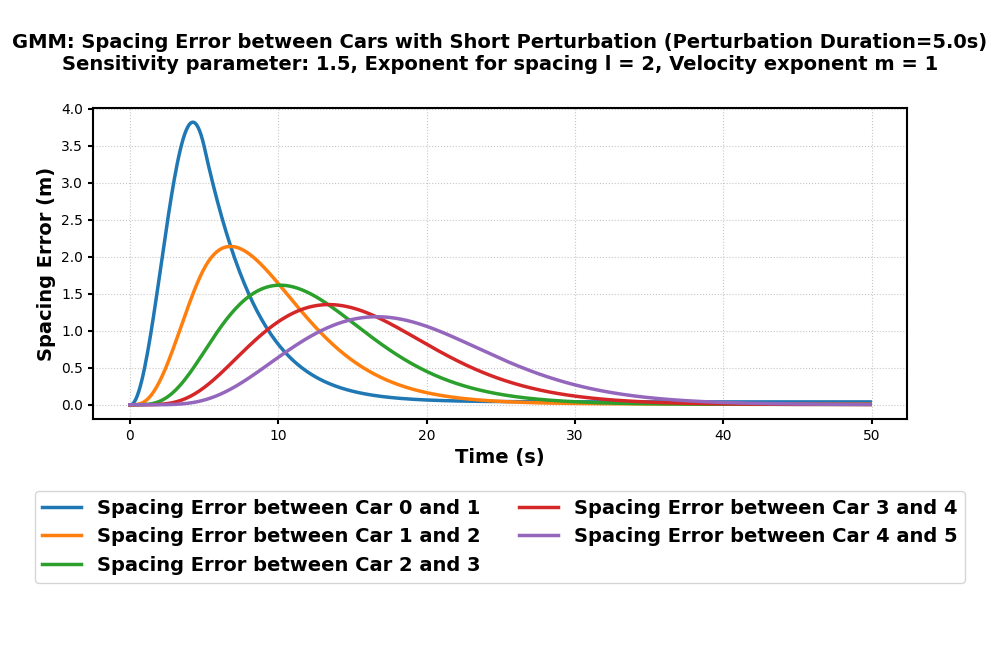}}
    \hfill
    \subfloat[GMM Model: Velocities between Vehicles after a Discrete Sinusoidal Perturbation for a Duration of $5$ s.\label{fig:gm_disc_vel}]{
        \includegraphics[width=0.48\linewidth]{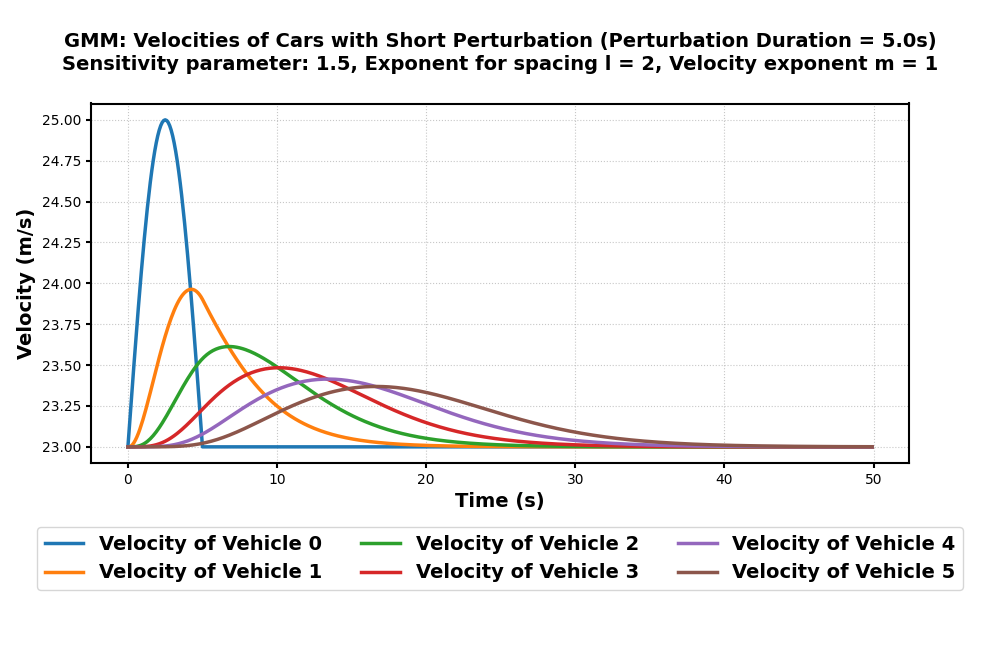}}
    \caption{GMM: Simulation Results after a Discrete Sinusoidal Perturbation for a Duration of $5$ s.}
    \label{fig:gm_disc}
\end{figure*}

The discrete perturbation response of the GMM is depicted by Figures~\ref{fig:gm_disc}\subref{fig:gm_disc_sp} and ~\ref{fig:gm_disc}\subref{fig:gm_disc_vel} with a perturbation duration of $5.0$ s (amplitude A = 2.0 m/s), a sensitivity parameter of $1.5$, spacing exponent $l = 2$, and velocity exponent $m = 1$. The figures illustrate the system's behavior over $50$ s. The spacing error graph in Figure~\ref{fig:gm_disc}\subref{fig:gm_disc_sp} indicates initial zero errors, with a peak deviation of about $3.5$ m between Cars 0 and 1 around $5$-$10$ s, followed by damped oscillations that stabilize near zero after $20$ s, with errors decreasing in amplitude from the front to the rear of the platoon. The velocity graph in Figure~\ref{fig:gm_disc}\subref{fig:gm_disc_vel} shows all vehicles starting at approximately $23.5$ m/s, with a sharp perturbation-induced peak around $5$-$10$ s (reaching up to $24.75$ m/s for Vehicle 0), followed by a rapid decline and stabilization near $23.0$ m/s after $20$ s, with the effect diminishing from the lead to the trailing vehicle, reflecting the influence of the sensitivity and exponent parameters on the system's stability and response.
\vspace{-0.5cm}
\subsection{CACC Response to Discrete Perturbations}
\begin{figure*}[htb!]
    \centering
    \subfloat[CACC Model: Spacing Error between Vehicles after a Discrete Sinusoidal Perturbation for a Duration of $25$ s. \label{fig:cacc_disc_sp}]{
        \includegraphics[width=0.48\linewidth]{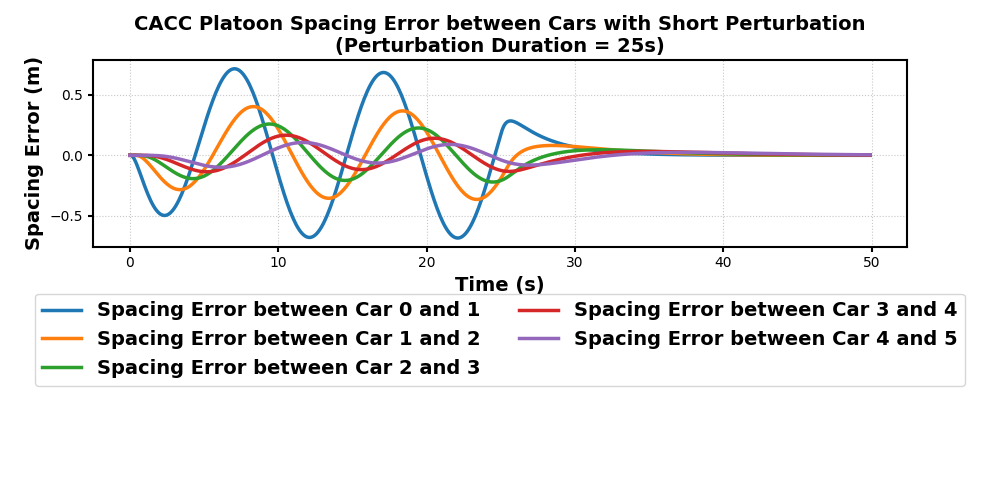}}
    \hfill
    \subfloat[CACC Model: Velocities between Vehicles after a Discrete Sinusoidal Perturbation for a Duration of $25$ s.\label{fig:cacc_disc_vel}]{
        \includegraphics[width=0.48\linewidth]{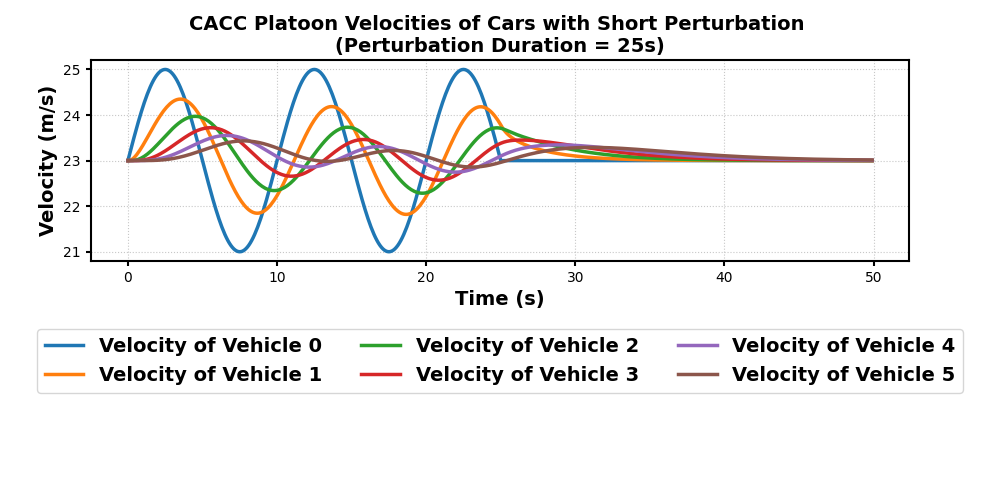}}
    \caption{CACC: Simulation Results after a Discrete Sinusoidal Perturbation for a Duration of $25$ s.}
    \label{fig:cacc_disc}
    \vspace{-0.2cm}
\end{figure*}

The discrete perturbation response for the five-vehicle platoon, as shown by Figures~\ref{fig:cacc_disc}\subref{fig:cacc_disc_sp} and ~\ref{fig:cacc_disc}\subref{fig:cacc_disc_vel} with a limited-time perturbation (amplitude $A = 2.0$ m/s), illustrates the system's dynamics over $50$ s. Figure~\ref{fig:cacc_disc}\subref{fig:cacc_disc_sp} depicts the spacing error for a limited-time perturbation (amplitude $A = 2.0$ m/s), and illustrates the system's dynamics over $50$ s. The spacing error graph shows initial near-zero errors, followed by oscillations peaking around $0.6$ m to $-0.6$ m between $5$-$20$ s, with the errors between Cars 0 and 1 exhibiting the largest deviations, damping out and stabilizing close to zero after $30$ s, indicating effective spacing control across the platoon as the perturbation effect diminishes. The velocity graph in Figure~\ref{fig:cacc_disc}\subref{fig:cacc_disc_vel} indicates that all vehicles start around $23.5$ m/s, with oscillations beginning at the perturbation's onset, peaking between $5$-$20$ s (reaching up to $24.5$ m/s for Vehicle 0), and gradually stabilizing near $23.0$ m/s after $30$ s, with the amplitude decreasing from the lead (Vehicle 0) to the trailing (Vehicle 5) vehicle.
\section{String Stability Non-Harmonic Perturbations: Experimental Study}
\subsection{IDM Response to Non-Harmonic Oscillations}
\begin{figure*}[htb!]
    \centering
    \subfloat[IDM: Spacing Error between Vehicles after a Non-Harmonic Perturbation. \label{fig:idm_nh_spac}]{
        \includegraphics[width=0.48\linewidth]{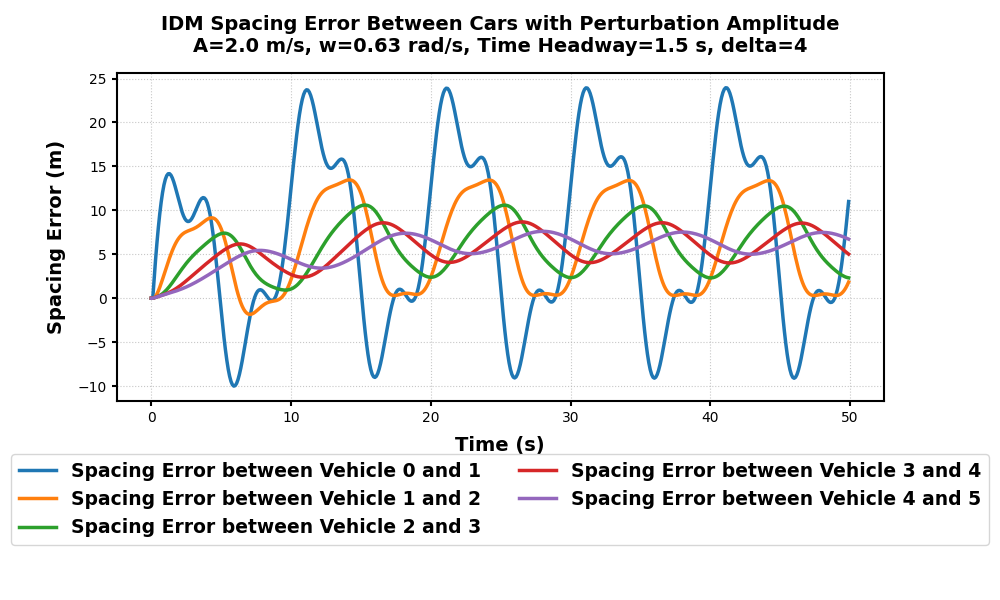}}
    \hfill
    \subfloat[IDM: Velocities between Vehicles after a Non-Harmonic Perturbation.\label{fig:idm_nh_vel}]{
        \includegraphics[width=0.48\linewidth]{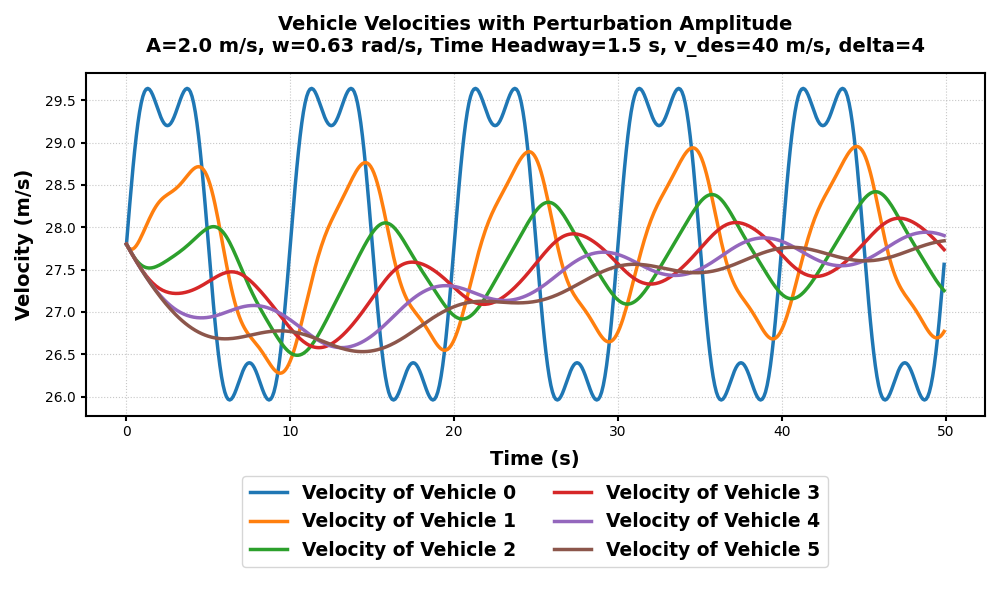}}
    \caption{IDM: Simulation Results after a Non-Harmonic Perturbation.}
    \label{fig:idm_nh}
    \vspace{-0.5cm}
\end{figure*}

Figures~\ref{fig:idm_nh}\subref{fig:idm_nh_spac} and~\ref{fig:idm_nh}\subref{fig:idm_nh_vel} demonstrate the response of the IDM to a non-harmonic perturbation with an amplitude of 2.0 m/s and a frequency of 0.63 rad/s. The velocity plot shows that the oscillations are consistent across vehicles, with a slight phase delay in downstream vehicles, confirming the propagation of the perturbation through the platoon. Meanwhile, the spacing error plot reveals persistent but stable oscillations between consecutive vehicles, suggesting that the IDM maintains string stability under these conditions—the perturbation does not amplify along the vehicle chain. However, the lack of complete damping implies that the system does not fully return to equilibrium, leaving residual fluctuations in both speed and spacing. This behavior highlights the trade-off in the IDM between stability and responsiveness, where the chosen parameters (time headway = $1.5 s$, $\delta$ = 4) prevent catastrophic growth of disturbances but do not eliminate them entirely.
\subsection{OVM Response to Non-Harmonic Oscillations}
\begin{figure*}[htb!]
    \centering
    \subfloat[OVM: Spacing Error between Vehicles after a Non-Harmonic Perturbation. \label{fig:ovm_nh_spac}]{
        \includegraphics[width=0.48\linewidth]{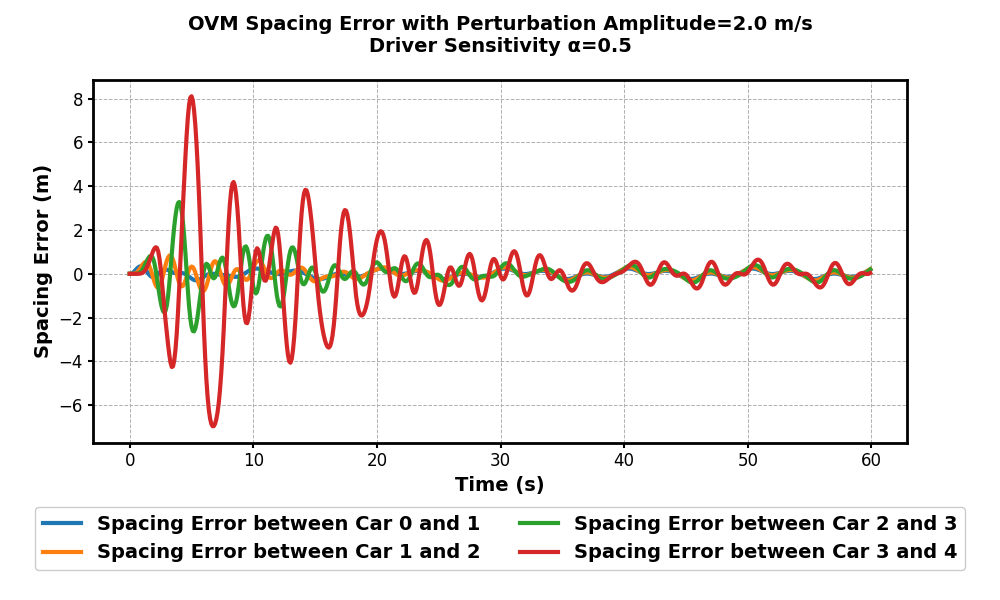}}
    \hfill
    \subfloat[OVM: Velocities between Vehicles after a Non-Harmonic Perturbation.\label{fig:ovm_nh_vel}]{
        \includegraphics[width=0.48\linewidth]{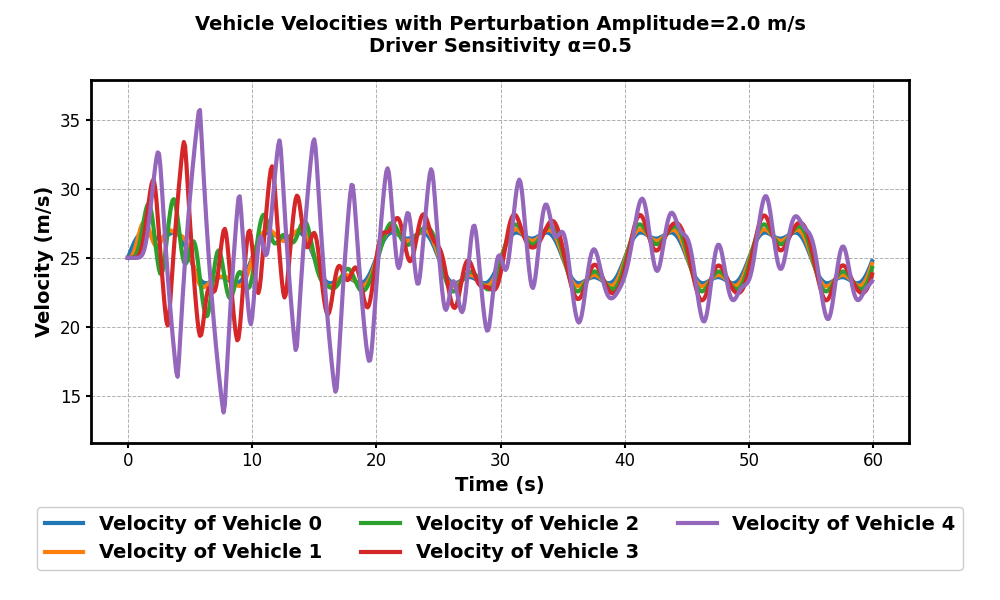}}
    \caption{OVM: Simulation Results after a Non-Harmonic Perturbation.}
    \label{fig:ovm_nh}
    \vspace{-0.5cm}
\end{figure*}

The graph shown by Figure~\ref{fig:ovm_nh} illustrates the response of the OVM to a non-harmonic perturbation with an amplitude of $2.0 \text{m/s}$ and a driver sensitivity parameter of $\alpha = 0.5$. In contrast to the IDM's lingering oscillations, the OVM demonstrates effective damping performance, with spacing errors fluctuating within a range of approximately $\pm4 \text{m}$, as shown in the OVM Spacing Error Figure~\ref{fig:ovm_nh}\subref{fig:ovm_nh_spac}, with a peak reaching $8 \text{m}$. Vehicle velocities, as depicted by Figure~\ref{fig:ovm_nh}\subref{fig:ovm_nh_vel}, remain relatively stable around $25 \text{m/s}$, with deviations reaching up to $35 \text{m/s}$ at times. The high driver sensitivity ($\alpha = 0.5$) plays a crucial role in this behavior, enabling aggressive correction of deviations while maintaining smooth traffic flow.
\subsection{GMM Response to Non-Harmonic Oscillations}
\begin{figure*}[htb!]
    \centering
    \subfloat[GMM: Spacing Error between Vehicles after a Non-Harmonic Perturbation. \label{fig:gm_nh_spac}]{
        \includegraphics[width=0.48\linewidth]{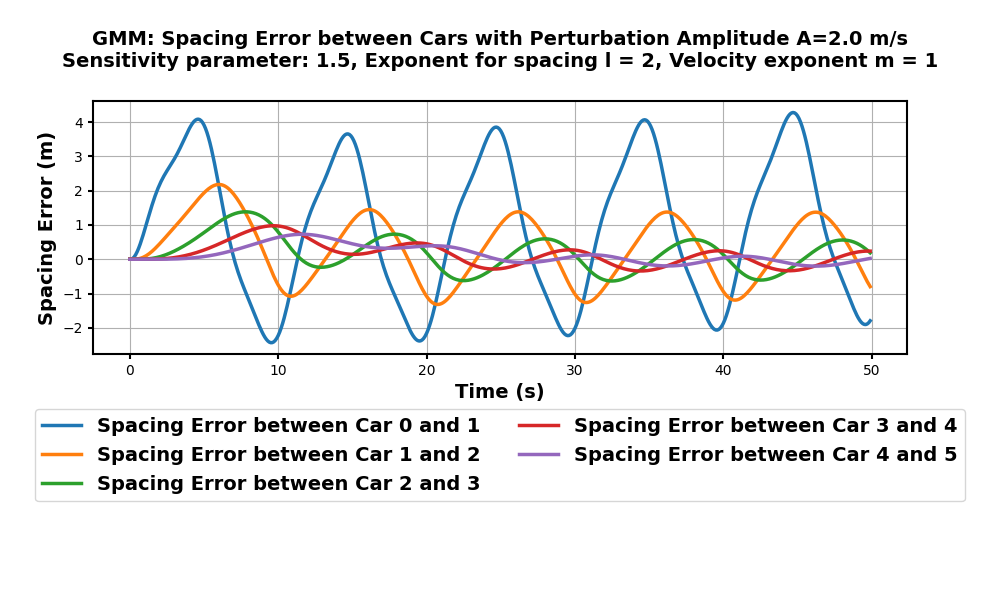}}
    \hfill
    \subfloat[GMM: Velocities between Vehicles after a Non-Harmonic Perturbation.\label{fig:gm_nh_vel}]{
        \includegraphics[width=0.48\linewidth]{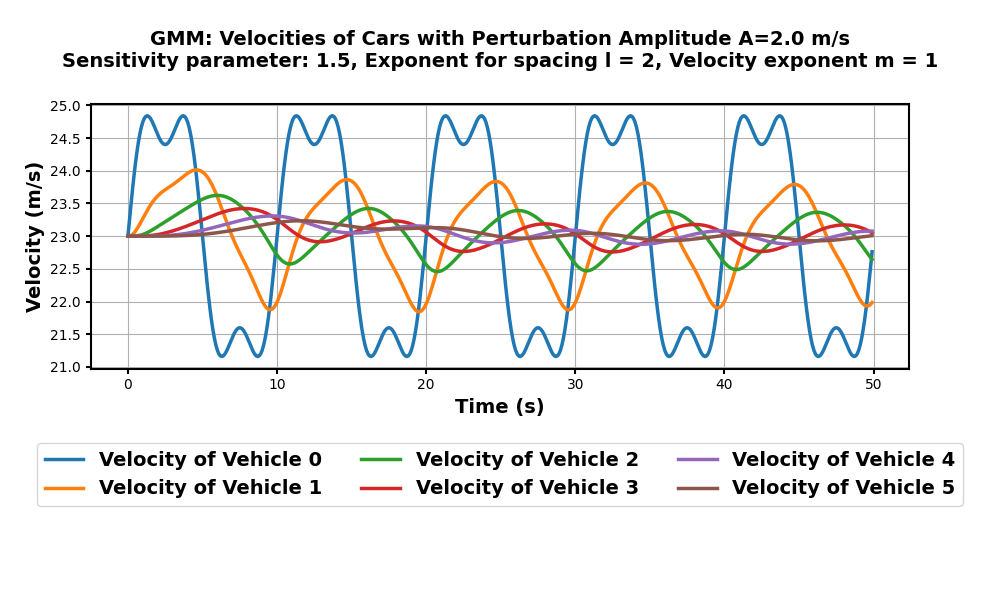}}
    \caption{GMM: Simulation Results after a Non-Harmonic Perturbation.}
    \label{fig:gm_nh}
\end{figure*}

The graphs depict the response of the GMM car-following model to a non-harmonic perturbation with an amplitude of $2.0$  m/s, using parameters including a sensitivity of $1.5$, spacing exponent $l=2$ and velocity exponent $m=1$. The spacing error plot in Figure~\ref{fig:gm_nh}\subref{fig:gm_nh_spac} shows deviations between consecutive vehicles (Cars 0-1, Cars 1-2, etc.), with fluctuations that remain bounded over time, indicating stable behavior without significant amplification of the perturbation. Meanwhile, the velocity plot in Figure~\ref{fig:gm_nh}\subref{fig:gm_nh_vel} reveals that all vehicles exhibit oscillations around a common speed range ($21-25$ m/s), with the perturbation propagating through the platoon while maintaining coherence. The GMM demonstrates string stability under these conditions, as neither spacing errors nor velocity deviations grow uncontrollably. However, the persistence of oscillations suggests that while the model effectively prevents disturbance amplification, it does not achieve perfect damping. This behavior highlights the model's trade-off between responsiveness and stability, with the chosen exponents and sensitivity parameter influencing its ability to mitigate perturbations.

\begin{figure*}[htb!]
    \centering
    \subfloat[CACC: Spacing Error between the Vehicles after a Non-Harmonic Perturbation with V2V Communication Delay of \(1.5\) s. \label{fig:cacc_space}]{
        \includegraphics[width=0.48\linewidth]{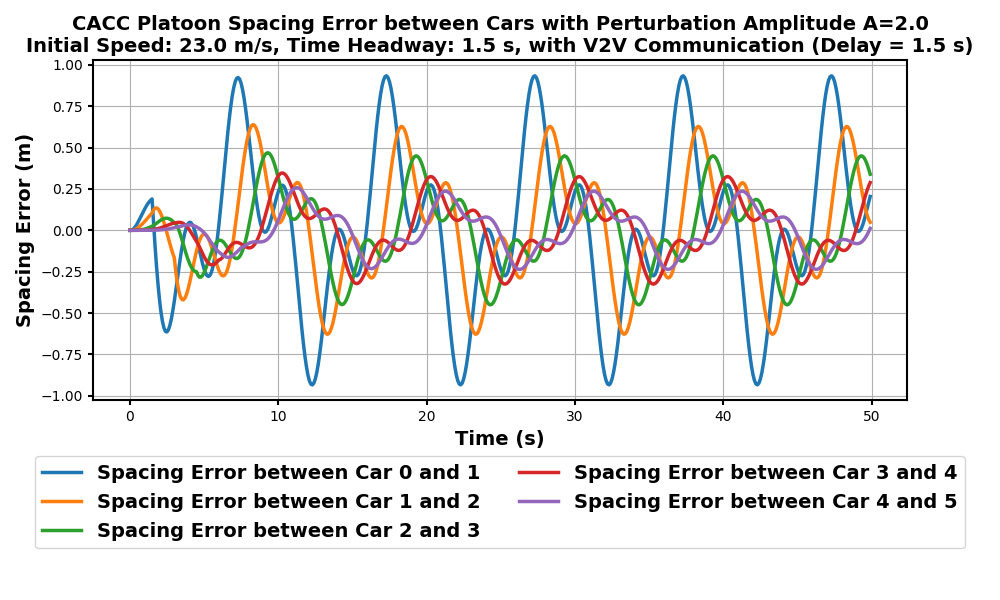}
    }
    \hfill
    \subfloat[CACC: Vehicles Velocities after a Non-Harmonic Perturbation with V2V Communication Delay of \(1.5\) s.\label{fig:cacc_vel}]{
        \includegraphics[width=0.48\linewidth]{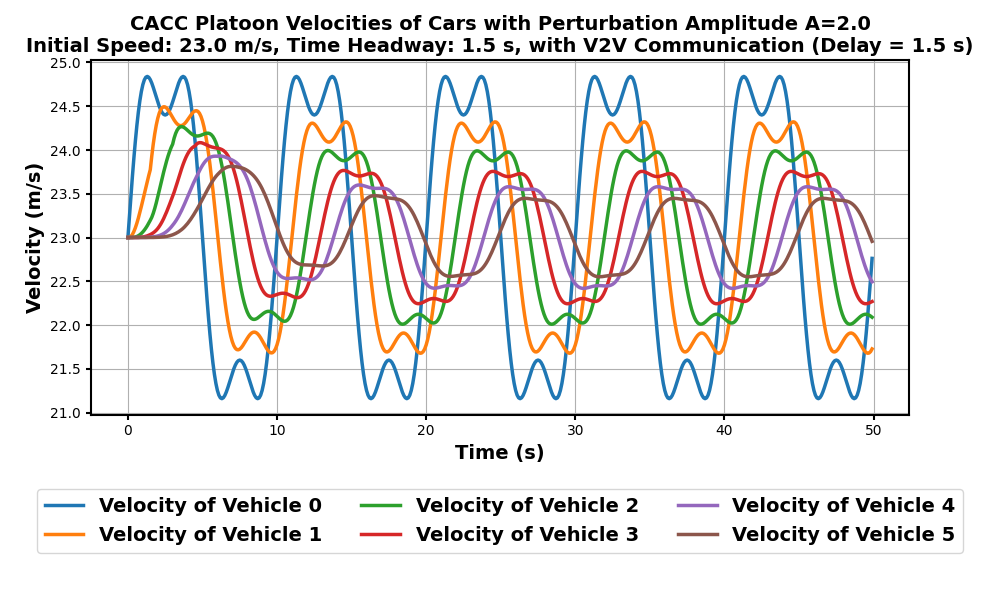}
    }
    \caption{CACC: Simulation Results after a Non-Harmonic Perturbation with V2V Communication Delay of \(1.5\) s}
    \label{fig:cacc_harm}
    \vspace{-0.5cm}
\end{figure*}
\subsection{CACC Response to Non-Harmonic Oscillations}
Non-harmonic oscillations—such as those resembling square or sawtooth waves—introduce complex frequency content composed of multiple Fourier components, which pose a significant challenge to car-following models such as CACC systems. Unlike simple harmonic disturbances, these signals demand that the control system manages a broader spectrum of frequencies without triggering instability or string amplification across the vehicle platoon. Additionally, if such oscillations originate from nonlinear dynamics—such as abrupt braking, actuator saturation, or tire slip—then linear control models typically used in CACC may become inadequate, resulting in degraded performance or safety risks. As depicted by Figure~\ref{fig:cacc_harm}, the velocity response plot in Figure~\ref{fig:cacc_harm}\subref{fig:cacc_vel} shows followers struggling to track the leader's abrupt changes, exhibiting overshoot and oscillations due to the $1.5s$ delay-induced phase lag, while the spacing error plot reveals dangerous downstream error amplification where errors grow progressively from lead to tail vehicles (Cars 0-1 to Cars 2-3), indicating clear string instability.

In this context, we apply a non-harmonic disturbance to the velocity of the lead vehicle in a five-car platoon to evaluate the performance and robustness of the CACC system in handling irregular, multi-frequency perturbations. The simulation results from Figure~\ref{fig:cacc_harm} reveal critical insights into the impact of non-harmonic oscillations (amplitude $A = 2.0$) on a 6-vehicle platoon operating under an initial speed of $23.0$ m/s, a time headway of $1.5$ s, and a V2V communication delay of $1.5$ s. Analyzing the spacing error propagation given by Figure~\ref{fig:cacc_harm}\subref{fig:cacc_space}, the errors between vehicles grow progressively worse from Cars 0–1 to Cars 2–3, indicating loss of string stability. In the velocity response depicted by Figure~\ref{fig:cacc_harm}\subref{fig:cacc_vel}, the leader (Vehicle 0) introduces non-harmonic perturbations, such as step changes or square-wave oscillations, causing followers (Vehicles 1–5) to exhibit overshoot and transient oscillations due to abrupt velocity changes. The $1.5$ s communication delay exacerbates this issue, creating a phase lag that amplifies spacing errors downstream. This amplification is driven by the non-harmonic nature of the perturbations, which excite higher-frequency modes and worsen errors.
\vfill

\end{document}